\documentclass[10pt,journal, compsoc]{IEEEtran}
\ifCLASSOPTIONcompsoc
\usepackage[nocompress]{cite}
\usepackage{amsmath,graphicx,mathrsfs,booktabs,type1cm,amssymb,threeparttable,multirow, booktabs, changepage, subfigure}
\else
\usepackage{cite}
\usepackage{amsmath,graphicx,mathrsfs,booktabs,type1cm,amssymb,threeparttable,multirow, booktabs, changepage, subfigure}
\fi
\usepackage{footnote}
\usepackage{color}
\usepackage{verbatim}
\usepackage{bm}
\usepackage{makecell, rotating}
\usepackage{enumerate}
\usepackage{amsthm}

\usepackage[ruled, linesnumbered]{algorithm2e}
\usepackage[colorlinks,linkcolor=blue]{hyperref}
\usepackage[table]{xcolor}

\ifCLASSINFOpdf
\else
\fi

\begin{document}
	
	\title{MERBench: A Unified Evaluation Benchmark \\ for Multimodal Emotion Recognition}

	\author{Zheng~Lian,~\IEEEmembership{}
		Licai~Sun,~\IEEEmembership{}
		Yong~Ren,~\IEEEmembership{}
		Hao~Gu,~\IEEEmembership{}
		Haiyang~Sun,~\IEEEmembership{}
		Lan~Chen,~\IEEEmembership{}
		Bin~Liu,~\IEEEmembership{}
		Jianhua~Tao~\IEEEmembership{} % <-this % stops a space

		% <-this % stops a space
		\IEEEcompsocitemizethanks{
			
			\IEEEcompsocthanksitem Zheng Lian, Lan Chen, and Bin Liu are with the State Key Laboratory of Multimodal Artificial Intelligence Systems, Institute of Automation, Chinese Academy of Sciences, Bejing, China.
			E-mail: lianzheng2016@ia.ac.cn, chenlan.2016.cn@gmail.com, liubin@nlpr.ia.ac.cn.
			\protect
			
			\IEEEcompsocthanksitem Licai Sun, Yong Ren, Hao Gu, and Haiyang Sun are with the School of Artificial Intelligence, University of Chinese Academy of Sciences, Beijing, China.
			E-mail: sunlicai2019@ia.ac.cn, renyong2020@ia.ac.cn, guhao2022@ia.ac.cn, sunhaiyang2021@ia.ac.cn.
			\protect
			
			\IEEEcompsocthanksitem Jianhua Tao is with Department of Automation, Tsinghua University, Bejing, China.
			E-mail: jhtao@tsinghua.edu.cn.
			\protect
		}
		
		% <-this % stops an unwanted space
		\thanks{Manuscript received \(\rm{xxxxxxxx}\); revised \(\rm{xxxxxxxx}\). (Corresponding author: Zheng Lian, Jianhua Tao)
		}
	}
	
	\markboth{}%
	{Shell \MakeLowercase{\textit{et al.}}: Bare Demo of IEEEtran.cls for Computer Society Journals}
	
	\IEEEtitleabstractindextext{%
		\begin{abstract}
			
			Multimodal emotion recognition plays a crucial role in enhancing user experience in human-computer interaction. Over the past few decades, researchers have proposed a series of algorithms and achieved impressive progress. Although each method shows its superior performance, different methods lack a fair comparison due to inconsistencies in feature extractors, evaluation manners, and experimental settings. These inconsistencies severely hinder the development of this field. Therefore, we build MERBench, a unified evaluation benchmark for multimodal emotion recognition. We aim to reveal the contribution of some important techniques employed in previous works, such as feature selection, multimodal fusion, robustness analysis, fine-tuning, pre-training, etc. We hope this benchmark can provide clear and comprehensive guidance for follow-up researchers. Based on the evaluation results of MERBench, we further point out some promising research directions. Additionally, we introduce a new emotion dataset MER2023, focusing on the Chinese language environment. This dataset can serve as a benchmark dataset for research on multi-label learning, noise robustness, and semi-supervised learning. We encourage the follow-up researchers to evaluate their algorithms under the same experimental setup as MERBench for fair comparisons. \textcolor[rgb]{0.93,0.0,0.47}{Our code is available at: https://github.com/zeroQiaoba/MERTools.}
			
		\end{abstract}
		
		}
	
	\maketitle
	\IEEEdisplaynontitleabstractindextext
	\IEEEpeerreviewmaketitle
	\IEEEraisesectionheading{\section{Introduction}\label{sec:introduction}}
	\IEEEPARstart{E}{motions} are complex and can be conveyed through multiple modalities, such as video, audio, and text \cite{picard2000affective}. Multimodal emotion recognition aims to integrate multi-source information to identify human emotional states \cite{zeng2007survey}. Due to its importance in human-computer interactions \cite{poria2017review, ma2020survey}, this task has become an active research topic.
	
	In recent years, researchers have proposed various algorithms and achieved noteworthy advancements \cite{kahou2016emonets, young2018recent}. Although each algorithm demonstrates its superior performance, it is difficult to fairly compare different algorithms due to their inconsistencies in feature extractors, evaluation manners, and experimental settings, which harms the development of this field. Take the feature extractor as an example. Different features contain distinct emotion-related information. Some researchers focus on handcrafted features \cite{schuller2010interspeech, zadeh2018multi}, while others exploit deep features \cite{lian2022smin, amiriparian2023muse}. Meanwhile, different deep features also perform differently in emotion recognition \cite{sun2020multi, lian2021investigation}. In addition to features, there are also some differences in evaluation methods. Taking one of the widely used datasets IEMOCAP \cite{busso2008iemocap} as an example, some use the leave-one-session-out strategy \cite{yang2021superb, lian2023gcnet} while some train on the first four sessions and evaluate on the last session \cite{poria2017context, majumder2019dialoguernn}. These inconsistencies make it difficult to directly compare the performance of different algorithms.
	
	Therefore, we build MERBench, a unified evaluation benchmark for multimodal emotion recognition. This benchmark involves primary datasets, features, and multimodal fusion strategies in this field. Under the same experimental setup, we explore some key problems, including how to select appropriate features for different datasets, how to determine multimodal fusion strategies, how to obtain better performance in cross-corpus settings, how to improve noise robustness, what is the impact of missing punctuation, whether the feature extractor needs to be pre-trained to adapt to downstream tasks, whether the feature extractor needs to be further fine-tuned with the classifier, etc. This paper aims to provide a comprehensive understanding of multimodal emotion recognition and provide guidance for follow-up researchers. Meanwhile, we hope that subsequent works can adopt the same experimental setup as our MERBench for fair comparisons.
	
	Additionally, this paper proposes a new Chinese emotion dataset MER2023. Compared with existing datasets \cite{yu2020ch, zhao2022m3ed}, we aim to provide a benchmark dataset for research on multi-label learning \cite{bendjoudi2021multi}, noise robustness \cite{lian2023gcnet}, and semi-supervised learning \cite{latif2020multi}, since these directions are currently hot topics in this field. Specifically, MER2023 contains three subsets: a multi-label subset for studying discrete and dimension label correlations, a noisy subset for evaluating noise robustness, and an unlabeled subset for studying semi-supervised learning. The main contribution of this paper can be summarized as follows:
	\begin{itemize}
		\item We build a unified evaluation benchmark for multimodal emotion recognition. To the best of our knowledge, this is the most comprehensive benchmark in this field, covering feature selection, multimodal fusion, cross-corpus performance, robustness analysis, language sensitivity analysis, etc. In this benchmark, we reproduce different methods under the same experimental setup for a fair comparison.
		
		\item We build MER2023, a Chinese emotion dataset designed to serve as a benchmark for evaluating multi-label learning, noise robustness, and semi-supervised learning in multimodal emotion recognition.
		
		\item Based on the evaluation results of MERBench, we point out some promising research directions in this field. We will also open-source the code to facilitate follow-up researchers to evaluate their algorithms under the same experimental setting as MERBench.
	\end{itemize}
	
	The rest of this paper is organized as follows: In Section \ref{sec2-related}, we briefly review some recent works. In Section \ref{sec3-mer2023} and Section \ref{sec4-merbaseline}, we introduce MER2023 and establish a baseline for this dataset. In Section \ref{sec5-merbench}, we build MERBench, an evaluation benchmark for multimodal emotion recognition. Finally, we conclude this paper and discuss future work in Section \ref{sec6-conclusion}.

	\section{Related Works}
	\label{sec2-related}
	This paper covers various research topics. In this section, we primarily review recent works related to corpus design, feature selection, and multimodal fusion.
	
	\subsection{Emotional Corpus}	
	Emotional corpora are the basis for training and evaluating emotion recognition algorithms. Current datasets gradually transition from laboratory-controlled to more challenging in-the-wild conditions. For example, the early dataset like IEMOCAP \cite{busso2008iemocap} is laboratory-controlled and contains conversations between two actors performing improvised or scripted scenarios. To mimic real-world conditions, researchers have directed their attention towards in-the-wild datasets, which are mainly collected from online websites, movies, and TV shows (e.g., CMU-MOSI \cite{zadeh2017tensor}, CMU-MOSEI \cite{zadeh2018multimodal}, MELD \cite{poria2019meld}, CH-SIMS \cite{yu2020ch}, and CH-SIMS v2 \cite{liu2022make}).
	
	Recent research mainly involves multi-label learning, semi-supervised learning, and noise robustness, which puts forward new requirements for dataset design. 1) Discrete and dimensional emotions are closely related \cite{hamann2012mapping}. For example, valence is a dimensional emotion that reflects the degree of pleasure \cite{wundt1912introduction}. For negative emotions (such as \emph{anger} and \emph{sadness}), the valence score should be less than 0; for positive emotions (such as \emph{happiness}), the valence score should be greater than 0 \cite{barrett1998discrete}. To study the impact of multi-label correlation, the dataset is required to provide both discrete and dimensional annotations; 2) Due to the high annotation cost, it is difficult to collect large amounts of samples with emotional labels. But training with limited data harms the generalization ability. To solve this problem, researchers exploit various pre-trained models, but they mainly focus on action recognition rather than facial expressions \cite{tong2022videomae}. This domain gap harms the performance of transfer learning \cite{weiss2016survey}. Therefore, the dataset is required to provide large amounts of unlabeled human-centered videos; 3) Noise and missing conditions increase the difficulty of emotion recognition \cite{parthasarathy2020training, ma2021smil}. However, due to the lack of benchmark datasets, existing works mainly rely on their own simulated missing data \cite{zhao2021missing, zhang2022deep}. For a fair comparison, a benchmark test set that focuses on realistic missing conditions is needed.
	
	Among existing datasets, some provide multi-label samples (such as IEMOCAP \cite{busso2008iemocap} and CMU-MOSEI \cite{zadeh2018multimodal}), some provide unlabeled samples (such as CH-SIMS v2 \cite{liu2022make}), and none of them provides a noise-corrupted test set. Therefore, we introduce MER2023, aiming to provide a benchmark dataset that can study these three directions simultaneously.

	\subsection{Unimodal Features}
	Different features lead to distinct results. To guide feature selection, we evaluate the performance of different features under the same experimental setup. This paper attempts to cover all typical features of each modality.
	
	\textbf{Acoustic Modality.}
	Paralinguistic information is important for understanding emotions \cite{chandrasekar2014automatic}. Previously, researchers explored various low-level descriptors (such as energy, duration, and pitch) and then used statistical functions to extract high-level features (e.g., IS09 \cite{schuller2009interspeech}, IS10 \cite{schuller2010interspeech}, and eGeMAPS \cite{trigeorgis2016adieu}). Besides handcrafted features, self-supervised models trained on large amounts of unlabeled data can also learn universal acoustic representations \cite{mohamed2022self}. Among them, wav2vec and wav2vec 2.0 are widely used due to their competitive performance. wav2vec \cite{schneider2019wav2vec} is a multi-layer convolutional network trained on the contrastive task, aiming to distinguish real future audio from negatives. Unlike wav2vec, wav2vec 2.0 \cite{baevski2020wav2vec} uses quantized representations for self-supervised learning. HUBERT \cite{hsu2021hubert} extends wav2vec 2.0 and exploits an offline clustering module to provide more accurate quantized features.

	\textbf{Lexical Modality.}
	The text also contains emotion-related clues \cite{hazarika2018icon, majumder2019dialoguernn}. Language models can learn universal lexical representations, which are beneficial for many tasks \cite{qiu2020pre}. Among all language models, BERT \cite{devlin2018bert} and its variants are widely utilized. BERT uses the ``masked language model (MLM)'' and ``next sentence prediction (NSP)'' objectives to learn word embeddings. MLM randomly masks words and attempts to predict the masked items based on their context; NSP tries to capture sentence relationships by distinguishing whether two sentences are continuous or not. To further improve its performance, RoBERTa \cite{liu2019roberta} proposes to remove the NSP objective in BERT and train with dynamic masking. XLNet \cite{yang2019xlnet} overcomes the pretrain-finetune discrepancy of BERT via autoregressive pre-training. DeBERTa \cite{he2020deberta} improves BERT by disentangling attention matrices on the content and position. To reduce computational costs, ALBERT \cite{lan2020albert} exploits parameter reduction techniques, while ELECTRA \cite{clark2020electra} replaces MLM with a compute-efficient task. Recently, large language models (LLMs) with massive parameters have attracted increasing attention due to their unprecedented performance. Therefore, we further evaluate some representative open-sourced LLMs, such as Llama \cite{touvron2023llama}, Falcon \cite{penedo2023refinedweb}, and BLOOM \cite{workshop2022bloom}. Due to GPU memory limitations, we evaluate the maximum 13B version of LLMs.

	\textbf{Visual Modality.} 
	Facial expressions are natural signals to convey emotions \cite{tian2001recognizing}. Compared with handcrafted features, deep features extracted from supervised models are useful for facial expression recognition \cite{li2020deep}. These models depend on the model structure (e.g., ResNet \cite{he2016deep}, SENet \cite{hu2018squeeze}, and MA-Net \cite{zhao2021learning}) and the training corpus (e.g., MS-Celeb-1M \cite{guo2016ms}, FER2013 \cite{goodfellow2013challenges}, and RAF-DB \cite{li2017reliable}). Different combinations will result in distinct performance. Recently, visual features extracted from weakly-supervised or self-supervised models have achieved remarkable results on many tasks \cite{jing2020self, schiappa2023self}. For example, CLIP \cite{radford2021learning} is a weakly-supervised model that learns universal representations by predicting correct image-text pairs. To further improve its performance, DINOv2 \cite{oquab2023dinov2} combines different techniques to increase data and model size. Besides weakly-supervised models, self-supervised models (such as VideoMAE \cite{tong2022videomae}) can also serve as powerful visual encoders.

	\subsection{Multimodal Fusion}
	Multimodal fusion aims to integrate multi-view cues to produce more emotionally discriminative features \cite{atrey2010multimodal, gandhi2023multimodal}. Current methods can be roughly divided into utterance-level and sequence-level algorithms. This paper evaluates some representative algorithms in these two categories.
	
	\textbf{Utterance-level Fusion.} An intuitive idea is to compress different modalities to the unified utterance level and then fuse them. For example, TFN \cite{zadeh2017tensor} utilizes a 3-fold Cartesian product to aggregate utterance-level features. To reduce computational complexity, LMF \cite{liu2018efficient} employs low-rank decomposition to approximate the high-order tensor in TFN. To address multimodal heterogeneity, MISA \cite{hazarika2020misa} decomposes each modality into modality-invariant and modality-specific features. MMIM \cite{han2021improving} reduces the loss of emotion-related information by maximizing mutual information.
	
	\textbf{Sequence-level Fusion.} Although utterance-level fusion achieves superior performance, compressing features to the utterance level inevitably loses some important information. Consequently, researchers propose sequence-level fusion algorithms. For example, MFN \cite{zadeh2018memory} introduces a multi-view sequence learning framework to capture both view-specific and cross-view interactions. GMFN \cite{zadeh2018multimodal} extends MFN by applying a dynamical fusion graph, placing more emphasis on important modalities. MCTN \cite{pham2019found} learns joint representations through cyclic translation from source to target feature sequences. To handle multimodal heterogeneity, MFM \cite{tsai2018learning} factorizes features into multimodal discriminative factors and modality-specific factors. To capture long-term cross-modal interactions, MulT \cite{tsai2019multimodal} uses the Transformer architecture to dynamically align different modalities.

	\section{MER2023 Dataset}
	\label{sec3-mer2023}
	This paper introduces a new Chinese emotion dataset, MER2023, consisting of four subsets: Train$\&$Val for training and validation, and MER-MULTI, MER-NOISE, MER-SEMI for testing. In this section, we provide an in-depth description of the data collection, annotation, and splitting. Statistics for each subset are shown in Table \ref{Table1}.

	\subsection{Data Collection}
	\label{sec:3-1}
	
	\textbf{Video Clip Generation.}
	We collect a large number of movies and TV series from the Internet and split them into video clips. Considering that subtitles have relatively accurate timestamps, we try to use them for video segmentation. However, we observe that only a few videos have subtitle tracks and we use FFmpeg\footnote{\emph{https://ffmpeg.org/}} to separate these tracks. For other videos, we first use the optical character recognition (OCR) tool, EasyOCR\footnote{\emph{https://github.com/JaidedAI/EasyOCR}}, but processing all frames using OCR is time-consuming. Then, we turn to the automatic audio recognition (ASR) tool, Capcut\footnote{\emph{https://www.capcut.cn/}}. However, it only supports manual video upload and does not provide a command line for automatic processing. Finally, we use the voice activity detection (VAD) tool, Silero VAD\footnote{\emph{https://github.com/snakers4/silero-vad}}, and split videos based on the presence or absence of human speech. To make video clips contain relatively complete content, we further use Deep Speaker\footnote{\emph{https://github.com/philipperemy/deep-speaker}} to measure speaker similarity and merge consecutive clips from the same speaker.
	
	\textbf{Video Filter.}
	We use two filters to remove difficult-to-label video clips. (1) Videos that are too short may not contain complete content to express emotions; videos that are too long may contain compound emotions, weakening the consistency of annotations. Therefore, we remove clips that are too long or too short. (2) For multi-speaker videos, it is necessary to eliminate the interference of other speakers during annotation. For convenience, we only select single-speaker clips. Specifically, we first use the face detection tool, YuNet\footnote{\emph{https://github.com/ShiqiYu/libfacedetection}}, to ensure that most frames contain only one face. Then, we use the face recognition tool, face.evoLVe\footnote{\emph{https://github.com/ZhaoJ9014/face.evoLVe}}, to ensure that most faces belong to the same person.

	\begin{table}[t]
		\centering
		\caption{Statistics of the MER2023 dataset.}
		\label{Table1}
		\begin{tabular}{l|cc|c}
			\hline
			\multirow{2}{*}{Partition} & \multicolumn{2}{c|}{\# of samples}  & \multirow{2}{*}{Annotation} \\
			& labeled & unlabeled &\\
			\hline \hline
			Train$\&$Val 	& 3373 	& 0 	& Discrete, Valence \\ 
			\hline
			MER-MULTI 		& 411 	& 0 	& Discrete, Valence \\ 
			MER-NOISE 		& 412 	& 0 	& Discrete, Valence \\ 
			MER-SEMI 		& 834 	& 73148 & Discrete \\ 
			\hline
		\end{tabular}
	\end{table}

	\begin{figure*}[t]
		\centering
		\includegraphics[width=\linewidth]{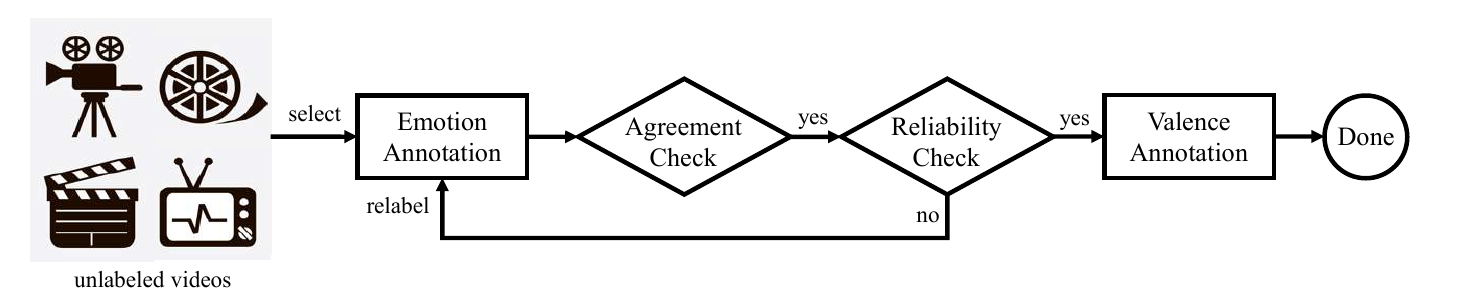}
		\caption{Pipeline of data annotation.}
		\label{Figure1}
	\end{figure*}

	\begin{figure}[t]
		\begin{center}
			\subfigure[Train$\&$Val]{
				\label{Figure2-1}
				\centering
				\includegraphics[width=0.22\linewidth, trim=30 0 30 0]{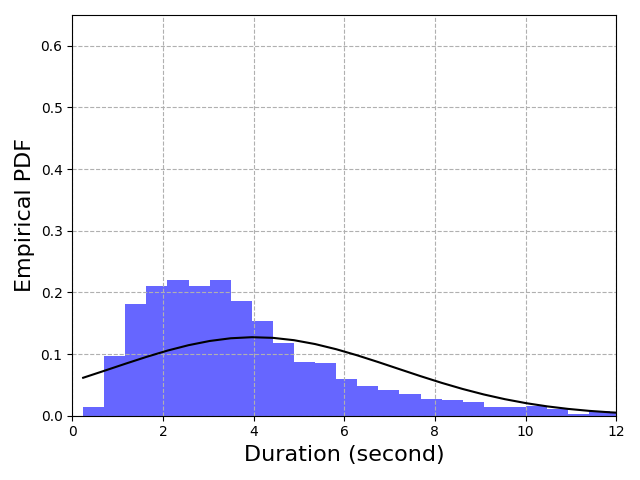}
			} 
			\subfigure[MULTI]{
				\label{Figure2-2}
				\centering
				\includegraphics[width=0.22\linewidth, trim=30 0 30 0]{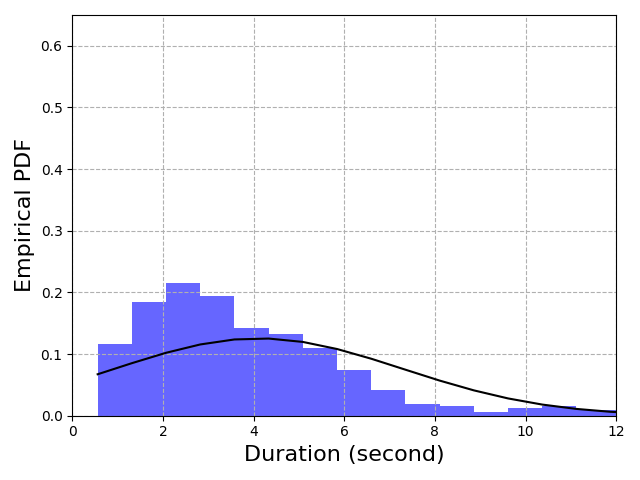}
			}
			\subfigure[NOISE]{
				\label{Figure2-3}
				\centering
				\includegraphics[width=0.22\linewidth, trim=30 0 30 0]{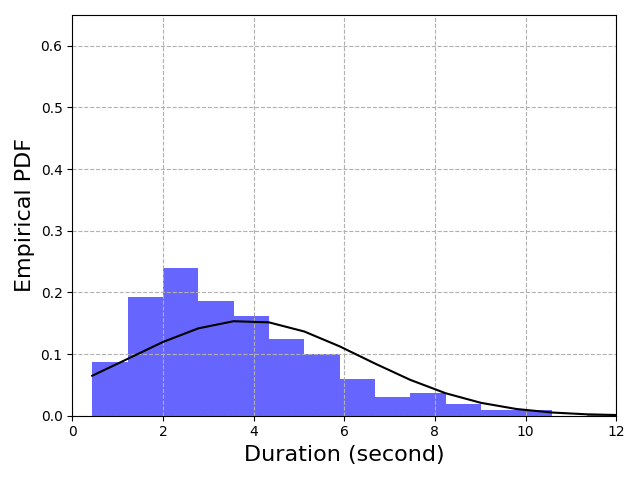}
			} 
			\subfigure[SEMI]{
				\label{Figure2-4}
				\centering
				\includegraphics[width=0.22\linewidth, trim=30 0 30 0]{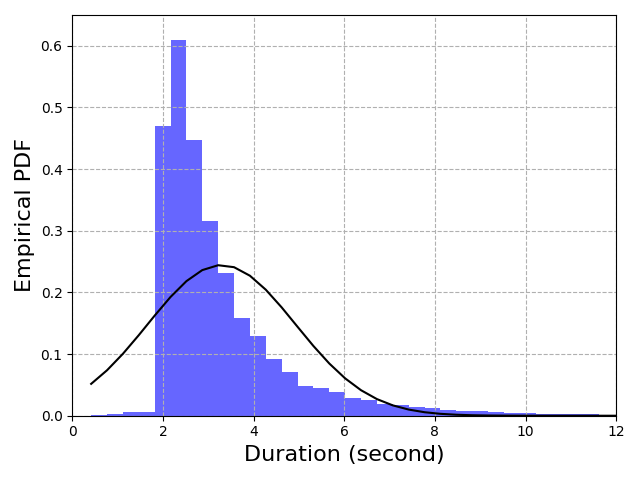}
			}
		\end{center}
		\caption{Empirical PDFs and estimated Gaussian models on sample lengths for different subsets.}
		\label{Figure2}
	\end{figure}

	\subsection{Data Annotation}
	\label{sec:3-2}
	The labeling process includes four steps: emotion annotation, agreement check, reliability check, and valence annotation. Fig. \ref{Figure1} shows the entire annotation pipeline.
	
	\textbf{Emotion Annotation.}
	Each video clip is labeled by at least three annotators using six categories: \emph{neutral}, \emph{anger}, \emph{happiness}, \emph{sadness}, \emph{worry}, and \emph{surprise}. If neither of them accurately describes the emotional state, two special categories are also provided (i.e., \emph{other} and \emph{unjudgeable}). Since most annotators specialize in affective computing, we have confidence in the trustworthiness of their annotation results.
	
	\textbf{Agreement Check.}
	To further enhance the annotation quality, we adopt a majority voting strategy. For samples without major emotions or with the major emotion belonging to \emph{other} or \emph{unjudgeable}, we treat them as unlabeled data.
	
	\textbf{Reliability Check.}
	In the reliability check, each (video, label) pair is labeled by two annotators using three categories: \emph{reliable}, \emph{unreliable}, and \emph{unjudgeable}. To pass this check, all annotators should assign \emph{reliable} to a sample. For unpassed samples, we relabel them since they have a certain degree of agreement. If these samples still fail to pass these checks after relabeling, they are treated as unlabeled data.
	
	\textbf{Valence Annotation.}
	For reliable samples, we hire six annotators to label valence using the self-assessment manikin \cite{bradley1994measuring}. To obtain high-quality annotations, we exclude the lowest and highest values and calculate the average of the remaining four annotations as the final valence score.

	\subsection{Data Splitting}
	\label{sec:3-3}
\begin{figure}[t]
		\begin{center}
			\subfigure[Train$\&$Val]{
				\label{Figure3-1}
				\centering
				\includegraphics[width=0.22\linewidth, trim=36 20 36 20]{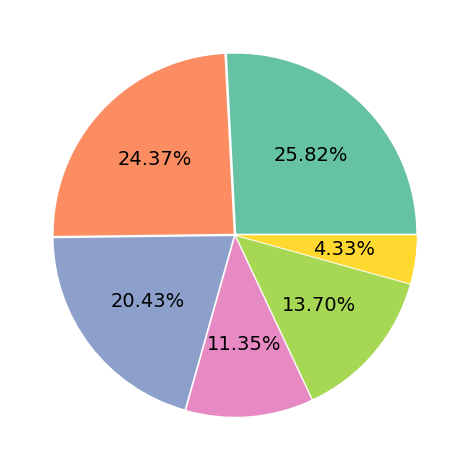}
			} 
			\subfigure[MULTI]{
				\label{Figure3-2}
				\centering
				\includegraphics[width=0.22\linewidth, trim=36 20 36 20]{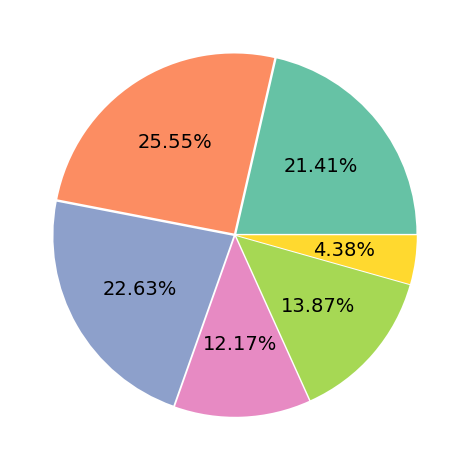}
			}
			\subfigure[NOISE]{
				\label{Figure3-3}
				\centering
				\includegraphics[width=0.22\linewidth, trim=36 20 36 20]{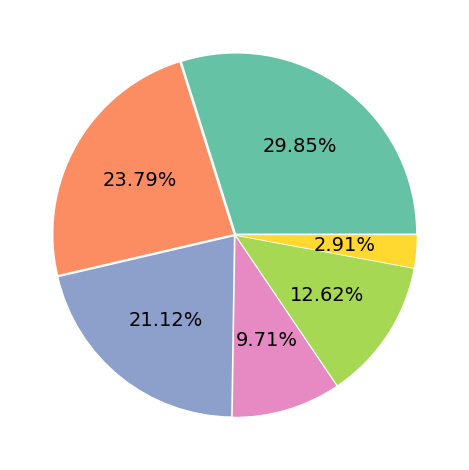}
			} 
			\subfigure[SEMI]{
				\label{Figure3-4}
				\centering
				\includegraphics[width=0.22\linewidth, trim=36 20 36 20]{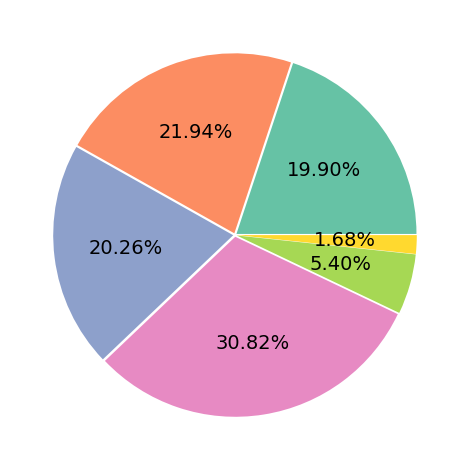}
			}
		\end{center}
		\caption{Distribution of discrete emotions for different subsets (\textcolor[rgb]{0.18039216, 0.45490196, 0.36862745}{neutral}, \textcolor[rgb]{0.76862745, 0.23137255, 0.01568627}{anger}, \textcolor[rgb]{0.25098039, 0.34509804, 0.54901961}{happiness}, \textcolor[rgb]{0.8, 0.16470588, 0.55686275}{sadness}, \textcolor[rgb]{0.38823529, 0.54901961, 0.1254902}{worry}, \textcolor[rgb]{0.72156863, 0.58431373, 0}{surprise}).}
		\label{Figure3}
	\end{figure}

	Due to the large amounts of collected samples and the high cost of emotion annotation, we only select a subset of samples for annotation (see Fig. \ref{Figure1}). The reliable samples are divided into three subsets: Train$\&$Val, MER-MULTI, and MER-NOISE. The remaining samples are treated as unlabeled data and form MER-SEMI. For MER-SEMI, we conduct additional annotation for samples likely to exhibit well-defined emotions (see Algorithm \ref{alg-1}). Statistics of these subsets are shown in Fig. \ref{Figure2} and Fig. \ref{Figure3}. We observe that most samples have durations ranging from 2 to 6 seconds. Meanwhile, the distribution of emotions is not well-balanced, with higher proportions of \emph{neutral}, \emph{anger}, \emph{happiness}, and \emph{sadness}, similar to previous datasets \cite{poria2019meld, zhao2022m3ed}.

	\begin{figure}[t]
		\begin{center}
			\subfigure[neutral]{
				\label{Figure4-1}
				\centering
				\includegraphics[width=0.3\linewidth, trim=15 0 15 0]{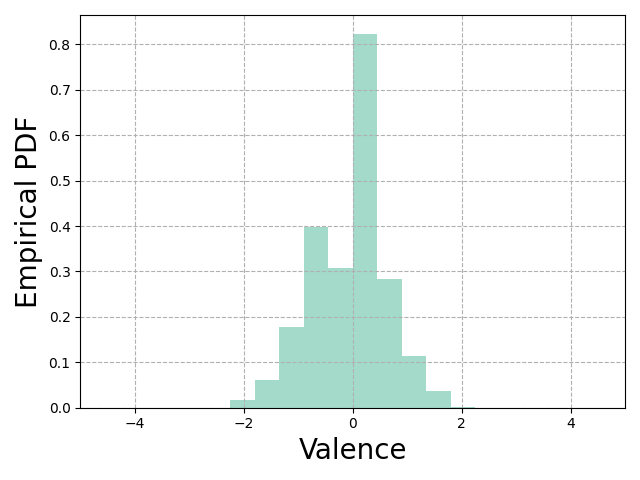}
			} 
			\subfigure[anger]{
				\label{Figure4-2}
				\centering
				\includegraphics[width=0.3\linewidth, trim=15 0 15 0]{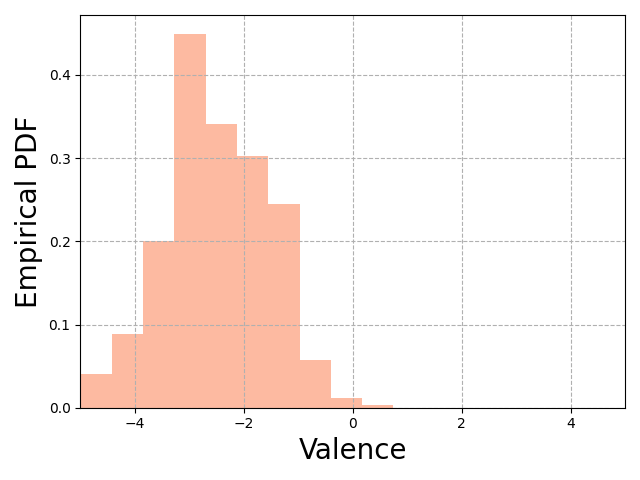}
			}
			\subfigure[happiness]{
				\label{Figure4-3}
				\centering
				\includegraphics[width=0.3\linewidth, trim=15 0 15 0]{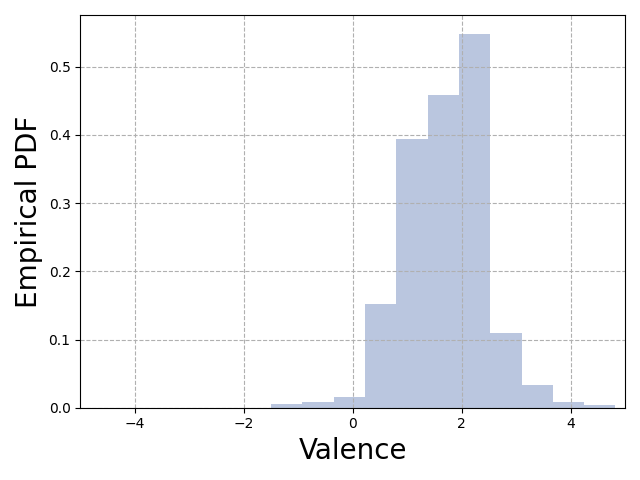}
			} 
			
			\subfigure[sadness]{
				\label{Figure4-4}
				\centering
				\includegraphics[width=0.3\linewidth, trim=15 0 15 0]{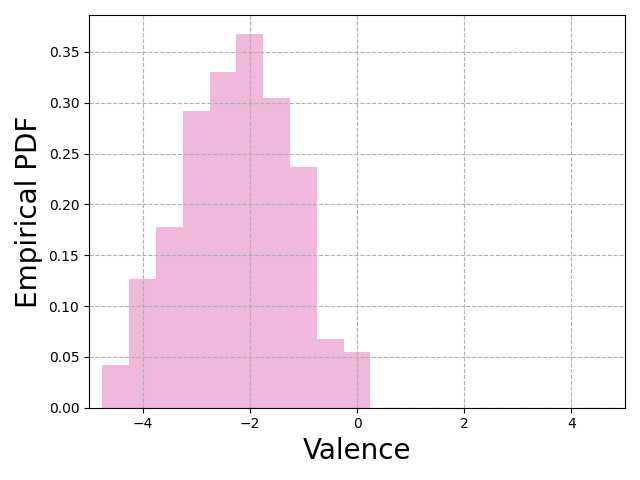}
			}
			\subfigure[worry]{
				\label{Figure4-5}
				\centering
				\includegraphics[width=0.3\linewidth, trim=15 0 15 0]{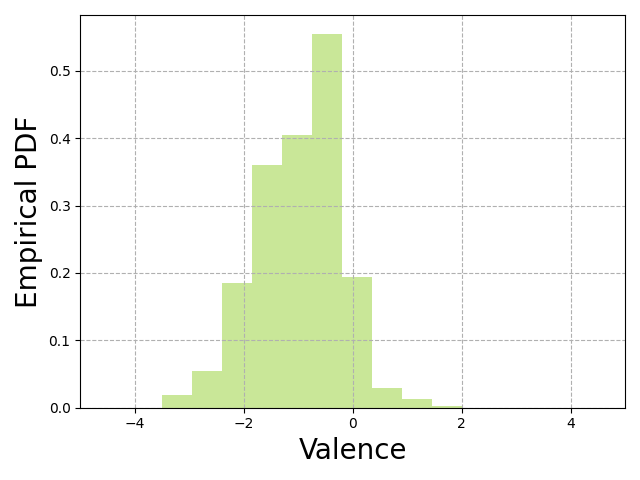}
			}
			\subfigure[surprise]{
				\label{Figure4-6}
				\centering
				\includegraphics[width=0.3\linewidth, trim=15 0 15 0]{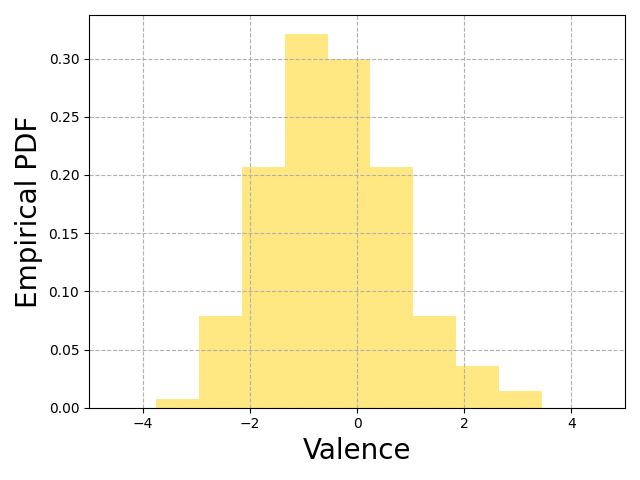}
			}
			
		\end{center}
		\caption{Empirical PDF on the valence for different discrete emotions. We calculate statistics using all valence-labeled samples.}
		\label{Figure4}
	\end{figure}
	
	\textbf{Train$\&$Val and MER-MULTI.} 
	These two subsets provide both discrete and dimensional annotations, and their relationship is shown in Fig. \ref{Figure4}. Valence represents the degree of pleasure and the value from small to large means the sentiment from negative to positive. From this figure, we observe that the valence distribution of different discrete labels is reasonable. For negative emotions (such as \emph{anger}, \emph{sadness}, and \emph{worry}), the valence is mostly less than 0; for positive emotions (such as \emph{happiness}), the valence is mostly greater than 0; the valence of \emph{neutral} is around 0; \emph{surprise} is a rather complex emotion that contains multiple meanings such as sadly surprised, angrily surprised, or happily surprised. Therefore, its valence ranges from negative to positive. These results ensure the quality of our labels and demonstrate the necessity of multi-label annotations, as they can help us distinguish some subtle emotional differences.

	\textbf{MER-NOISE.}
	This subset involves introducing noise to samples to assess noise robustness. For the audio modality, we randomly select noises from MUSAN \cite{snyder2015musan}, a noise corpus comprising three subsets: \emph{music}, \emph{noise}, and \emph{speech}. Among these subsets, \emph{music} may affect the original emotion of the input audio, while \emph{noise} also conveys emotions. For example, rain and thunder may evoke negative sentiments; a pleasant breeze may evoke positive sentiments. Therefore, we select noises from the \emph{speech} subset. Subsequently, we randomly choose a signal-to-noise ratio (SNR) between 5$\sim$10dB, adjust the noise amplitude according to SNR, crop the noise to the same length as the audio, and finally merge them. For the visual modality, blurring is a common technique for obtaining low-quality images, as network limitations in transmission often occur. To mimic this process, we first downsample the image to lose some details and then upsample the low-resolution image to maintain its original size. For each video, we randomly select a downsampling factor from $\{1, 2, 4\}$ and apply it to all frames. For convenience, this paper does not discuss the case where different frames have distinct downsampling factors.
	
	\textbf{MER-SEMI.}
	This subset consists of large amounts of unlabeled samples. To further annotate some of these samples, we utilize the maximum softmax probabilities \cite{hendrycks2017baseline} to measure confidence and only annotate samples with high confidence. We find this process helps reduce the difficulty of labeling. Suppose $f(\cdot)$ is a pre-trained emotion classifier, and $f_j(\cdot)$ represents the prediction result for the class $j$. For sample $x_i$, its confidence score is defined as $s_i = \mbox{max}_jf_j(x_i)$ \cite{hendrycks2017baseline}. After obtaining the confidence score, previous works often determine a threshold, and samples with scores higher than the threshold are considered reliable \cite{hendrycks2017baseline}. However, an inherent class imbalance exists in the corpus (see Fig. \ref{Figure3}). It is unreasonable to select a fixed threshold for different classes. Inspired by long-tail learning \cite{karim2022unicon}, we select the sample whose confidence falls in the highest $\eta$ portion for each class to form a class-specific threshold. The more detailed procedure is presented in Algorithm \ref{alg-1}.

	\begin{algorithm}[t]
		\caption{Selection strategy in MER-SEMI}
		\label{alg-1}
		\KwIn{The model $f$ pre-trained on the labeled samples $\mathcal{D}_l=\{ \left(x_{i}, e_{i}\right)\}_{i=1}^{N_l}$, the unlabeled set $\tilde{\mathcal{D}}_u=\{ \tilde{x}_{i}\}_{i=1}^{N_u}$, the quantile $\eta$.}
		\KwOut{The unlabeled subset $\tilde{\mathcal{S}}_u \subseteq \tilde{\mathcal{D}}_u$ for annotation.}
		\BlankLine
		
		Create an empty list $G_j$ for each label $j \in [1, C]$;
		
		\For{$(x_{i}, e_{i}) \in \mathcal{D}_l$}{
			
			Gain the predicted label $\hat{e}_i = \mbox{argmax}_{j}{f_j\left(x_i\right)}$;
			
			Calculate the confidence score $s_i = \max_{j}{f_j\left(x_i\right)}$;
			
			\If{$\hat{e}_i = e_i$}{
				
				Store the value $s_i$ into $G_{e_i}$;
				
			}
			
		}
		
		\For{$j = 1, \cdots, C$}{
			
			$\lambda_j \leftarrow \eta$-quantile of the list $G_j$;
			
		}
		
		~\\
		
		Create an empty set $\tilde{\mathcal{S}}_u$;
		
		\For{$\tilde{x}_{i} \in \tilde{\mathcal{D}}_u$}{
			
			Gain the predicted label $\hat{e}_i = \mbox{argmax}_{j}{f_j\left(\tilde{x}_{i}\right)}$;
			
			Calculate the confidence score $s_i = \max_{j}{f_j\left(\tilde{x}_{i}\right)}$;
			
			\If{$s_i > \lambda_{\hat{e}_i}$}{
				
				Store the sample $\tilde{x}_{i}$ into $\tilde{\mathcal{S}}_u$;
				
			}
			
		}
	\end{algorithm}

	\section{MER2023 Baselines}
	\label{sec4-merbaseline}
	This section first formulates the problem definition and introduces the notations used. Subsequently, we introduce the data preprocessing, model structure, and implementation details of the baseline. Finally, we present unimodal and multimodal results and provide an in-depth discussion.

	\subsection{Problem Definition and Notations}
	Let us define a labeled dataset $\mathcal{D}_l=\{(x_i, e_i, v_i)\}_{i=1}^{N_l}$ and an unlabeled dataset $\tilde{\mathcal{D}}_u=\{\tilde{x}_i\}_{i=1}^{N_u}$, where $e_i \in \{1,2,\cdots C\}$ and $v_i \in [-5, 5]$ represent the discrete category and valence of the sample $x_i$. Here, $N_l$, $N_u$, and $C$ denote the number of labeled samples, unlabeled samples, and discrete classes, respectively. In this paper, we aim to predict discrete and dimensional emotions from audiovisual recordings. 
	
	\subsection{Data Preprocessing}
	For the visual modality, we first crop and align faces via OpenFace\footnote{\emph{https://github.com/cmusatyalab/openface}} \cite{baltruvsaitis2016openface}. Then, we use visual encoders to extract frame-level or segment-level features, followed by average pooling to compress them to the video level. For the audio modality, we use FFmpeg to separate the audio from the video and unify the audio format to 16kHz and mono. For the lexical modality, we first extract subtitles using WeNet\footnote{\emph{https://github.com/wenet-e2e/wenet}} \cite{yao2021wenet}, an open-source ASR toolkit. But the timestamps in video segmentation are not always accurate, causing the audio signal loss in some samples and resulting in ASR errors. Additionally, we observe that some videos also include subtitles. Therefore, we further use EasyOCR to extract subtitles from video frames. Then, we manually merge the ASR and OCR outputs to generate the final subtitles.

	After data preprocessing, for each sample $x_i$, we extract acoustic features $f_i^a \in \mathbb{R}^{d_a}$, lexical features $f_i^l \in \mathbb{R}^{d_l}$, and visual features $f_i^v \in \mathbb{R}^{d_v}$, where $\{d_m\}_{m \in \{a,l,v\}}$ is the feature dimension for each modality. Table \ref{Table18} (see Appendix) presents the model cards of the main encoders involved in this paper. Since MER2023 is a Chinese dataset, we choose lexical encoders that support Chinese. Additionally, we evaluate the performance of acoustic encoders trained in different languages. More analysis on feature selection is available in the MERBench (see Section \ref{sec5-merbench}).

	\begin{table*}[t]
		\centering
		\renewcommand\tabcolsep{5pt}
		\caption{Unimodal results on MER2023. Besides five-fold cross-validation results on Train$\&$Val, we also report test results on three subsets.}
		\label{Table2}
		\begin{tabular}{l|cc|cc|cc|c||cc}
			\hline
			\multirow{2}{*}{Feature} & \multicolumn{2}{c|}{Train$\&$Val} & \multicolumn{2}{c|}{MER-MULTI} & \multicolumn{2}{c|}{MER-NOISE} & {MER-SEMI} & \multicolumn{2}{c}{Average}\\
			& WAF $(\uparrow)$ & MSE $(\downarrow)$ & WAF $(\uparrow)$ & MSE $(\downarrow)$ & WAF $(\uparrow)$ & MSE $(\downarrow)$ & WAF $(\uparrow)$ & WAF $(\uparrow)$ & MSE $(\downarrow)$ \\
			
			\hline
			\multicolumn{10}{c}{Visual Modality} \\
			\hline
			
			ResNet-MSCeleb&43.04$\pm$0.39 & 2.21$\pm$0.01&43.10$\pm$1.11 & 2.24$\pm$0.02&40.26$\pm$0.90 & 2.13$\pm$0.04&31.27$\pm$1.24 & 39.42 & 2.19 \\
			ResNet-ImageNet&46.07$\pm$0.31 & 2.20$\pm$0.01&45.29$\pm$0.77 & 1.95$\pm$0.02&44.60$\pm$0.41 & 1.80$\pm$0.01&41.61$\pm$0.52 & 44.39 & 1.98 \\
			VideoMAE-base&47.77$\pm$0.93 & 2.19$\pm$0.02&46.72$\pm$0.75 & 2.04$\pm$0.02&47.81$\pm$0.51 & 1.98$\pm$0.01&45.70$\pm$0.36 & 47.00 & 2.07 \\
			EmoNet&47.34$\pm$0.34 & 1.98$\pm$0.00&47.66$\pm$0.31 & 2.00$\pm$0.01&44.95$\pm$0.37 & 1.96$\pm$0.01&49.20$\pm$0.46 & 47.29 & 1.98 \\
			VideoMAE-large&53.15$\pm$0.24 & 1.89$\pm$0.02&53.26$\pm$0.68 & 1.80$\pm$0.01&53.74$\pm$0.19 & 1.50$\pm$0.01&55.42$\pm$0.30 & 53.89 & 1.73 \\
			SENet-FER2013&56.28$\pm$0.26 & 1.42$\pm$0.01&56.92$\pm$0.55 &\textcolor[rgb]{1,0.36,0.36}{\textbf{ 1.27$\pm$0.01}}&55.99$\pm$0.43 & 1.81$\pm$0.01&50.99$\pm$0.14 & 55.05 & 1.50 \\
			DINOv2-large&54.30$\pm$0.75 & 1.63$\pm$0.01&54.85$\pm$0.46 & 1.52$\pm$0.01&53.56$\pm$0.77 & 1.58$\pm$0.02&57.65$\pm$0.35 & 55.09 & 1.57 \\
			ResNet-FER2013&57.13$\pm$0.32 & 1.39$\pm$0.01&57.43$\pm$0.59 &\textcolor[rgb]{1,0.72,0.72}{\textbf{ 1.30$\pm$0.01}}&58.09$\pm$0.25 &\textcolor[rgb]{1,0.72,0.72}{\textbf{ 1.38$\pm$0.01}}&52.68$\pm$0.64 & 56.33 &\textcolor[rgb]{1,0.36,0.36}{\textbf{1.36}}\\
			MANet-RAFDB&58.63$\pm$0.29 & 1.40$\pm$0.01&57.46$\pm$0.50 & 1.44$\pm$0.02&57.01$\pm$0.20 & 1.44$\pm$0.01&58.65$\pm$0.50 & 57.94 & 1.43 \\
			CLIP-base&\textcolor[rgb]{1,0.72,0.72}{\textbf{58.84$\pm$0.41 }}&\textcolor[rgb]{1,0.36,0.36}{\textbf{ 1.37$\pm$0.01}}&\textcolor[rgb]{1,0.36,0.36}{\textbf{60.12$\pm$0.85 }}& 1.39$\pm$0.01&\textcolor[rgb]{1,0.72,0.72}{\textbf{60.69$\pm$0.61 }}&\textcolor[rgb]{1,0.36,0.36}{\textbf{ 1.38$\pm$0.02}}&\textcolor[rgb]{1,0.72,0.72}{\textbf{62.21$\pm$0.40 }}&\textcolor[rgb]{1,0.72,0.72}{\textbf{ 60.46 }}&\textcolor[rgb]{1,0.72,0.72}{\textbf{1.38}}\\
			EVA-02-base&\textcolor[rgb]{1,0.36,0.36}{\textbf{59.61$\pm$0.24 }}&\textcolor[rgb]{1,0.72,0.72}{\textbf{ 1.39$\pm$0.01}}&\textcolor[rgb]{1,0.72,0.72}{\textbf{59.10$\pm$0.26 }}& 1.39$\pm$0.01&\textcolor[rgb]{1,0.36,0.36}{\textbf{62.27$\pm$0.39 }}& 1.45$\pm$0.02&\textcolor[rgb]{1,0.36,0.36}{\textbf{62.48$\pm$0.48 }}&\textcolor[rgb]{1,0.36,0.36}{\textbf{ 60.87 }}& 1.41 \\
			CLIP-large&\textcolor[rgb]{1,0,0}{\textbf{65.17$\pm$0.18 }}&\textcolor[rgb]{1,0,0}{\textbf{ 1.07$\pm$0.01}}&\textcolor[rgb]{1,0,0}{\textbf{62.69$\pm$0.23 }}&\textcolor[rgb]{1,0,0}{\textbf{ 1.12$\pm$0.01}}&\textcolor[rgb]{1,0,0}{\textbf{64.00$\pm$0.42 }}&\textcolor[rgb]{1,0,0}{\textbf{ 1.11$\pm$0.00}}&\textcolor[rgb]{1,0,0}{\textbf{70.23$\pm$0.32 }}&\textcolor[rgb]{1,0,0}{\textbf{ 65.52 }}&\textcolor[rgb]{1,0,0}{\textbf{1.10}}\\

			\hline
			\multicolumn{10}{c}{Acoustic Modality} \\
			\hline
			
			eGeMAPS&19.71$\pm$0.59 & 3.39$\pm$0.11&18.86$\pm$1.27 & 3.00$\pm$0.01&23.39$\pm$1.44 & 2.42$\pm$0.02&16.52$\pm$1.26 & 19.62 & 2.94 \\
			IS09&31.08$\pm$0.76 & 2.86$\pm$0.02&27.05$\pm$1.35 & 2.75$\pm$0.02&28.67$\pm$1.77 & 2.23$\pm$0.01&22.44$\pm$1.37 & 27.31 & 2.61 \\
			VGGish&45.92$\pm$0.34 & 2.60$\pm$0.01&40.45$\pm$0.13 & 2.54$\pm$0.01&41.52$\pm$0.24 & 2.28$\pm$0.01&55.57$\pm$0.49 & 45.87 & 2.47 \\
			data2vec-base&43.88$\pm$0.21 & 2.60$\pm$0.01&45.13$\pm$0.39 & 2.31$\pm$0.01&36.00$\pm$0.45 & 2.36$\pm$0.02&58.49$\pm$0.30 & 45.87 & 2.42 \\
			wav2vec-large&50.63$\pm$0.19 & 2.19$\pm$0.02&53.11$\pm$0.45 & 1.99$\pm$0.01&43.67$\pm$0.35 & 2.08$\pm$0.02&62.79$\pm$0.40 & 52.55 & 2.09 \\
			WavLM-base&51.44$\pm$0.24 & 2.21$\pm$0.01&50.11$\pm$0.21 & 2.03$\pm$0.00&42.60$\pm$0.47 & 2.13$\pm$0.02&70.30$\pm$0.21 & 53.61 & 2.12 \\
			Whisper-base&51.42$\pm$0.21 & 2.13$\pm$0.01&56.94$\pm$0.29 & 1.71$\pm$0.01&42.98$\pm$0.20 & 2.31$\pm$0.01&69.89$\pm$0.18 & 55.31 & 2.05 \\
			Whisper-large&56.41$\pm$0.42 & 1.57$\pm$0.01&57.63$\pm$0.84 & 1.49$\pm$0.01&45.64$\pm$0.73 & 2.16$\pm$0.02&75.24$\pm$0.41 & 58.73 & 1.74 \\
			WavLM-large&58.02$\pm$0.27 & 1.83$\pm$0.00&59.99$\pm$0.09 & 1.63$\pm$0.01&46.51$\pm$0.50 & 2.09$\pm$0.02&\textcolor[rgb]{1,0.72,0.72}{\textbf{77.63$\pm$0.43 }}& 60.54 & 1.85 \\
			wav2vec 2.0-large&59.39$\pm$0.27 &\textcolor[rgb]{1,0.72,0.72}{\textbf{ 1.49$\pm$0.01}}&\textcolor[rgb]{1,0.72,0.72}{\textbf{61.59$\pm$0.42 }}&\textcolor[rgb]{1,0.72,0.72}{\textbf{ 1.44$\pm$0.01}}&50.60$\pm$0.68 &\textcolor[rgb]{1,0.36,0.36}{\textbf{ 1.82$\pm$0.01}}&76.18$\pm$0.16 & 61.94 &\textcolor[rgb]{1,0.36,0.36}{\textbf{1.59}}\\
			wav2vec 2.0-base&\textcolor[rgb]{1,0.72,0.72}{\textbf{60.18$\pm$0.24 }}& 1.55$\pm$0.01&60.83$\pm$0.36 & 1.53$\pm$0.01&\textcolor[rgb]{1,0.36,0.36}{\textbf{52.66$\pm$0.20 }}&\textcolor[rgb]{1,0.72,0.72}{\textbf{ 1.84$\pm$0.03}}&76.02$\pm$0.34 &\textcolor[rgb]{1,0.72,0.72}{\textbf{ 62.42 }}& 1.64 \\
			HUBERT-base&\textcolor[rgb]{1,0.36,0.36}{\textbf{65.54$\pm$0.37 }}&\textcolor[rgb]{1,0.36,0.36}{\textbf{ 1.26$\pm$0.01}}&\textcolor[rgb]{1,0.36,0.36}{\textbf{67.48$\pm$0.18 }}&\textcolor[rgb]{1,0.36,0.36}{\textbf{ 1.18$\pm$0.01}}&\textcolor[rgb]{1,0.72,0.72}{\textbf{51.40$\pm$0.82 }}& 2.39$\pm$0.05&\textcolor[rgb]{1,0.36,0.36}{\textbf{83.77$\pm$0.26 }}&\textcolor[rgb]{1,0.36,0.36}{\textbf{ 67.05 }}&\textcolor[rgb]{1,0.72,0.72}{\textbf{1.61}}\\
			HUBERT-large&\textcolor[rgb]{1,0,0}{\textbf{70.16$\pm$0.16 }}&\textcolor[rgb]{1,0,0}{\textbf{ 1.01$\pm$0.00}}&\textcolor[rgb]{1,0,0}{\textbf{69.91$\pm$0.58 }}&\textcolor[rgb]{1,0,0}{\textbf{ 0.97$\pm$0.01}}&\textcolor[rgb]{1,0,0}{\textbf{65.07$\pm$0.83 }}&\textcolor[rgb]{1,0,0}{\textbf{ 1.31$\pm$0.03}}&\textcolor[rgb]{1,0,0}{\textbf{86.04$\pm$0.15 }}&\textcolor[rgb]{1,0,0}{\textbf{ 72.79 }}&\textcolor[rgb]{1,0,0}{\textbf{1.10}}\\
			
			\hline
			\multicolumn{10}{c}{Lexical Modality} \\
			\hline
			
			OPT-13B&40.84$\pm$0.13 & 2.57$\pm$0.01&45.78$\pm$0.19 & 2.34$\pm$0.01&45.37$\pm$0.24 & 2.04$\pm$0.00&38.98$\pm$0.51 & 42.74 & 2.32 \\
			ALBERT-small&41.73$\pm$0.41 & 2.55$\pm$0.01&46.43$\pm$0.29 & 2.41$\pm$0.01&43.54$\pm$0.45 & 2.00$\pm$0.01&42.80$\pm$0.33 & 43.63 & 2.32 \\
			Llama-13B&43.87$\pm$0.20 & 2.44$\pm$0.01&49.15$\pm$0.63 & 2.29$\pm$0.01&46.03$\pm$0.20 & 1.92$\pm$0.01&41.82$\pm$0.34 & 45.22 & 2.22 \\
			Vicuna-13B&43.89$\pm$0.11 & 2.45$\pm$0.01&49.35$\pm$0.28 & 2.27$\pm$0.01&48.24$\pm$0.28 & 1.82$\pm$0.01&44.11$\pm$0.45 & 46.40 & 2.18 \\
			XLNet-base&46.11$\pm$0.30 & 2.33$\pm$0.01&50.01$\pm$0.46 & 2.21$\pm$0.01&45.40$\pm$0.37 & 1.90$\pm$0.01&48.68$\pm$0.25 & 47.55 & 2.15 \\
			StableLM-7B&45.40$\pm$0.20 & 2.33$\pm$0.00&51.10$\pm$0.30 & 2.17$\pm$0.01&48.72$\pm$0.51 & 1.91$\pm$0.01&46.09$\pm$0.99 & 47.83 & 2.13 \\
			Sentence-BERT&45.85$\pm$0.18 & 2.28$\pm$0.00&51.74$\pm$0.65 & 2.13$\pm$0.00&46.83$\pm$0.35 & 1.85$\pm$0.01&47.20$\pm$0.23 & 47.91 & 2.09 \\
			DeBERTa-large&44.79$\pm$0.28 & 2.39$\pm$0.02&52.48$\pm$0.20 & 2.14$\pm$0.01&47.12$\pm$0.40 & 1.98$\pm$0.00&47.40$\pm$0.70 & 47.95 & 2.17 \\
			Llama2-13B&45.95$\pm$0.13 & 2.34$\pm$0.01&51.63$\pm$0.43 & 2.25$\pm$0.01&49.47$\pm$0.35 & 1.83$\pm$0.01&46.58$\pm$0.89 & 48.41 & 2.14 \\
			PERT-base&47.23$\pm$0.17 & 2.23$\pm$0.01&52.75$\pm$0.48 & 2.04$\pm$0.01&49.53$\pm$0.16 & 1.74$\pm$0.00&52.53$\pm$0.65 & 50.51 & 2.00 \\
			MOSS-7B&48.31$\pm$0.19 & 2.15$\pm$0.01&54.35$\pm$0.28 & 1.88$\pm$0.01&50.02$\pm$0.32 & 1.80$\pm$0.02&49.96$\pm$0.95 & 50.66 & 1.94 \\
			ELECTRA-base&48.06$\pm$0.18 & 2.23$\pm$0.01&54.61$\pm$0.33 & 2.02$\pm$0.01&48.87$\pm$0.18 &\textcolor[rgb]{1,0.72,0.72}{\textbf{ 1.69$\pm$0.01}}&52.54$\pm$0.57 & 51.02 & 1.98 \\
			Falcon-7B&48.97$\pm$0.24 & 2.16$\pm$0.01&54.45$\pm$0.30 & 1.89$\pm$0.01&49.83$\pm$0.44 & 1.77$\pm$0.01&50.96$\pm$0.63 & 51.05 & 1.94 \\
			LERT-base&49.44$\pm$0.08 & 2.17$\pm$0.01&57.30$\pm$0.46 & 1.94$\pm$0.00&49.60$\pm$0.46 & 1.74$\pm$0.00&53.32$\pm$0.85 & 52.41 & 1.95 \\
			ChatGLM2-6B&\textcolor[rgb]{1,0.36,0.36}{\textbf{50.70$\pm$0.21 }}&\textcolor[rgb]{1,0.72,0.72}{\textbf{ 2.08$\pm$0.01}}&56.86$\pm$0.55 &\textcolor[rgb]{1,0.72,0.72}{\textbf{ 1.84$\pm$0.01}}&\textcolor[rgb]{1,0.36,0.36}{\textbf{51.63$\pm$0.60 }}& 1.70$\pm$0.01&51.38$\pm$0.74 & 52.64 &\textcolor[rgb]{1,0.72,0.72}{\textbf{1.87}}\\
			MacBERT-base&49.25$\pm$0.16 & 2.16$\pm$0.01&56.38$\pm$0.32 & 1.94$\pm$0.00&49.92$\pm$0.33 & 1.72$\pm$0.01&\textcolor[rgb]{1,0.36,0.36}{\textbf{55.40$\pm$0.48 }}& 52.74 & 1.94 \\
			RoBERTa-large&50.02$\pm$0.07 & 2.09$\pm$0.01&\textcolor[rgb]{1,0.72,0.72}{\textbf{58.22$\pm$0.69 }}& 1.89$\pm$0.01&50.15$\pm$0.33 &\textcolor[rgb]{1,0.36,0.36}{\textbf{ 1.69$\pm$0.01}}&\textcolor[rgb]{1,0.72,0.72}{\textbf{55.36$\pm$0.39 }}&\textcolor[rgb]{1,0.72,0.72}{\textbf{ 53.44 }}& 1.89 \\
			BLOOM-7B&\textcolor[rgb]{1,0.72,0.72}{\textbf{50.48$\pm$0.14 }}&\textcolor[rgb]{1,0.36,0.36}{\textbf{ 2.06$\pm$0.01}}&\textcolor[rgb]{1,0.36,0.36}{\textbf{59.23$\pm$0.37 }}&\textcolor[rgb]{1,0.36,0.36}{\textbf{ 1.81$\pm$0.01}}&\textcolor[rgb]{1,0.72,0.72}{\textbf{50.17$\pm$0.24 }}& 1.71$\pm$0.01&\textcolor[rgb]{1,0,0}{\textbf{56.15$\pm$0.45 }}&\textcolor[rgb]{1,0.36,0.36}{\textbf{ 54.01 }}&\textcolor[rgb]{1,0.36,0.36}{\textbf{1.86}}\\
			Baichuan-13B&\textcolor[rgb]{1,0,0}{\textbf{52.65$\pm$0.09 }}&\textcolor[rgb]{1,0,0}{\textbf{ 1.94$\pm$0.01}}&\textcolor[rgb]{1,0,0}{\textbf{62.54$\pm$0.69 }}&\textcolor[rgb]{1,0,0}{\textbf{ 1.63$\pm$0.01}}&\textcolor[rgb]{1,0,0}{\textbf{51.97$\pm$0.72 }}&\textcolor[rgb]{1,0,0}{\textbf{ 1.60$\pm$0.01}}&54.23$\pm$0.92 &\textcolor[rgb]{1,0,0}{\textbf{ 55.35 }}&\textcolor[rgb]{1,0,0}{\textbf{1.72}}\\

			\hline
		\end{tabular}
	\end{table*}

	\begin{table*}[t]
		\centering
		\renewcommand\tabcolsep{6pt}
		\caption{Multimodal results on MER2023. We select acoustic features from HUBERT-base (HB) and HUBERT-large (HL), lexical features from BLOOM-7B (BL) and Baichuan-13B (BA), and visual features from EVA-02-base (EB) and CLIP-large (CL). Here, ``A'', ``L'', and ``V'' represent the acoustic, lexical, and visual modalities, respectively.}
		\label{Table3}
		\begin{tabular}{ccc|cc|cc|cc|c||cc}
			\hline
			\multirow{2}{*}{A} & \multirow{2}{*}{L} & \multirow{2}{*}{V} & \multicolumn{2}{c|}{Train$\&$Val} & \multicolumn{2}{c|}{MER-MULTI} & \multicolumn{2}{c|}{MER-NOISE} & {MER-SEMI} & \multicolumn{2}{c}{Average} \\
			& & & WAF $(\uparrow)$ & MSE $(\downarrow)$ & WAF $(\uparrow)$ & MSE $(\downarrow)$ & WAF $(\uparrow)$ & MSE $(\downarrow)$ & WAF $(\uparrow)$ & WAF $(\uparrow)$ & MSE $(\downarrow)$ \\
			\hline
			\multicolumn{12}{c}{Unimodal Results} \\
			\hline
			HB &--- &--- & 65.54$\pm$0.37 & 1.26$\pm$0.01 & 67.48$\pm$0.18 & 1.18$\pm$0.01 & 51.40$\pm$0.82 & 2.39$\pm$0.05& 83.77$\pm$0.26 & 67.05 & 1.61 \\
			HL &--- &--- &\textbf{70.16$\pm$0.16}&\textbf{1.01$\pm$0.00}&\textbf{69.91$\pm$0.58}&\textbf{0.97$\pm$0.01}&\textbf{65.07$\pm$0.83}& 1.31$\pm$0.03 &\textbf{86.04$\pm$0.15}&\textbf{72.79}& 1.10 \\
			--- & BL &--- & 50.48$\pm$0.14 & 2.06$\pm$0.01 & 59.23$\pm$0.37 & 1.81$\pm$0.01 &50.17$\pm$0.24 & 1.71$\pm$0.01& 56.15$\pm$0.45 & 54.01 & 1.86 \\
			--- & BA &--- & 52.65$\pm$0.09 & 1.94$\pm$0.01 & 62.54$\pm$0.69 & 1.63$\pm$0.01 & 51.97$\pm$0.72 & 1.60$\pm$0.01 &54.23$\pm$0.92 & 55.35 & 1.72 \\
			--- & --- & EB & 59.61$\pm$0.24 & 1.39$\pm$0.01 & 59.10$\pm$0.26 & 1.39$\pm$0.01& 62.27$\pm$0.39 & 1.45$\pm$0.02& 62.48$\pm$0.48 & 60.87 & 1.41 \\
			--- & --- & CL & 65.17$\pm$0.18 & 1.07$\pm$0.01 & 62.69$\pm$0.23 & 1.12$\pm$0.01 & 64.00$\pm$0.42 &\textbf{1.11$\pm$0.00}& 70.23$\pm$0.32 & 65.52 &\textbf{1.10}\\
			
			\hline
			\multicolumn{12}{c}{Multimodal Results}\\
			\hline
			
			HB &BL &EB &73.24$\pm$0.27 & 0.81$\pm$0.00&78.86$\pm$0.32 & 0.74$\pm$0.01&72.87$\pm$0.86 & 0.97$\pm$0.02&87.10$\pm$0.37 & 78.02 & 0.84 \\
			HB &BL &CL &74.23$\pm$0.19 & 0.71$\pm$0.01&79.59$\pm$0.80 & 0.71$\pm$0.00&73.13$\pm$0.09 & 0.82$\pm$0.03&87.84$\pm$0.15 & 78.70 & 0.75 \\
			HB &BA &EB &73.83$\pm$0.12 & 0.78$\pm$0.01&80.38$\pm$0.25 & 0.70$\pm$0.01&72.23$\pm$0.46 & 0.93$\pm$0.01&86.35$\pm$0.30 & 78.20 & 0.80 \\
			HB &BA &CL &75.05$\pm$0.29 & 0.69$\pm$0.01&81.82$\pm$0.28 & 0.67$\pm$0.01&75.00$\pm$0.52 & 0.81$\pm$0.02&87.45$\pm$0.24 & 79.83 & 0.72 \\
			HL &BL &EB &74.77$\pm$0.34 & 0.73$\pm$0.00&80.92$\pm$0.36 & 0.69$\pm$0.01&75.30$\pm$0.24 & 0.76$\pm$0.01&88.48$\pm$0.14 & 79.87 & 0.73 \\
			HL &BL &CL &75.42$\pm$0.13 & 0.66$\pm$0.00&81.67$\pm$0.42 & 0.65$\pm$0.01&75.80$\pm$0.25 & 0.65$\pm$0.01&\textbf{89.66$\pm$0.18}& 80.64 & 0.65 \\
			HL &BA &EB &75.00$\pm$0.08 & 0.73$\pm$0.00&81.79$\pm$0.54 & 0.64$\pm$0.01&75.04$\pm$0.49 & 0.74$\pm$0.02&88.39$\pm$0.16 & 80.06 & 0.70 \\
			HL &BA &CL &\textbf{75.95$\pm$0.12}&\textbf{0.64$\pm$0.01}&\textbf{83.02$\pm$0.31}&\textbf{0.62$\pm$0.01}&\textbf{77.08$\pm$0.46}&\textbf{0.63$\pm$0.01}&89.29$\pm$0.36 &\textbf{81.34}&\textbf{0.63}\\
			
			\hline
			
		\end{tabular}
	\end{table*}

	\subsection{Model Structure}
	\label{sec:model}
	For unimodal features, we utilize the fully-connected layers to extract hidden representations, followed by a multi-task framework to predict both discrete and dimensional emotions simultaneously:
	\begin{equation}
	h_i^m =\mbox{ReLU}\left(f_i^mW_m^h + b_m^h\right), m \in \{a, l, v\},
	\end{equation}
	\begin{equation}
	\label{eq-v}
	\hat{v}_i = h_i^mW_m^v + b_m^v, m \in \{a,l,v\},
	\end{equation}
	\begin{equation}
	\label{eq-e}
	\hat{e}_i = \mbox{softmax}\left(h_i^mW_m^e + b_m^e\right), m \in \{a,l,v\},
	\end{equation}
	where $h_i^m \in \mathbb{R}^{h}$ is the hidden feature for each modality. $\hat{v}_i \in \mathbb{R}^{1}$ and $\hat{e}_i \in \mathbb{R}^{C}$ are the estimated valence and emotion probabilities. $W_m^h \in \mathbb{R}^{d_m \times h}$, $b_m^h \in \mathbb{R}^{h}$, $W_m^v \in \mathbb{R}^{h \times 1}$, $b_m^v \in \mathbb{R}^{1}$, $W_m^e \in \mathbb{R}^{h \times C}$, and $b_m^e \in \mathbb{R}^{C}$ are trainable parameters.

	For multimodal features, different modalities contribute differently to emotion recognition. Therefore, we compute importance scores $\alpha_i \in \mathbb{R}^{3 \times 1}$ for each modality and exploit weighted fusion to obtain multimodal features:
	\begin{equation}
	h_i = \mbox{Concat}\left(h_i^a, h_i^l, h_i^v\right),
	\end{equation}
	\begin{equation}
	\alpha_i = \mbox{softmax}\left(h_i^TW_\alpha+b_\alpha\right).
	\end{equation}
	Then, multimodal features $z_i = h_i\alpha_i$ are fed into Eqs. \ref{eq-v}$\sim$\ref{eq-e} to estimate discrete and dimension emotions simultaneously. Here, $W_\alpha \in \mathbb{R}^{h \times 1}$ and $b_\alpha \in \mathbb{R}^{3}$ are trainable parameters.
	
	During training, we choose the cross-entropy loss $\mathcal{L}_e$ for classification and the mean squared error (MSE) $\mathcal{L}_v$ for regression. We combine them into a joint objective and optimize all trainable parameters in an end-to-end manner.
	\begin{equation}
	\mathcal{L}_e = -\frac{1}{N_l}\sum_{i=1}^{N_l}e_i\log(\hat{e}_i),
	\end{equation}
	\begin{equation}
	\mathcal{L}_v = \frac{1}{N_l}\sum_{i=1}^{N_l} \left( v_i - \hat{v}_i \right)^2,
	\end{equation}
	\begin{equation}
	\mathcal{L} = \mathcal{L}_e + \mathcal{L}_v.
	\end{equation}

	\subsection{Implementation Details}
	There is mainly one user-specific parameter in the baseline, i.e., the dimension of latent representations $h$. We select $h$ from $\{64, 128, 256\}$. For the Train$\&$Val subset, we employ five-fold cross-validation for hyperparameter tuning. During training, we use the Adam \cite{kingma2015adam} optimizer and choose the learning rate from $\{10^{-3}, 10^{-4}\}$. We set the maximum number of epochs to $100$ and the weight decay to $10^{-5}$. Dropout \cite{srivastava2014dropout} is also employed, and we select the rate from $\{0.2, 0.3, 0.4, 0.5\}$ to alleviate the over-fitting problem. To mitigate randomness, we run each experiment six times and report the average result along with the standard deviation.

	\subsection{Results and Discussion}
	Tables \ref{Table2}$\sim$\ref{Table3} present unimodal and multimodal results. In addition to five-fold cross-validation results on Train$\&$Val, we also report test results on the MER-MULTI, MER-NOISE, and MER-SEMI subsets. For discrete emotions, considering the inherent emotion class imbalance in the dataset (see Fig. \ref{Figure3}), we choose the weighted average F-score (WAF) as the evaluation metric. For dimensional emotions, we use MSE as the evaluation metric, consistent with previous works \cite{kollias2020analysing}. Higher $\mbox{WAF}$ and lower $\mbox{MSE}$ indicate better performance. In Tables \ref{Table2}$\sim$\ref{Table3}, the standard deviation is relatively small, which eliminates the influence of randomness and ensures the reliability of our conclusions.
	
	Table \ref{Table2} summarizes the unimodal results. For the audio modality, deep features consistently outperform handcrafted features. Since emotion-related acoustic features vary across datasets, using a fixed set of handcrafted features may limit performance. In contrast, deep features can capture more universal acoustic representations for emotion recognition. For the visual modality, models trained on facial expression recognition (e.g., FER2013 \cite{goodfellow2013challenges} and RAF-DB \cite{li2017reliable}) generally perform better than those trained on face recognition (e.g., MS-Celeb-1M \cite{guo2016ms}) and object recognition (e.g., ImageNet \cite{deng2009imagenet}). These results demonstrate the importance of task similarity in transfer learning. Meanwhile, the audio modality can achieve better performance than the visual and lexical modalities, which indicates that our dataset focuses more on audio to express emotions.
	
	In Table \ref{Table3}, we select several well-performing unimodal features and report their fusion results. Experimental results demonstrate that multimodal fusion consistently improves performance. The reason lies in the fact that emotions can be conveyed through multiple modalities. The integration of multimodal information allows the model to better comprehend the video content and accurately recognize emotions.
	
	In Tables \ref{Table2}$\sim$\ref{Table3}, models that perform well on Train$\&$Val generally exhibit good performance on other subsets. These results demonstrate that our system does not overfit training data and has a good generalization ability to unseen data. Meanwhile, MER-SEMI achieves better performance than the other two subsets. The reason lies in our confidence-based selection strategy (in Algorithm \ref{alg-1}), which selects samples with well-defined emotions that can be easily recognized by emotion recognition systems.
	
	Meanwhile, we observe that models performing well on dimensional emotions generally achieve good performance on discrete emotions (see Tables \ref{Table2}$\sim$\ref{Table3}). These results reflect a high correlation between discrete and dimensional labels, consistent with the phenomenon in Fig. \ref{Figure4}. Compared with MER-MULTI, our system performs worse on MER-NOISE, and the acoustic modality degrades more than the visual modality. These results indicate that adding noise to audio has a larger impact than blurring video frames.

	\section{MERBench}
	\label{sec5-merbench}
	%This section further builds MERBench, an evaluation benchmark for multimodal emotion recognition. We first introduce the datasets involved in the benchmark. Then, we try to answer some key questions in this field, including feature selection, multimodal fusion, cross-corpus performance, robustness analysis, fine-tuning necessity analysis, etc.

	\begin{table}[t]
		\centering
		\renewcommand\tabcolsep{3.6pt}
		\caption{Statistics for different multimodal emotion datasets. ``LOSO'' means the leave-one-session-out strategy and ``Default'' means using the official training/validation/test splits provided by the dataset.}
		\label{Table4}
		\begin{tabular}{l|cccc}
			\hline
			Dataset & Lang. & \# Samples & Evaluation & Label\\
			\hline
			IEMOCAP(four) 	& English & 5531 	& LOSO 		& Discrete 	 	\\
			IEMOCAP(six)	& English & 7433 	& LOSO 		& Discrete 	 	\\
			MELD 			& English & 13708	& Default 	& Discrete 		\\ 
			CMU-MOSI 		& English & 2199 	& Default 	& Dimensional 	\\ 
			CMU-MOSEI 		& English & 22856	& Default 	& Dimensional 	\\ 
			CH-SIMS 		& Chinese & 2281 	& Default 	& Dimensional 	\\ 
			CH-SIMS v2 		& Chinese & 4403 	& Default 	& Dimensional 	\\ 
			MER-MULTI 		& Chinese & 3784 	& Default 	& Discrete 		\\ 
			\hline
		\end{tabular}
	\end{table}

	\begin{table*}[t]
		\centering
		\renewcommand\tabcolsep{3.6pt}
		\caption{Unimodal benchmark. In this table, we report unimodal results for all datasets under the same experimental setup.}
		\label{Table5}
		\begin{tabular}{l|cccccccc|c}
			\hline
			{Feature} & {MER-MULTI} & {CMU-MOSI} & {CMU-MOSEI} & {CH-SIMS} & {CH-SIMS v2} & {MELD} & {\begin{tabular}[c]{@{}c@{}}IEMOCAP \\ (four-class)\end{tabular}} & {\begin{tabular}[c]{@{}c@{}}IEMOCAP \\ (six-class) \end{tabular}} & Mean \\
			
			\hline
			\multicolumn{10}{c}{Visual Modality} \\
			\hline
			
			ResNet-MSCeleb&43.10$\pm$1.11&42.55$\pm$3.08&67.22$\pm$0.15&73.84$\pm$0.72&66.42$\pm$0.44&31.39$\pm$0.03&35.95$\pm$0.21&24.78$\pm$0.30& 48.16 \\
			ResNet-ImageNet&45.29$\pm$0.77&52.09$\pm$0.60&68.53$\pm$0.12&71.61$\pm$0.69&66.68$\pm$0.56&33.98$\pm$0.52&36.26$\pm$0.16&24.21$\pm$0.10& 49.83 \\
			EmoNet&47.66$\pm$0.31&48.47$\pm$0.80&65.87$\pm$0.14&78.64$\pm$0.48&75.37$\pm$0.35&34.59$\pm$0.55&37.09$\pm$0.50&24.15$\pm$0.38& 51.48 \\
			VideoMAE-base&46.72$\pm$0.75&\textcolor[rgb]{1,0,0}{\textbf{56.97$\pm$0.64}}&\textcolor[rgb]{1,0.36,0.36}{\textbf{70.40$\pm$0.05}}&68.95$\pm$0.69&70.11$\pm$0.26&36.74$\pm$0.20&39.15$\pm$0.18&26.76$\pm$0.17& 51.98 \\
			VideoMAE-large&53.26$\pm$0.68&54.60$\pm$0.78&\textcolor[rgb]{1,0,0}{\textbf{70.78$\pm$0.03}}&72.44$\pm$0.21&73.97$\pm$0.45&37.85$\pm$0.17&40.17$\pm$0.27&27.59$\pm$0.19& 53.83 \\
			SENet-FER2013&56.92$\pm$0.55&52.31$\pm$0.77&66.80$\pm$0.26&\textcolor[rgb]{1,0.72,0.72}{\textbf{80.69$\pm$0.19}}&\textcolor[rgb]{1,0.36,0.36}{\textbf{78.25$\pm$0.23}}&37.56$\pm$0.16&35.96$\pm$0.22&23.67$\pm$0.21& 54.02 \\
			ResNet-FER2013&57.43$\pm$0.59&50.33$\pm$1.95&66.37$\pm$0.24&78.32$\pm$0.26&74.57$\pm$0.61&\textcolor[rgb]{1,0.36,0.36}{\textbf{39.73$\pm$0.10}}&40.63$\pm$0.12&26.72$\pm$0.21& 54.26 \\
			DINOv2-large&54.85$\pm$0.46&54.26$\pm$1.95&69.20$\pm$0.30&78.27$\pm$0.73&76.59$\pm$0.63&37.97$\pm$0.28&\textcolor[rgb]{1,0,0}{\textbf{47.88$\pm$0.18}}&\textcolor[rgb]{1,0.72,0.72}{\textbf{31.72$\pm$0.18}}& 56.34 \\
			EVA-02-base&\textcolor[rgb]{1,0.72,0.72}{\textbf{59.10$\pm$0.26}}&52.22$\pm$0.23&70.04$\pm$0.15&79.55$\pm$0.25&77.70$\pm$0.31&37.94$\pm$0.15&45.91$\pm$0.18&30.17$\pm$0.10& 56.58 \\
			CLIP-base&\textcolor[rgb]{1,0.36,0.36}{\textbf{60.12$\pm$0.85}}&55.04$\pm$0.97&\textcolor[rgb]{1,0.72,0.72}{\textbf{70.24$\pm$0.07}}&80.13$\pm$0.11&\textcolor[rgb]{1,0.72,0.72}{\textbf{77.97$\pm$0.19}}&\textcolor[rgb]{1,0.72,0.72}{\textbf{38.59$\pm$0.33}}&44.52$\pm$0.21&28.70$\pm$0.08& \textcolor[rgb]{1,0.72,0.72}{\textbf{56.91}} \\
			MANet-RAFDB&57.46$\pm$0.50&\textcolor[rgb]{1,0.36,0.36}{\textbf{56.84$\pm$1.67}}&68.11$\pm$0.38&\textcolor[rgb]{1,0.36,0.36}{\textbf{81.15$\pm$0.16}}&76.86$\pm$0.10&38.24$\pm$0.24&\textcolor[rgb]{1,0.36,0.36}{\textbf{47.50$\pm$0.08}}&\textcolor[rgb]{1,0.36,0.36}{\textbf{31.75$\pm$0.25}}& \textcolor[rgb]{1,0.36,0.36}{\textbf{57.24}} \\
			CLIP-large&\textcolor[rgb]{1,0,0}{\textbf{62.69$\pm$0.23}}&\textcolor[rgb]{1,0.72,0.72}{\textbf{56.72$\pm$1.02}}&70.12$\pm$0.13&\textcolor[rgb]{1,0,0}{\textbf{82.62$\pm$0.31}}&\textcolor[rgb]{1,0,0}{\textbf{80.31$\pm$0.27}}&\textcolor[rgb]{1,0,0}{\textbf{40.03$\pm$0.14}}&\textcolor[rgb]{1,0.72,0.72}{\textbf{46.88$\pm$0.33}}&\textcolor[rgb]{1,0,0}{\textbf{31.84$\pm$0.13}}& \textcolor[rgb]{1,0,0}{\textbf{58.90}} \\
			
			\hline
			\multicolumn{10}{c}{Acoustic Modality} \\
			\hline
			
			eGeMAPS&18.86$\pm$1.27&55.40$\pm$0.79&56.93$\pm$0.52&58.83$\pm$1.01&54.92$\pm$0.38&37.14$\pm$0.45&49.70$\pm$0.30&34.99$\pm$0.14& 45.85 \\
			IS09&27.05$\pm$1.35&54.55$\pm$1.27&61.68$\pm$0.33&58.43$\pm$0.94&56.97$\pm$0.34&35.02$\pm$0.41&52.01$\pm$0.39&37.61$\pm$0.06& 47.91 \\
			VGGish&40.45$\pm$0.13&54.05$\pm$0.77&71.21$\pm$0.11&64.92$\pm$0.53&61.21$\pm$0.39&40.18$\pm$0.18&57.44$\pm$0.13&40.88$\pm$0.13& 53.79 \\
			wav2vec-large&53.11$\pm$0.45&58.03$\pm$0.31&69.30$\pm$0.35&67.67$\pm$0.27&65.22$\pm$0.66&42.44$\pm$0.24&60.89$\pm$0.09&45.61$\pm$0.20& 57.78 \\
			data2vec-base&45.13$\pm$0.39&\textcolor[rgb]{1,0.72,0.72}{\textbf{66.24$\pm$0.45}}&73.11$\pm$0.12&65.00$\pm$0.30&59.87$\pm$0.57&44.30$\pm$0.34&64.08$\pm$0.12&48.22$\pm$0.28& 58.24 \\
			wav2vec 2.0-large&\textcolor[rgb]{1,0.72,0.72}{\textbf{61.59$\pm$0.42}}&57.58$\pm$0.39&71.51$\pm$0.17&70.25$\pm$0.88&56.29$\pm$1.25&39.02$\pm$0.37&65.83$\pm$0.13&49.73$\pm$0.10& 58.98 \\
			wav2vec 2.0-base&60.83$\pm$0.36&55.29$\pm$0.51&71.85$\pm$0.13&73.00$\pm$0.54&68.26$\pm$0.30&45.10$\pm$0.18&64.10$\pm$0.14&47.76$\pm$0.06& 60.77 \\
			WavLM-base &50.11$\pm$0.21&64.79$\pm$0.97&74.28$\pm$0.24&68.10$\pm$0.48&66.82$\pm$0.43&45.79$\pm$0.40&66.26$\pm$0.17&51.12$\pm$0.09& 60.91 \\
			Whisper-base &56.94$\pm$0.29&\textcolor[rgb]{1,0.36,0.36}{\textbf{66.81$\pm$0.81}}&\textcolor[rgb]{1,0.72,0.72}{\textbf{75.12$\pm$0.18}}&69.26$\pm$0.53&65.90$\pm$0.15&\textcolor[rgb]{1,0.72,0.72}{\textbf{46.94$\pm$0.31}}&65.96$\pm$0.09&50.96$\pm$0.21& 62.24 \\
			HUBERT-base&\textcolor[rgb]{1,0.36,0.36}{\textbf{67.48$\pm$0.18}}&58.05$\pm$0.51&73.50$\pm$0.11&\textcolor[rgb]{1,0.36,0.36}{\textbf{78.72$\pm$0.21}}&\textcolor[rgb]{1,0.36,0.36}{\textbf{75.36$\pm$0.31}}&\textcolor[rgb]{1,0.36,0.36}{\textbf{48.16$\pm$0.21}}&67.23$\pm$0.11&51.83$\pm$0.19& 65.04 \\
			WavLM-large&59.99$\pm$0.09&\textcolor[rgb]{1,0,0}{\textbf{69.55$\pm$0.49}}&\textcolor[rgb]{1,0,0}{\textbf{77.87$\pm$0.12}}&71.57$\pm$0.53&68.36$\pm$0.16&45.84$\pm$0.23&\textcolor[rgb]{1,0,0}{\textbf{71.48$\pm$0.11}}&\textcolor[rgb]{1,0,0}{\textbf{56.13$\pm$0.23}}& \textcolor[rgb]{1,0.72,0.72}{\textbf{65.10}} \\
			Whisper-large&57.63$\pm$0.84&65.25$\pm$0.51&\textcolor[rgb]{1,0.36,0.36}{\textbf{76.72$\pm$0.39}}&\textcolor[rgb]{1,0.72,0.72}{\textbf{76.08$\pm$0.54}}&\textcolor[rgb]{1,0.72,0.72}{\textbf{74.28$\pm$0.59}}&\textcolor[rgb]{1,0,0}{\textbf{49.17$\pm$0.15}}&\textcolor[rgb]{1,0.36,0.36}{\textbf{69.68$\pm$0.24}}&\textcolor[rgb]{1,0.72,0.72}{\textbf{52.72$\pm$0.40}}& \textcolor[rgb]{1,0.36,0.36}{\textbf{65.19}} \\
			HUBERT-large&\textcolor[rgb]{1,0,0}{\textbf{69.91$\pm$0.58}}&59.21$\pm$1.05&73.07$\pm$0.13&\textcolor[rgb]{1,0,0}{\textbf{82.49$\pm$0.42}}&\textcolor[rgb]{1,0,0}{\textbf{78.65$\pm$0.33}}&46.54$\pm$0.13&\textcolor[rgb]{1,0.72,0.72}{\textbf{69.67$\pm$0.15}}&\textcolor[rgb]{1,0.36,0.36}{\textbf{53.41$\pm$0.10}}& \textcolor[rgb]{1,0,0}{\textbf{66.62}} \\
			
			\hline
			\multicolumn{10}{c}{Lexical Modality} \\
			\hline
			
			OPT-13B&45.78$\pm$0.19&74.56$\pm$0.27&79.70$\pm$0.17&72.55$\pm$0.61&70.12$\pm$0.21&57.49$\pm$0.13&61.70$\pm$0.05&48.05$\pm$0.08& 63.75 \\
			ALBERT-small&46.43$\pm$0.29&74.16$\pm$0.63&79.50$\pm$0.24&73.27$\pm$0.23&72.30$\pm$0.22&56.42$\pm$0.12&61.14$\pm$0.14&47.43$\pm$0.23& 63.83 \\
			XLNet-base&50.01$\pm$0.46&78.89$\pm$0.22&82.13$\pm$0.14&74.28$\pm$0.47&77.31$\pm$0.11&57.21$\pm$0.08&62.29$\pm$0.18&47.62$\pm$0.05& 66.22 \\
			Llama-13B&49.15$\pm$0.63&78.07$\pm$0.17&83.04$\pm$0.15&75.72$\pm$0.16&73.70$\pm$0.27&57.62$\pm$0.12&64.24$\pm$0.14&50.01$\pm$0.16& 66.44 \\
			Vicuna-13B&49.35$\pm$0.28&78.77$\pm$0.34&82.99$\pm$0.15&76.72$\pm$0.58&74.54$\pm$0.34&58.51$\pm$0.26&65.32$\pm$0.12&50.77$\pm$0.14& 67.12 \\
			DeBERTa-large&52.48$\pm$0.20&80.01$\pm$0.31&82.36$\pm$0.09&77.83$\pm$0.53&76.05$\pm$0.20&57.04$\pm$0.15&62.93$\pm$0.13&48.33$\pm$0.17& 67.13 \\
			StableLM-7B&51.10$\pm$0.30&78.05$\pm$0.31&83.32$\pm$0.18&75.89$\pm$0.42&74.31$\pm$0.31&58.20$\pm$0.16&65.52$\pm$0.12&51.19$\pm$0.08& 67.20 \\
			MOSS-7B&54.35$\pm$0.28&76.34$\pm$0.54&82.91$\pm$0.19&76.40$\pm$0.49&77.82$\pm$0.24&57.76$\pm$0.10&63.98$\pm$0.06&48.78$\pm$0.03& 67.29 \\
			Llama2-13B&51.63$\pm$0.43&78.62$\pm$0.16&83.62$\pm$0.09&75.46$\pm$0.49&75.37$\pm$0.20&58.09$\pm$0.12&65.74$\pm$0.14&50.75$\pm$0.06& 67.41 \\
			PERT-base&52.75$\pm$0.48&81.19$\pm$0.08&83.36$\pm$0.12&78.02$\pm$0.54&77.76$\pm$0.13&57.20$\pm$0.16&64.73$\pm$0.05&49.44$\pm$0.06& 68.06 \\
			ELECTRA-base&54.61$\pm$0.33&81.51$\pm$0.19&83.38$\pm$0.08&78.86$\pm$0.46&79.21$\pm$0.15&59.03$\pm$0.27&62.79$\pm$0.08&49.03$\pm$0.08& 68.55 \\
			Falcon-7B&54.45$\pm$0.30&79.50$\pm$0.67&83.75$\pm$0.13&79.43$\pm$0.62&77.42$\pm$0.26&59.03$\pm$0.11&66.13$\pm$0.16&50.97$\pm$0.08& 68.83 \\
			ChatGLM2-6B&56.86$\pm$0.55&81.09$\pm$0.24&84.02$\pm$0.07&79.17$\pm$0.22&78.37$\pm$0.20&\textcolor[rgb]{1,0.36,0.36}{\textbf{59.69$\pm$0.21}}&66.25$\pm$0.15&\textcolor[rgb]{1,0.72,0.72}{\textbf{51.31$\pm$0.06}}& 69.59 \\
			MacBERT-base&56.38$\pm$0.32&\textcolor[rgb]{1,0.36,0.36}{\textbf{83.89$\pm$0.19}}&84.03$\pm$0.25&80.24$\pm$0.26&79.73$\pm$0.10&57.04$\pm$0.14&65.41$\pm$0.04&50.18$\pm$0.08& 69.61 \\
			Sentence-BERT&51.74$\pm$0.65&\textcolor[rgb]{1,0,0}{\textbf{85.91$\pm$0.32}}&\textcolor[rgb]{1,0,0}{\textbf{85.48$\pm$0.07}}&75.68$\pm$0.23&77.40$\pm$0.51&\textcolor[rgb]{1,0,0}{\textbf{60.56$\pm$0.28}}&\textcolor[rgb]{1,0,0}{\textbf{68.10$\pm$0.18}}&\textcolor[rgb]{1,0,0}{\textbf{52.42$\pm$0.03}}& 69.66 \\
			LERT-base&57.30$\pm$0.46&83.17$\pm$0.08&\textcolor[rgb]{1,0.72,0.72}{\textbf{84.33$\pm$0.11}}&79.23$\pm$0.95&79.94$\pm$0.18&58.86$\pm$0.24&65.54$\pm$0.04&50.88$\pm$0.14& 69.90 \\
			BLOOM-7B&\textcolor[rgb]{1,0.36,0.36}{\textbf{59.23$\pm$0.37}}&79.72$\pm$0.30&83.90$\pm$0.24&\textcolor[rgb]{1,0.36,0.36}{\textbf{82.53$\pm$0.27}}&\textcolor[rgb]{1,0.36,0.36}{\textbf{80.64$\pm$0.12}}&\textcolor[rgb]{1,0.72,0.72}{\textbf{59.42$\pm$0.32}}&\textcolor[rgb]{1,0.72,0.72}{\textbf{66.76$\pm$0.12}}&51.06$\pm$0.12& \textcolor[rgb]{1,0.72,0.72}{\textbf{70.41}} \\
			RoBERTa-large&\textcolor[rgb]{1,0.72,0.72}{\textbf{58.22$\pm$0.69}}&\textcolor[rgb]{1,0.72,0.72}{\textbf{83.17$\pm$0.30}}&84.11$\pm$0.11&\textcolor[rgb]{1,0,0}{\textbf{82.56$\pm$0.40}}&\textcolor[rgb]{1,0.72,0.72}{\textbf{80.04$\pm$0.23}}&59.20$\pm$0.24&66.21$\pm$0.18&50.98$\pm$0.09& \textcolor[rgb]{1,0.36,0.36}{\textbf{70.56}} \\
			Baichuan-13B&\textcolor[rgb]{1,0,0}{\textbf{62.54$\pm$0.69}}&79.37$\pm$0.39&\textcolor[rgb]{1,0.36,0.36}{\textbf{84.42$\pm$0.05}}&\textcolor[rgb]{1,0.72,0.72}{\textbf{81.99$\pm$0.50}}&\textcolor[rgb]{1,0,0}{\textbf{80.93$\pm$0.19}}&58.27$\pm$0.08&\textcolor[rgb]{1,0.36,0.36}{\textbf{67.01$\pm$0.14}}&\textcolor[rgb]{1,0.36,0.36}{\textbf{51.70$\pm$0.13}}& \textcolor[rgb]{1,0,0}{\textbf{70.78}} \\

			\hline
		\end{tabular}
	\end{table*}

	\subsection{Corpus Description}
	In addition to MER2023, we also include mainstream datasets for multimodal emotion recognition. The statistical information for all corpora is presented in Table \ref{Table4}.
	
	IEMOCAP \cite{busso2008iemocap} consists of five sessions, each with two actors performing improvised or scripted conversations. We adopt two popular label processing methods, resulting in four-class \cite{poria2017context, hazarika2018conversational} and six-class \cite{majumder2019dialoguernn, mai2020modality} versions. CMU-MOSI \cite{zadeh2017tensor} contains movie review videos from online websites. CMU-MOSEI \cite{zadeh2018multimodal} extends CMU-MOSI with more samples and a greater variety of topics. CH-SIMS \cite{yu2020ch} and CH-SIMS v2 \cite{liu2022make} are Chinese datasets collected from movies, TV series, and shows. MELD \cite{poria2019meld} extends EmotionLines \cite{chen2018emotionlines} with multimodal information. For MER2023, we focus on the MER-MULTI subset in this benchmark. 
	
	Since IEMOCAP does not provide official data splitting, we perform five-fold cross-validation using the leave-one-session-out strategy. For other datasets, they provide training/validation/test sets and we follow the default data splitting method to ensure a fair comparison. Among these datasets, IEMOCAP, MELD, and MER-MULTI provide discrete labels, while the remaining datasets utilize continuous sentiment scores. For the latter, we focus on the negative/positive classification task, where positive and negative classes are assigned for $< 0$ and $> 0$ scores, respectively. In this way, all datasets are transformed into categorical datasets, and we use WAF as the evaluation metric.

	\begin{table*}[t]
		\centering
		\renewcommand\tabcolsep{3.6pt}
		\caption{Importance of domain compatibility for visual encoders. In this table, we select VideoMAE and train it on different datasets.}
		\label{Table6}
		\begin{tabular}{cc|cccccccc}
			\hline
			{VideoMAE} &Training Data& {MER-MULTI} & {CMU-MOSI} & {CMU-MOSEI} & {CH-SIMS} & {CH-SIMS v2} & {MELD} & {\begin{tabular}[c]{@{}c@{}}IEMOCAP \\ (four-class)\end{tabular}} & {\begin{tabular}[c]{@{}c@{}}IEMOCAP \\ (six-class) \end{tabular}} \\
			
			\hline
			\rowcolor{lightgray}
			base&Kinetics-400 &46.72$\pm$0.75& 56.97$\pm$0.64 & 70.40$\pm$0.05 &68.95$\pm$0.69&70.11$\pm$0.26&36.74$\pm$0.20&39.15$\pm$0.18&26.76$\pm$0.17 \\
			base&MER-SEMI &60.86$\pm$0.61&58.67$\pm$0.67&70.61$\pm$0.31&81.37$\pm$0.36&79.12$\pm$0.32&38.82$\pm$0.24&48.32$\pm$0.14&33.44$\pm$0.20 \\
			base&VoxCeleb2 &61.76$\pm$0.32&63.34$\pm$0.51&69.83$\pm$0.20&78.33$\pm$0.15&77.52$\pm$0.27&40.05$\pm$0.20&46.97$\pm$0.36&32.23$\pm$0.11 \\
			\hline
			
			\rowcolor{lightgray}
			large&Kinetics-400 &53.26$\pm$0.68&54.60$\pm$0.78&70.78$\pm$0.03&72.44$\pm$0.21&73.97$\pm$0.45&37.85$\pm$0.17&40.17$\pm$0.27&27.59$\pm$0.19 \\
			large&MER-SEMI &59.89$\pm$0.37&61.02$\pm$0.49&70.46$\pm$0.16&81.36$\pm$0.22&77.51$\pm$0.21&39.17$\pm$0.16&48.12$\pm$0.16&33.49$\pm$0.06 \\
			large&VoxCeleb2 &58.35$\pm$0.32&60.50$\pm$0.57&70.94$\pm$0.21&77.14$\pm$0.25&76.13$\pm$0.17&39.48$\pm$0.19&48.66$\pm$0.52&33.65$\pm$0.20 \\
			
			\hline
			
		\end{tabular}
	\end{table*}

	\begin{table*}[t]
		\centering
		\renewcommand\tabcolsep{3pt}
		\caption{Importance of language matching for acoustic encoders. To generate audio in the target language, we use ChatGPT 3.5 for translation and the TTS system for audio synthesis. We test encoders trained in different languages, including English (EN), Chinese (CH), and multilingual (MULTI).}
		\label{Table7}
		\begin{tabular}{cc|cccccccc}
			\hline
			{\begin{tabular}[c]{@{}c@{}} Encoder (Lang.) \end{tabular}} & {TTS Lang.} & {MER-MULTI} & {CMU-MOSI} & {CMU-MOSEI} & {CH-SIMS} & {CH-SIMS v2} & {MELD} & {\begin{tabular}[c]{@{}c@{}}IEMOCAP \\ (four-class)\end{tabular}} & {\begin{tabular}[c]{@{}c@{}}IEMOCAP \\ (six-class) \end{tabular}} \\
			\hline
			
			\multirow{2}{*}{\begin{tabular}[c]{@{}c@{}} data2vec-base\\(EN) \end{tabular}} & EN & \textbf{35.03$\pm$0.10} & \textbf{69.53$\pm$0.44} & \textbf{72.82$\pm$0.28} & \textbf{65.23$\pm$0.68} & \textbf{64.13$\pm$0.15} & \textbf{46.14$\pm$0.27} & \textbf{59.77$\pm$0.27} & \textbf{45.67$\pm$0.13} \\
			& CH &28.61$\pm$0.40&60.54$\pm$0.71&66.40$\pm$0.18&62.24$\pm$0.52&58.89$\pm$0.52&42.01$\pm$0.53&48.01$\pm$0.61&36.33$\pm$0.15 \\
			
			\hline
			
			\multirow{2}{*}{\begin{tabular}[c]{@{}c@{}} WavLM-large\\(MULTI, mainly EN) \end{tabular}} & EN & \textbf{39.21$\pm$0.59} & \textbf{71.52$\pm$0.39} & \textbf{77.23$\pm$0.27} & \textbf{67.78$\pm$0.28} & \textbf{67.27$\pm$0.36} & \textbf{47.60$\pm$0.08} & \textbf{60.53$\pm$0.23} & \textbf{47.25$\pm$0.14} \\
			& CH &31.33$\pm$0.24&61.99$\pm$0.37&68.38$\pm$0.36&63.84$\pm$0.46&61.08$\pm$0.45&43.03$\pm$0.28&51.97$\pm$0.19&39.85$\pm$0.24 \\
			
			\hline
			
			\multirow{2}{*}{\begin{tabular}[c]{@{}c@{}}HUBERT-large\\(CH) \end{tabular}} & EN &29.52$\pm$0.35&64.80$\pm$0.56&68.99$\pm$0.14&62.03$\pm$0.56&55.89$\pm$0.41&42.57$\pm$0.36&50.77$\pm$0.48&39.13$\pm$0.20 \\
			& CH & \textbf{42.82$\pm$0.38} & \textbf{70.16$\pm$0.51} & \textbf{77.22$\pm$0.29} & \textbf{72.78$\pm$0.41} & \textbf{73.47$\pm$0.23} & \textbf{47.88$\pm$0.31} & \textbf{58.70$\pm$0.31} & \textbf{44.99$\pm$0.12} \\
			
			\hline
			
		\end{tabular}
	\end{table*}

	\subsection{Unimodal Benchmark}
	In this section, we establish the unimodal benchmark for all datasets and report results in Table \ref{Table5}. We hope this benchmark can provide guidance for feature selection and point the way to developing powerful feature extractors.
	
	\textbf{Visual Modality.}
	We evaluate some representative fully-, weakly-, and self-supervised visual encoders. A fully-supervised model relies on both the model structure and the training corpus. As shown in Table \ref{Table5}, models trained for facial expression recognition tend to exhibit superior performance than object recognition and face recognition. These findings demonstrate the importance of domain compatibility between training corpora and downstream tasks.
	
	Although different datasets exhibit preferences for distinct visual encoders, their best features consistently come from weakly- or self-supervised models. These results indicate that weakly- or self-supervised models can learn universal visual representations, which are also useful for emotion recognition. Notably, the model cards in Table \ref{Table18} (see Appendix) reveal that current encoders are often trained on action recognition datasets (e.g., Kinetics-400 \cite{kay2017kinetics}) or web images (e.g., LVD-142M \cite{oquab2023dinov2}). A heuristic idea is to further narrow the domain gap by training on human-centric videos. In Table \ref{Table6}, we choose a self-supervised model, VideoMAE, and train it on human-centric corpora, MER-SEMI and VoxCeleb2 \cite{chung2018voxceleb2}. Consistent with previous findings \cite{sun2023mae}, this approach significantly improves performance, which suggests future directions for visual encoders. In emotion recognition, a good visual encoder should focus on weakly- or self-supervised learning and train on large amounts of human-centric videos.
	
	\textbf{Audio Modality.}
	In Table \ref{Table5}, we observe that acoustic encoders trained on Chinese data (e.g., HUBERT-large) generally perform well on Chinese emotion corpora (e.g., MER-MULTI and CH-SIMS), while acoustic encoders trained on English data (e.g., WavLM-large) generally perform well on English emotion corpora (e.g., CMU-MOSI and CMU-MOSEI). These results suggest that acoustic encoders are language-sensitive. To uncover the underlying reasons behind this phenomenon, we conduct more experiments.
	
	Audio contains linguistic and paralinguistic information. Considering that linguistic information is language-sensitive, a heuristic conjecture is that the reason behind the language sensitivity of acoustic encoders is that they can implicitly capture linguistic information. To prove this guess, we use neutral voice to synthesize audio, eliminating the interference of emotion-related paralinguistic information. To generate audio in the target language, we use ChatGPT 3.5 for translation and the TTS system for audio generation. Experimental results are shown in Table \ref{Table7}. In this table, we test the performance of acoustic encoders trained in different languages. We observe that language-matching encoders consistently achieve better performance. These results show that the acoustic encoder can implicitly capture linguistic information, leading to its language sensitivity.
	
\begin{figure*}[t]
		\begin{center}
			\subfigure[MER-MULTI (CH)]{
				\label{Figure5-1}
				\centering
				\includegraphics[width=0.22\linewidth, trim=0 0 0 0]{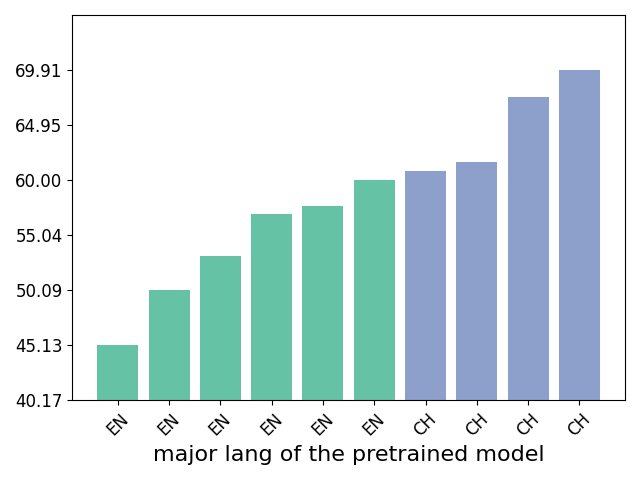}
			} 
			\subfigure[CH-SIMS (CH)]{
				\label{Figure5-4}
				\centering
				\includegraphics[width=0.22\linewidth, trim=0 0 0 0]{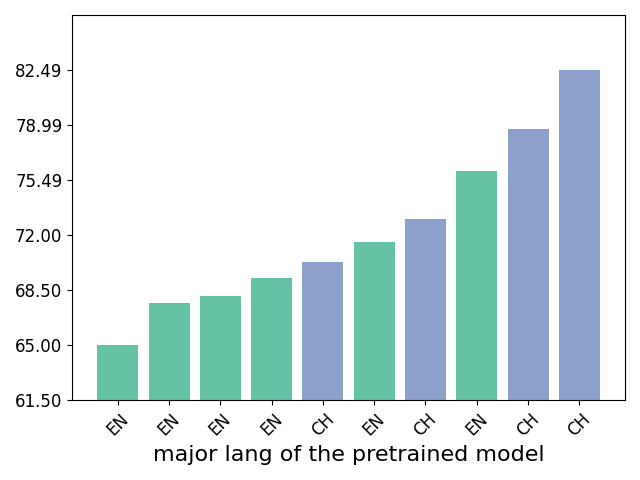}
			} 
			\subfigure[CH-SIMS v2 (CH)]{
				\label{Figure5-5}
				\centering
				\includegraphics[width=0.22\linewidth, trim=0 0 0 0]{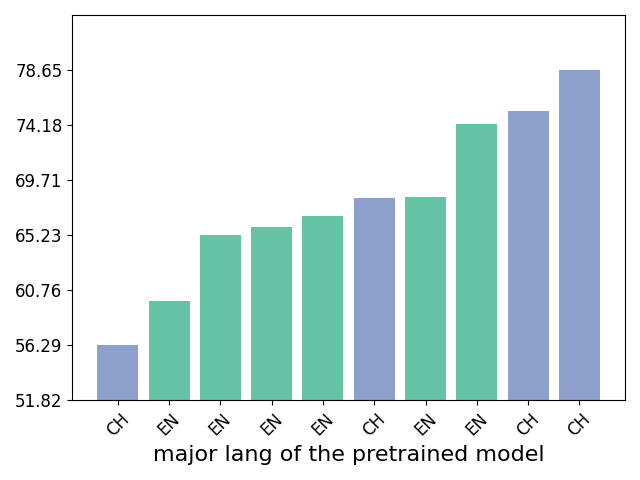}
			}
			\subfigure[CMU-MOSI (EN)]{
				\label{Figure5-2}
				\centering
				\includegraphics[width=0.22\linewidth, trim=0 0 0 0]{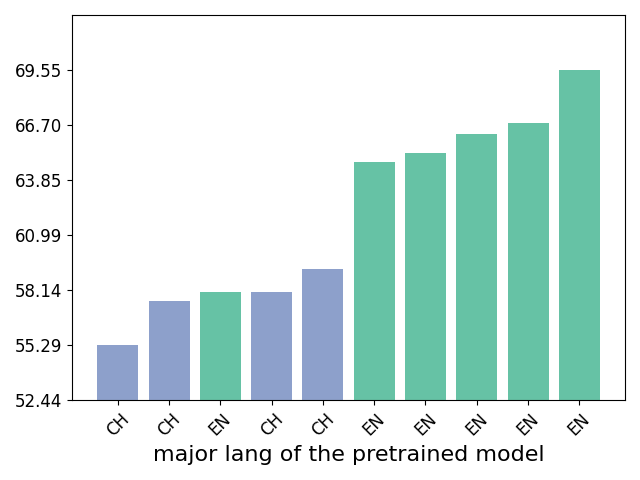}
			}
			\subfigure[CMU-MOSEI (EN)]{
				\label{Figure5-3}
				\centering
				\includegraphics[width=0.22\linewidth, trim=0 0 0 0]{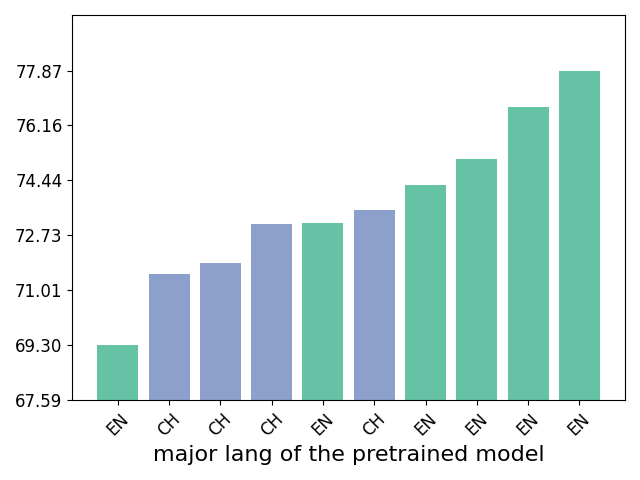}
			}
			\subfigure[MELD (EN)]{
				\label{Figure5-6}
				\centering
				\includegraphics[width=0.22\linewidth, trim=0 0 0 0]{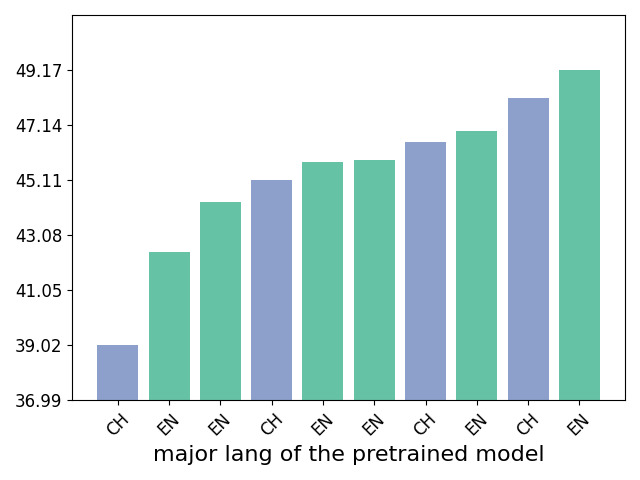}
			}
			\subfigure[IEMOCAP(four) (EN)]{
				\label{Figure5-7}
				\centering
				\includegraphics[width=0.22\linewidth, trim=0 0 0 0]{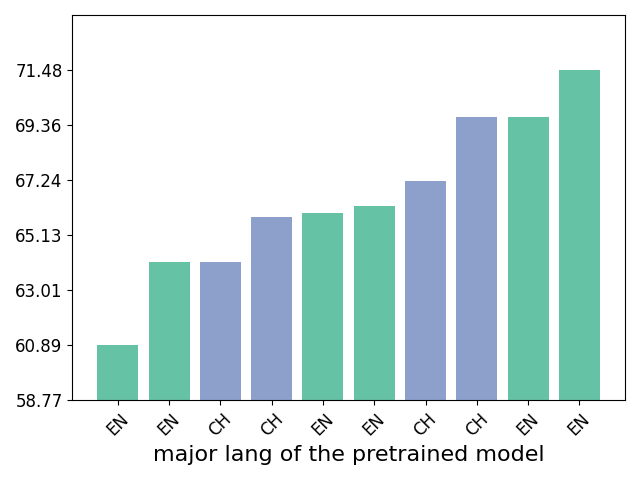}
			}
			\subfigure[IEMOCAP(six) (EN)]{
				\label{Figure5-8}
				\centering
				\includegraphics[width=0.22\linewidth, trim=0 0 0 0]{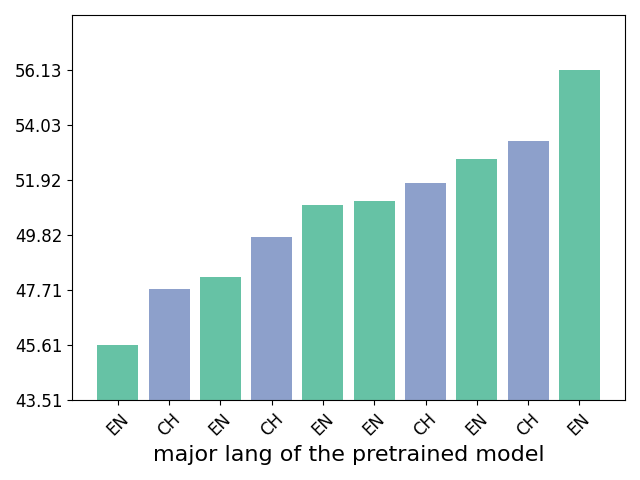}
			}
		\end{center}
		\caption{Impact of language matching for acoustic encoders. In this table, we reveal the relationship between the primary training language of the acoustic encoder and the input language.}
		\label{Figure5}
	\end{figure*}

	Fig. \ref{Figure5} further reveals the relationship between the major training language of the acoustic encoder and the input language. Although language-matching encoders generally result in better performance, there are some exceptions. The reason lies in that audio conveys emotions through linguistic and paralinguistic information. Although language-matching encoders can capture more linguistic information, good encoders can capture more emotion-related paralinguistic information and achieve better performance.
	
	In summary, acoustic encoders can capture both linguistic and paralinguistic information. To capture linguistic information, it is best to use a language-matching encoder. To capture paralinguistic information, it is best to train the encoder with more expressive audio rather than just neutral audio. If you want a multilingual universal acoustic encoder, we recommend that the encoder should be trained on large amounts of expressive multilingual audio.

	\textbf{Lexical Modality.}
	Different lexical encoders support distinct languages. In Table \ref{Table5}, we focus on lexical encoders that support Chinese (LLMs are generally trained on multilingual corpora containing Chinese). For English emotion datasets, we use ChatGPT 3.5 to translate them into Chinese. We observe some interesting results in Table \ref{Table5}. For emotion recognition, Baichuan-13B can achieve very promising results, but some powerful LLMs (such as Llama-13B and OPT-13B) perform poorly. To figure out the reason behind this phenomenon, we further investigate the impact of language matching on lexical encoders.
	
	Table \ref{Table8} evaluates lexical encoders that support multiple languages, but each encoder has distinct major languages. In this table, we exploit ChatGPT 3.5 to translate the input text into the target language. For Baichuan-13B and BLOOM-7B, despite powerful translation systems like ChatGPT 3.5, emotion-related information is still lost during translation, resulting in performance degradation. Differently, Llama2-13B demonstrates superior results when translating MER-MULTI from Chinese to English. The reason lies in that although Llama2-13B supports Chinese, its primary training language is English, allowing it to better comprehend English. Therefore, we should focus on the primary language of lexical encoders, preferably aligning it with the input language and avoiding the use of translation systems.

	\begin{table*}[t]
		\centering
		\renewcommand\tabcolsep{3pt}
		\caption{Importance of language matching for lexical encoders. We choose encoders that support multiple languages but have different major languages. We use ChatGPT 3.5 to translate the input into the target language. $\dag$ denotes the result using translation and bold indicates the best result.}
		\label{Table8}
		\begin{tabular}{ccc|cccccccc}
			\hline
			\multirow{2}{*}{Feature} & \multicolumn{2}{c|}{Major Lang.} & \multirow{2}{*}{MER-MULTI} & \multirow{2}{*}{CMU-MOSI} & \multirow{2}{*}{CMU-MOSEI} & \multirow{2}{*}{CH-SIMS} & \multirow{2}{*}{CH-SIMS v2} & \multirow{2}{*}{MELD} & \multirow{2}{*}{\begin{tabular}[c]{@{}c@{}}IEMOCAP \\ (four-class)\end{tabular}} & \multirow{2}{*}{\begin{tabular}[c]{@{}c@{}}IEMOCAP \\ (six-class) \end{tabular}} \\
			&Train & Test &&&&&&& \\
			\hline
			Baichuan-13B & EN, CH & EN&55.56$\pm$0.35$^\dag$&\textbf{82.31$\pm$0.58}&\textbf{85.09$\pm$0.20}&76.41$\pm$0.29$^\dag$&78.59$\pm$0.35$^\dag$&\textbf{62.02$\pm$0.16}&\textbf{69.43$\pm$0.14}&\textbf{54.45$\pm$0.21} \\			
			Baichuan-13B & EN, CH & CH&\textbf{62.54$\pm$0.69} &79.37$\pm$0.39$^\dag$& 84.42$\pm$0.05$^\dag$ & \textbf{81.99$\pm$0.50} & \textbf{80.93$\pm$0.19} &58.27$\pm$0.08$^\dag$& 67.01$\pm$0.14$^\dag$ & 51.70$\pm$0.13$^\dag$ \\

			\hline			
			
			BLOOM-7B & MULTI & EN&53.35$\pm$0.80$^\dag$&\textbf{79.72$\pm$0.22}&\textbf{84.37$\pm$0.14}&78.13$\pm$0.47$^\dag$&79.23$\pm$0.24$^\dag$&\textbf{61.85$\pm$0.28}&\textbf{69.32$\pm$0.06}&\textbf{54.20$\pm$0.12} \\
			BLOOM-7B & MULTI & CH & \textbf{59.23$\pm$0.37} &79.72$\pm$0.30$^\dag$&83.90$\pm$0.24$^\dag$& \textbf{82.53$\pm$0.27} & \textbf{80.64$\pm$0.12} & 59.42$\pm$0.32$^\dag$ & 66.76$\pm$0.12$^\dag$ &51.06$\pm$0.12$^\dag$ \\
			
			\hline
			
			Llama2-13B  & EN & EN&\textbf{53.77$\pm$0.29}$^\dag$&\textbf{81.81$\pm$0.30}&\textbf{85.29$\pm$0.14}&\textbf{77.75$\pm$0.32}$^\dag$&\textbf{78.42$\pm$0.23}$^\dag$&\textbf{62.23$\pm$0.29}&\textbf{70.28$\pm$0.11}&\textbf{54.64$\pm$0.16} \\
			Llama2-13B  & EN & CH&51.63$\pm$0.43&78.62$\pm$0.16$^\dag$&83.62$\pm$0.09$^\dag$&75.46$\pm$0.49&75.37$\pm$0.20&58.09$\pm$0.12$^\dag$&65.74$\pm$0.14$^\dag$&50.75$\pm$0.06$^\dag$ \\
			
			\hline
			
		\end{tabular}
	\end{table*}

	\begin{table*}[t]
		\centering
		\renewcommand\tabcolsep{5pt}
		\caption{Multimodal benchmark. $\diamondsuit$ and $\heartsuit$ represent utterance-level and sequence-level fusion algorithms. We test two feature sets: a medium-performance set (SENet-FER2013, wav2vec 2.0-base, DeBERTa-large) and a high-performance set (CLIP-large, HUBERT-large, Baichuan-13B).}
		\label{Table9}
		\begin{tabular}{l|cccccccc|c}
			\hline
			{Fusion} & {MER-MULTI} & {CMU-MOSI} & {CMU-MOSEI} & {CH-SIMS} & {CH-SIMS v2} & {MELD} & {\begin{tabular}[c]{@{}c@{}}IEMOCAP \\ (four-class)\end{tabular}} & {\begin{tabular}[c]{@{}c@{}}IEMOCAP \\ (six-class) \end{tabular}} & Mean \\
			
			\hline
			\multicolumn{10}{c}{Medium-Performance Set} \\
			\hline
			
			$\heartsuit$MCTN&49.41$\pm$0.23&79.74$\pm$0.36&\textcolor[rgb]{1,0,0}{\textbf{84.56$\pm$0.04}}&76.97$\pm$0.62&75.29$\pm$0.09&56.31$\pm$0.56&63.08$\pm$0.36&48.66$\pm$0.16 & 66.75 \\
			$\heartsuit$MFM&71.02$\pm$0.13&\textcolor[rgb]{1,0.72,0.72}{\textbf{79.86$\pm$0.16}}&82.21$\pm$0.35&86.44$\pm$0.90&83.84$\pm$0.36&54.55$\pm$0.20&70.84$\pm$0.21&54.12$\pm$0.02 & 72.86 \\
			$\diamondsuit$MISA&73.75$\pm$0.35&79.46$\pm$0.29&82.91$\pm$0.13&87.92$\pm$0.28&84.24$\pm$0.20&58.40$\pm$0.46&\textcolor[rgb]{1,0.36,0.36}{\textbf{72.84$\pm$0.16}}&45.48$\pm$0.15 & 73.12 \\
			$\heartsuit$GMFN&\textcolor[rgb]{1,0.72,0.72}{\textbf{74.84$\pm$0.17}}&79.56$\pm$0.05&82.64$\pm$0.05&87.53$\pm$0.08&83.97$\pm$0.11&56.73$\pm$0.11&71.22$\pm$0.12&55.14$\pm$0.39 & 73.96 \\
			$\heartsuit$MFN&74.78$\pm$0.21&\textcolor[rgb]{1,0.36,0.36}{\textbf{80.16$\pm$0.20}}&82.50$\pm$0.03&87.34$\pm$0.87&83.87$\pm$0.21&57.80$\pm$0.29&72.53$\pm$0.12&56.02$\pm$0.18 & 74.37 \\
			$\diamondsuit$MMIM&72.26$\pm$0.67&79.58$\pm$0.48&82.89$\pm$0.15&87.81$\pm$0.35&84.26$\pm$0.41&\textcolor[rgb]{1,0.36,0.36}{\textbf{59.03$\pm$0.40}}&\textcolor[rgb]{1,0.72,0.72}{\textbf{72.71$\pm$0.28}}&\textcolor[rgb]{1,0.36,0.36}{\textbf{56.69$\pm$0.18 }}& 74.40 \\
			$\heartsuit$MulT&\textcolor[rgb]{1,0,0}{\textbf{76.47$\pm$0.26}}&\textcolor[rgb]{1,0,0}{\textbf{80.83$\pm$1.01}}&\textcolor[rgb]{1,0.36,0.36}{\textbf{83.66$\pm$0.04}}&86.80$\pm$0.09&\textcolor[rgb]{1,0.36,0.36}{\textbf{85.15$\pm$0.05}}&57.63$\pm$0.44&71.13$\pm$0.04&55.07$\pm$0.02 & 74.59 \\
			$\diamondsuit$LMF&74.44$\pm$0.15&79.63$\pm$0.15&83.08$\pm$0.22&\textcolor[rgb]{1,0.36,0.36}{\textbf{88.41$\pm$0.39}}&84.52$\pm$0.21&58.24$\pm$0.39&72.14$\pm$0.24&\textcolor[rgb]{1,0.72,0.72}{\textbf{56.52$\pm$0.11 }}&\textcolor[rgb]{1,0.72,0.72}{\textbf{74.62}}\\
			$\diamondsuit$TFN&\textcolor[rgb]{1,0.36,0.36}{\textbf{75.25$\pm$0.46}}&79.37$\pm$0.36&83.10$\pm$0.12&\textcolor[rgb]{1,0,0}{\textbf{88.96$\pm$0.24}}&\textcolor[rgb]{1,0.72,0.72}{\textbf{84.60$\pm$0.24}}&\textcolor[rgb]{1,0.72,0.72}{\textbf{58.58$\pm$0.33}}&72.36$\pm$0.08&56.34$\pm$0.18 &\textcolor[rgb]{1,0.36,0.36}{\textbf{74.82}}\\
			$\diamondsuit$Attention&74.42$\pm$0.45&79.43$\pm$0.15&\textcolor[rgb]{1,0.72,0.72}{\textbf{83.19$\pm$0.14}}&\textcolor[rgb]{1,0.72,0.72}{\textbf{88.15$\pm$0.44}}&\textcolor[rgb]{1,0,0}{\textbf{85.21$\pm$0.28}}&\textcolor[rgb]{1,0,0}{\textbf{59.16$\pm$0.30}}&\textcolor[rgb]{1,0,0}{\textbf{73.00$\pm$0.20}}&\textcolor[rgb]{1,0,0}{\textbf{56.70$\pm$0.13 }}&\textcolor[rgb]{1,0,0}{\textbf{74.91}}\\
			
			\hline
			\multicolumn{10}{c}{High-Performance Set} \\
			\hline
			
			$\heartsuit$MCTN&42.85$\pm$0.28&78.85$\pm$0.30&83.87$\pm$0.02&67.69$\pm$10.6&78.94$\pm$0.26&48.96$\pm$0.57&62.06$\pm$1.80&43.34$\pm$2.23 & 63.32 \\
			$\heartsuit$MFM&70.48$\pm$0.62&78.34$\pm$1.23&83.67$\pm$0.45&87.14$\pm$0.47&87.80$\pm$0.08&51.45$\pm$0.22&70.83$\pm$0.06&51.47$\pm$0.86 & 72.65 \\
			$\heartsuit$GMFN&77.00$\pm$0.41&76.15$\pm$0.05&82.48$\pm$0.39&87.11$\pm$0.02&\textcolor[rgb]{1,0.72,0.72}{\textbf{88.35$\pm$0.45}}&56.58$\pm$0.61&73.69$\pm$0.25&55.52$\pm$0.04 & 74.61 \\
			$\heartsuit$MFN&77.40$\pm$0.03&77.78$\pm$0.27&82.58$\pm$0.02&87.00$\pm$0.13&87.86$\pm$0.28&57.33$\pm$0.15&72.70$\pm$0.03&56.15$\pm$0.29 & 74.85 \\
			$\diamondsuit$MISA&82.42$\pm$0.70&79.08$\pm$0.35&85.01$\pm$0.27&89.82$\pm$0.22&87.13$\pm$0.43&57.80$\pm$0.51&\textcolor[rgb]{1,0,0}{\textbf{78.27$\pm$0.09}}&48.43$\pm$0.24 & 75.99 \\
			$\diamondsuit$MMIM&81.01$\pm$0.41&79.38$\pm$0.42&84.64$\pm$0.21&88.68$\pm$0.47&87.20$\pm$0.16&58.63$\pm$0.50&77.94$\pm$0.20&\textcolor[rgb]{1,0.72,0.72}{\textbf{60.54$\pm$0.19 }}& 77.25 \\
			$\diamondsuit$TFN&82.52$\pm$0.48&\textcolor[rgb]{1,0.36,0.36}{\textbf{79.63$\pm$0.24}}&\textcolor[rgb]{1,0,0}{\textbf{85.36$\pm$0.15}}&\textcolor[rgb]{1,0.72,0.72}{\textbf{90.56$\pm$0.12}}&87.85$\pm$0.25&\textcolor[rgb]{1,0.36,0.36}{\textbf{58.82$\pm$0.34}}&77.38$\pm$0.12&59.43$\pm$0.27 & 77.69 \\
			$\heartsuit$MulT&\textcolor[rgb]{1,0.72,0.72}{\textbf{82.94$\pm$0.44}}&\textcolor[rgb]{1,0,0}{\textbf{81.41$\pm$0.72}}&84.88$\pm$0.06&\textcolor[rgb]{1,0,0}{\textbf{91.07$\pm$0.34}}&\textcolor[rgb]{1,0,0}{\textbf{89.30$\pm$0.82}}&\textcolor[rgb]{1,0.72,0.72}{\textbf{58.68$\pm$0.14}}&75.81$\pm$0.15&58.88$\pm$0.03 &\textcolor[rgb]{1,0.72,0.72}{\textbf{77.87}}\\
			$\diamondsuit$LMF&\textcolor[rgb]{1,0,0}{\textbf{83.44$\pm$0.69}}&79.11$\pm$0.57&\textcolor[rgb]{1,0.36,0.36}{\textbf{85.22$\pm$0.29}}&\textcolor[rgb]{1,0.36,0.36}{\textbf{90.84$\pm$0.86}}&87.94$\pm$0.06&58.62$\pm$0.23&\textcolor[rgb]{1,0.72,0.72}{\textbf{78.14$\pm$0.13}}&\textcolor[rgb]{1,0,0}{\textbf{60.95$\pm$0.18 }}&\textcolor[rgb]{1,0.36,0.36}{\textbf{78.03}}\\
			$\diamondsuit$Attention&\textcolor[rgb]{1,0.36,0.36}{\textbf{83.03$\pm$0.16}}&\textcolor[rgb]{1,0.72,0.72}{\textbf{79.40$\pm$0.20}}&\textcolor[rgb]{1,0.72,0.72}{\textbf{85.05$\pm$0.23}}&90.50$\pm$0.36&\textcolor[rgb]{1,0.36,0.36}{\textbf{88.48$\pm$0.32}}&\textcolor[rgb]{1,0,0}{\textbf{59.41$\pm$0.26}}&\textcolor[rgb]{1,0.36,0.36}{\textbf{78.20$\pm$0.12}}&\textcolor[rgb]{1,0.36,0.36}{\textbf{60.90$\pm$0.22 }}&\textcolor[rgb]{1,0,0}{\textbf{78.12}}\\
			
			\hline
		\end{tabular}
	\end{table*}

	\begin{table*}[t]
		\centering
		\caption{Impact of feature number in multimodal fusion. ``Top2'' means we select the top-2 features for each modality (their ranking is based on the overall results in Table \ref{Table5}). In this table, we exploit the attention mechanism for multimodal fusion.}
		\label{Table10}
		\begin{tabular}{c|cccccccc|c}
			\hline
			{\# Top} & {MER-MULTI} & {CMU-MOSI} & {CMU-MOSEI} & {CH-SIMS} & {CH-SIMS v2} & {MELD} & {\begin{tabular}[c]{@{}c@{}}IEMOCAP \\ (four-class)\end{tabular}} & {\begin{tabular}[c]{@{}c@{}}IEMOCAP \\ (six-class) \end{tabular}} & Mean \\
			
			\hline
			
			Top1 & 83.03$\pm$0.16 & 79.40$\pm$0.20 & 85.05$\pm$0.23 &90.50$\pm$0.36& 88.48$\pm$0.32 & 59.41$\pm$0.26 & 78.20$\pm$0.12 & 60.90$\pm$0.22 & 78.12 \\
			Top2&\textcolor[rgb]{1,0.72,0.72}{\textbf{84.46$\pm$0.68}}&82.84$\pm$0.27&\textcolor[rgb]{1,0.72,0.72}{\textbf{85.83$\pm$0.10}}&\textcolor[rgb]{1,0.36,0.36}{\textbf{91.57$\pm$0.43}}&\textcolor[rgb]{1,0.36,0.36}{\textbf{88.58$\pm$0.33}}&60.19$\pm$0.08&78.58$\pm$0.04&60.75$\pm$0.16 & 79.10 \\
			Top3&\textcolor[rgb]{1,0,0}{\textbf{84.75$\pm$0.26}}&83.33$\pm$0.34&85.46$\pm$0.16&91.10$\pm$0.60&\textcolor[rgb]{1,0,0}{\textbf{88.78$\pm$0.41}}&60.23$\pm$0.29&\textcolor[rgb]{1,0.36,0.36}{\textbf{79.14$\pm$0.15}}&61.53$\pm$0.12 & 79.29 \\
			Top4&\textcolor[rgb]{1,0.36,0.36}{\textbf{84.65$\pm$0.47}}&\textcolor[rgb]{1,0.72,0.72}{\textbf{83.58$\pm$0.31}}&85.71$\pm$0.13&\textcolor[rgb]{1,0,0}{\textbf{91.75$\pm$0.29}}&88.23$\pm$0.36&\textcolor[rgb]{1,0.72,0.72}{\textbf{61.48$\pm$0.39}}&78.83$\pm$0.06&\textcolor[rgb]{1,0.72,0.72}{\textbf{61.59$\pm$0.27 }}&\textcolor[rgb]{1,0.72,0.72}{\textbf{79.48}}\\
			Top5&83.53$\pm$0.35&\textcolor[rgb]{1,0,0}{\textbf{85.96$\pm$0.25}}&\textcolor[rgb]{1,0.36,0.36}{\textbf{86.43$\pm$0.05}}&\textcolor[rgb]{1,0.72,0.72}{\textbf{91.17$\pm$0.51}}&87.94$\pm$0.47&\textcolor[rgb]{1,0.36,0.36}{\textbf{61.90$\pm$0.23}}&\textcolor[rgb]{1,0.72,0.72}{\textbf{79.11$\pm$0.07}}&\textcolor[rgb]{1,0,0}{\textbf{61.84$\pm$0.11 }}&\textcolor[rgb]{1,0.36,0.36}{\textbf{79.73}}\\
			Top6&83.85$\pm$0.12&\textcolor[rgb]{1,0.36,0.36}{\textbf{85.94$\pm$0.29}}&\textcolor[rgb]{1,0,0}{\textbf{86.80$\pm$0.12}}&91.06$\pm$0.27&\textcolor[rgb]{1,0.72,0.72}{\textbf{88.57$\pm$0.32}}&\textcolor[rgb]{1,0,0}{\textbf{62.21$\pm$0.32}}&\textcolor[rgb]{1,0,0}{\textbf{79.15$\pm$0.06}}&\textcolor[rgb]{1,0.36,0.36}{\textbf{61.80$\pm$0.19 }}&\textcolor[rgb]{1,0,0}{\textbf{79.92}}\\
			
			\hline
		\end{tabular}
	\end{table*}

	\subsection{Multimodal Benchmark}
	\label{sec:multibench}
	In this section, we evaluate mainstream fusion strategies, including utterance-level algorithms (e.g., TFN \cite{zadeh2017tensor}, LMF \cite{liu2018efficient}, MISA \cite{hazarika2020misa}, MMIM \cite{han2021improving}, Attention in Section \ref{sec:model}) and sequence-level algorithms (MFN \cite{zadeh2018memory}, GMFN \cite{zadeh2018multimodal}, MCTN \cite{pham2019found}, MFM \cite{tsai2018learning}, MulT \cite{tsai2019multimodal}). Since the choice of unimodal features seriously affects multimodal results, we compare the performance of different fusion algorithms under two feature sets: a medium-performance set (SENet-FER2013, wav2vec 2.0-base, DeBERTa-large) and a high-performance set (CLIP-large, HUBERT-large, Baichuan-13B).
	
	In Table \ref{Table9}, different combinations of features and datasets prefer distinct fusion algorithms. These results show the necessity of each algorithm and we should choose the best one for each combination. According to the overall results, MulT \cite{tsai2019multimodal} is the best sequence-level algorithm. The reason lies in that MulT does not rely on hard alignment but uses Transformer to capture dynamic alignment, which can exploit long-term cross-modal interactions and achieve better performance. Furthermore, the attention mechanism achieves the best overall results, indicating that complex fusion algorithms can easily cause overfitting due to the small size of emotional datasets. The simple but effective attention mechanism can achieve relatively good performance among all algorithms. Therefore, we use the attention mechanism as the default fusion strategy in the following experiments.
	
	Table \ref{Table10} illustrates the impact of feature number in multimodal fusion. Specifically, ``Top2'' denotes the selection of the top-2 features for each modality (i.e., MANet-RAFDB, CLIP-large, Whisper-large, HUBERT-large, RoBERTa-large, and Baichuan-13B). From Table \ref{Table10}, we observe that increasing the number of features generally enhances the results. However, incorporating poorly-performing features can also have a detrimental effect. Therefore, it is necessary to adjust the feature number for each dataset.
	
\begin{figure}[t]
		\begin{center}
			\subfigure[\scriptsize MER-MULTI]{
				\label{Figure6-1}
				\centering
				\includegraphics[width=0.22\linewidth, trim=20 23 20 23]{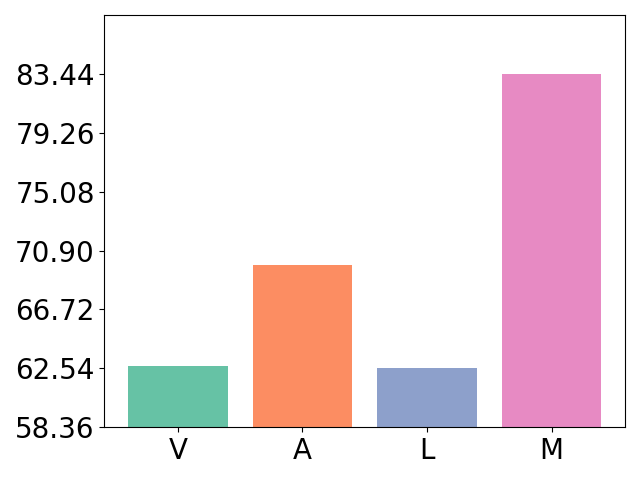}
			} 
			\subfigure[\scriptsize CMU-MOSI]{
				\label{Figure6-2}
				\centering
				\includegraphics[width=0.22\linewidth, trim=20 23 20 23]{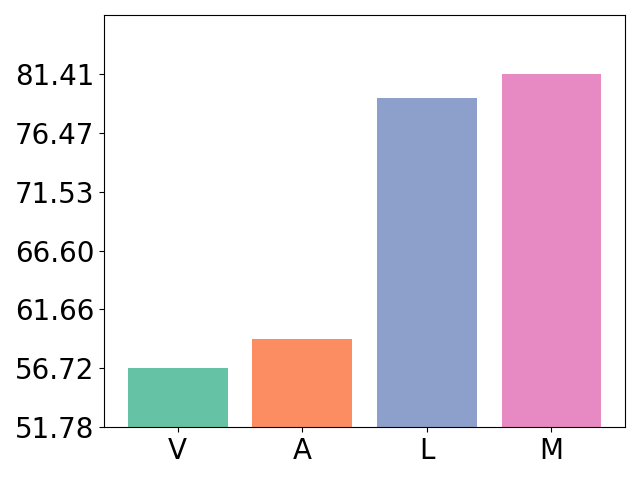}
			}
			\subfigure[\scriptsize CMU-MOSEI]{
				\label{Figure6-3}
				\centering
				\includegraphics[width=0.22\linewidth, trim=20 23 20 23]{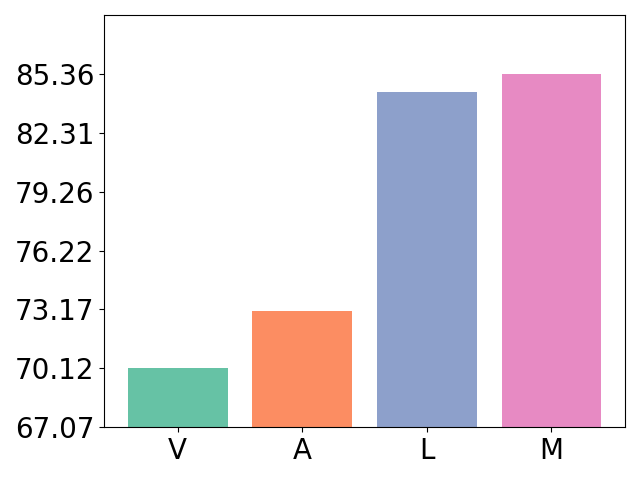}
			}
			\subfigure[\scriptsize CH-SIMS]{
				\label{Figure6-4}
				\centering
				\includegraphics[width=0.22\linewidth, trim=20 23 20 23]{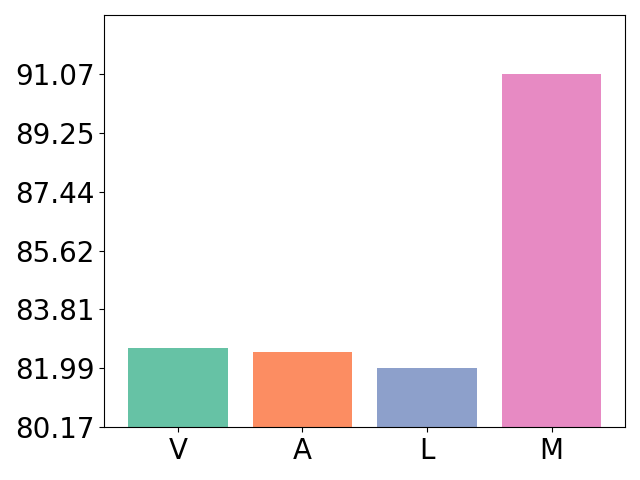}
			} 
			\subfigure[\scriptsize CH-SIMS v2]{
				\label{Figure6-5}
				\centering
				\includegraphics[width=0.22\linewidth, trim=20 23 20 23]{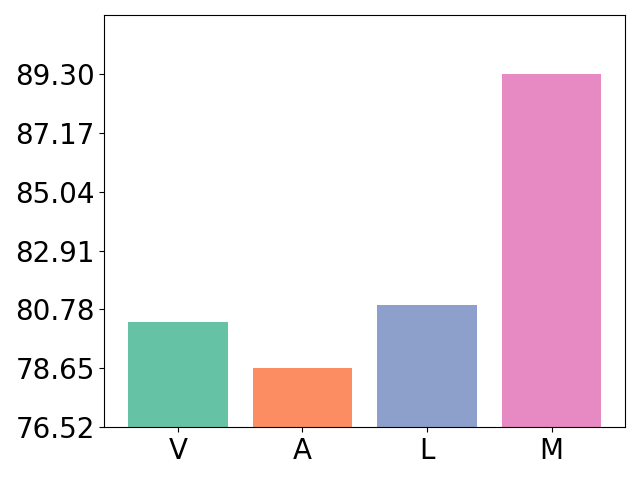}
			}
			\subfigure[\scriptsize MELD]{
				\label{Figure6-6}
				\centering
				\includegraphics[width=0.22\linewidth, trim=20 23 20 23]{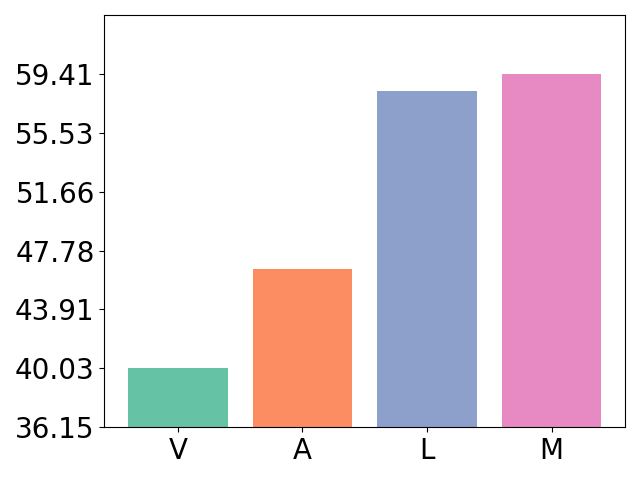}
			}
			\subfigure[\scriptsize IEMOCAP-4]{
				\label{Figure6-7}
				\centering
				\includegraphics[width=0.22\linewidth, trim=20 23 20 23]{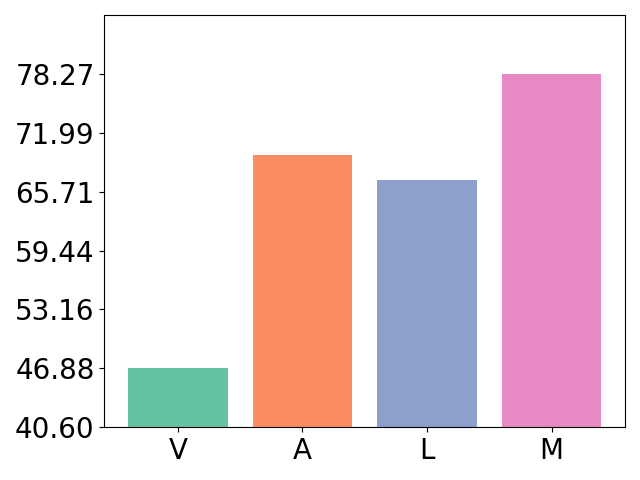}
			}
			\subfigure[\scriptsize IEMOCAP-6]{
				\label{Figure6-8}
				\centering
				\includegraphics[width=0.22\linewidth, trim=20 23 20 23]{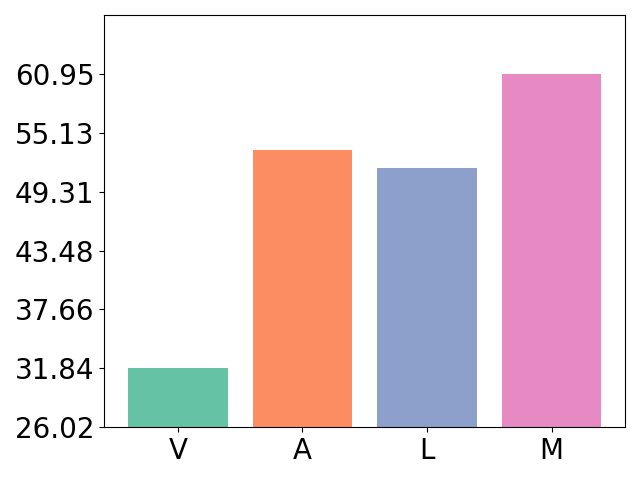}
			}
		\end{center}
		\caption{Modality preference analysis. Here, ``V'', ``A'', ``L'', and ``M'' represent visual, acoustic, lexical, and multimodal results, respectively.}
		\label{Figure6}
	\end{figure}

	Fig. \ref{Figure6} shows unimodal and multimodal results for each dataset. Interestingly, different datasets convey emotions in distinct ways. For example, MER-MULTI mainly conveys emotions through audio, while CMU-MOSI relies more on textual cues for emotional expression. In other words, although multimodal datasets provide information across all modalities, they may not be equally suitable for unimodal research. For example, it is not suitable to conduct visual emotion recognition on IEMOCAP as it conveys fewer emotions through video. Similarly, some datasets are not suitable for multimodal fusion research. For example, CMU-MOSI, CMU-MOSEI, and MELD heavily emphasize the lexical modality, making it challenging for advanced fusion algorithms to showcase their advantages.
	
	In summary, the choice of unimodal features significantly impacts multimodal results. Hence, different fusion algorithms should be compared under the same feature set. Although the attention mechanism may not consistently yield the optimal result, it can achieve relatively good performance among all algorithms. Additionally, not all multimodal emotion datasets are equally suitable for research on unimodal emotion recognition and multimodal fusion.
	
	\begin{figure*}[t]
		\begin{center}
			\subfigure[\scriptsize MOSEI$\rightarrow$MOSI]{
				\label{Figure7-1}
				\centering
				\includegraphics[width=0.15\linewidth, trim=20 0 0 0]{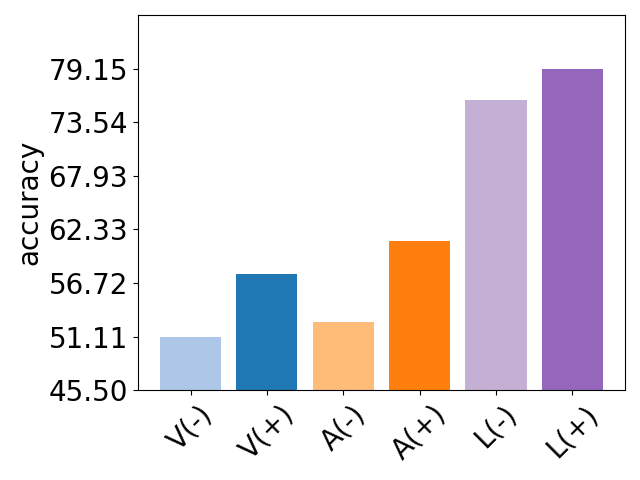}
			} 
			\subfigure[\scriptsize MOSEI$\rightarrow$SIMS]{
				\label{Figure7-2}
				\centering
				\includegraphics[width=0.15\linewidth, trim=20 0 0 0]{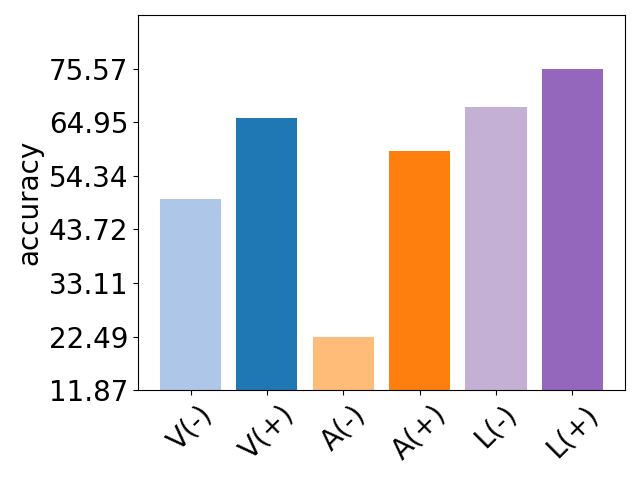}
			}
			\subfigure[\scriptsize MOSEI$\rightarrow$SIMS v2]{
				\label{Figure7-3}
				\centering
				\includegraphics[width=0.15\linewidth, trim=20 0 0 0]{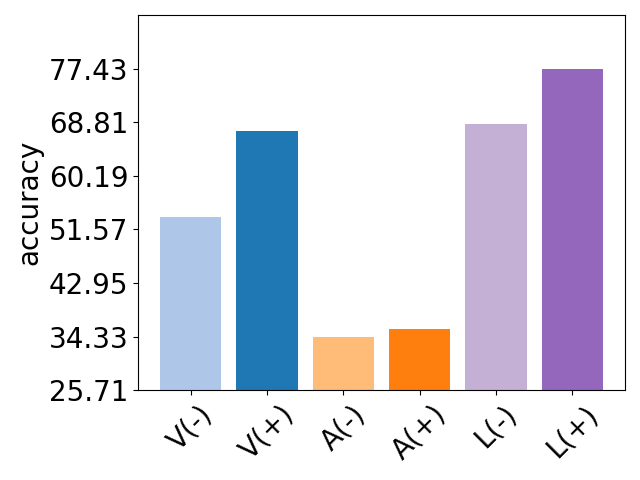}
			}
			\subfigure[\scriptsize MOSI$\rightarrow$MOSEI]{
				\label{Figure7-4}
				\centering
				\includegraphics[width=0.15\linewidth, trim=20 0 0 0]{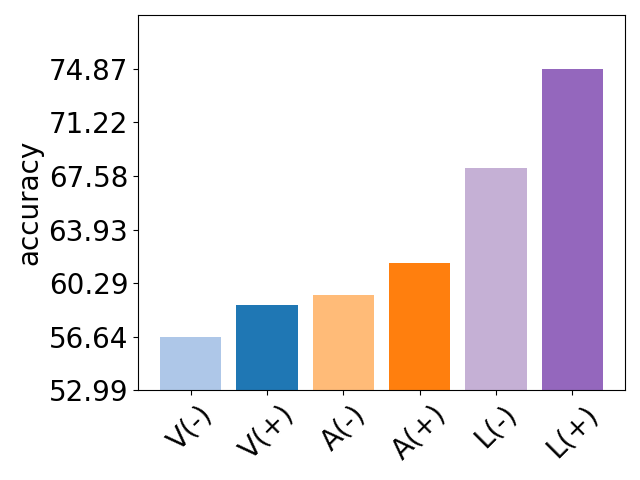}
			} 
			\subfigure[\scriptsize MOSI$\rightarrow$SIMS]{
				\label{Figure7-5}
				\centering
				\includegraphics[width=0.15\linewidth, trim=20 0 0 0]{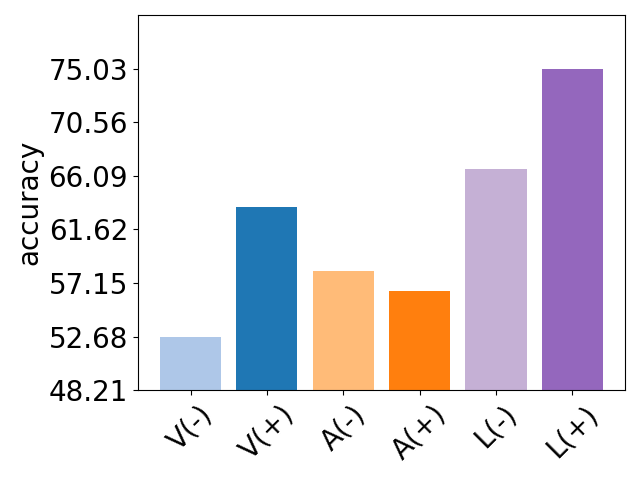}
			}
			\subfigure[\scriptsize MOSI$\rightarrow$SIMS v2]{
				\label{Figure7-6}
				\centering
				\includegraphics[width=0.15\linewidth, trim=20 0 0 0]{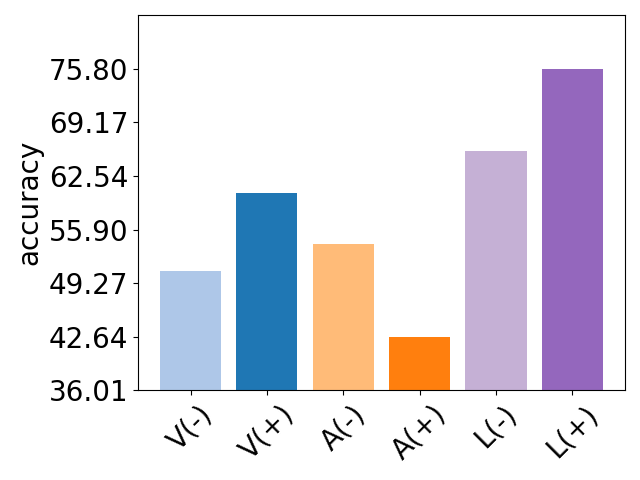}
			}
			\subfigure[\scriptsize SIMS$\rightarrow$MOSEI]{
				\label{Figure7-7}
				\centering
				\includegraphics[width=0.15\linewidth, trim=20 0 0 0]{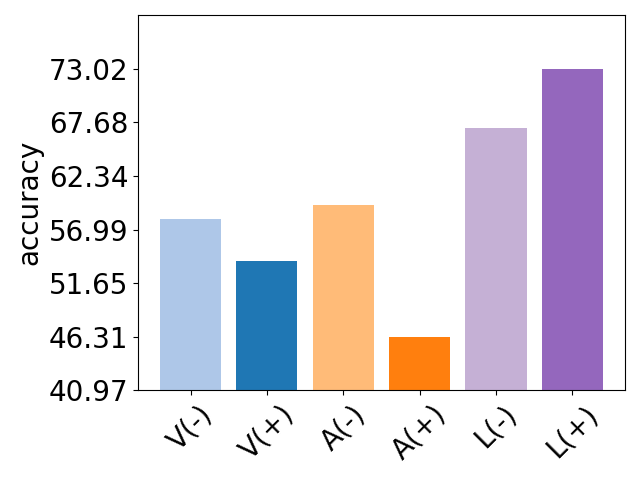}
			}
			\subfigure[\scriptsize SIMS$\rightarrow$MOSI]{
				\label{Figure7-8}
				\centering
				\includegraphics[width=0.15\linewidth, trim=20 0 0 0]{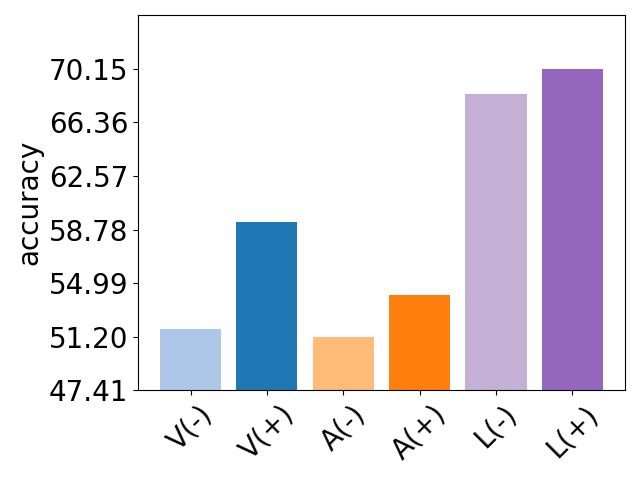}
			}
			\subfigure[\scriptsize SIMS$\rightarrow$SIMS v2]{
				\label{Figure7-9}
				\centering
				\includegraphics[width=0.15\linewidth, trim=20 0 0 0]{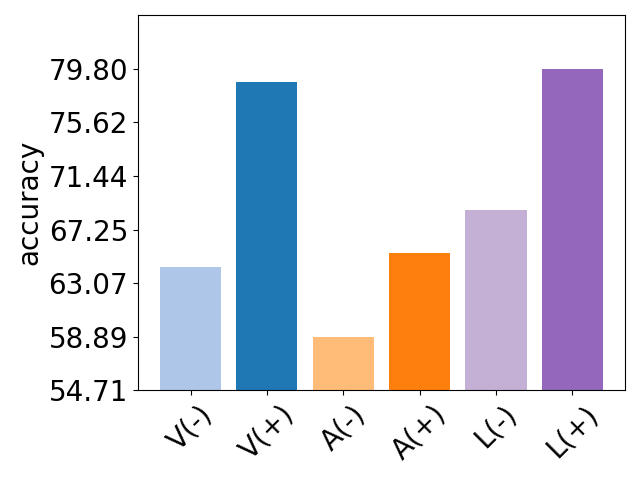}
			}
			\subfigure[\scriptsize SIMS v2$\rightarrow$MOSEI]{
				\label{Figure7-10}
				\centering
				\includegraphics[width=0.15\linewidth, trim=20 0 0 0]{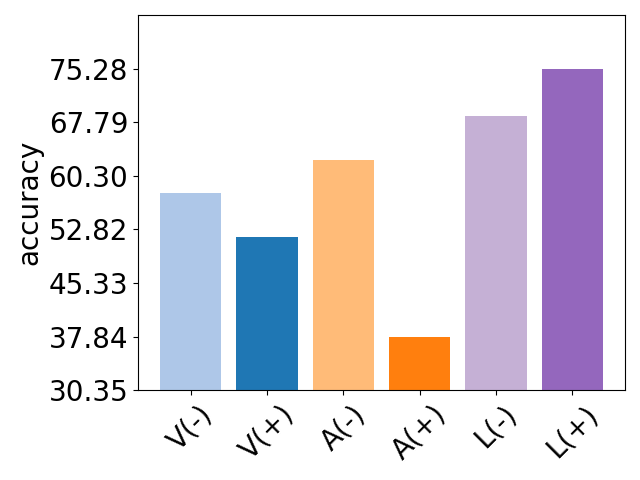}
			}
			\subfigure[\scriptsize SIMS v2$\rightarrow$MOSI]{
				\label{Figure7-11}
				\centering
				\includegraphics[width=0.15\linewidth, trim=20 0 0 0]{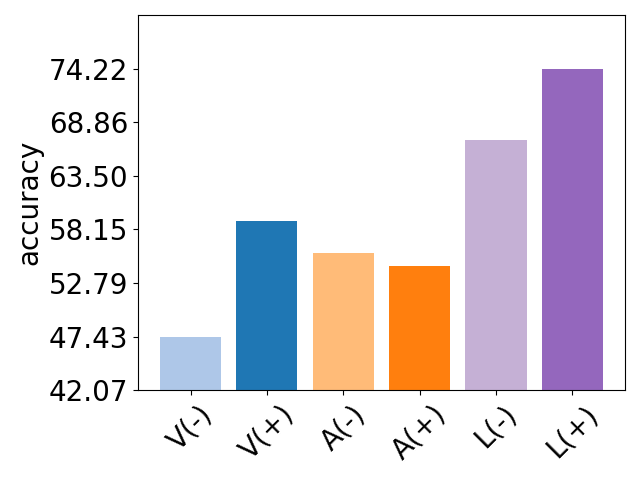}
			}
			\subfigure[\scriptsize SIMS v2$\rightarrow$SIMS]{
				\label{Figure7-12}
				\centering
				\includegraphics[width=0.15\linewidth, trim=20 0 0 0]{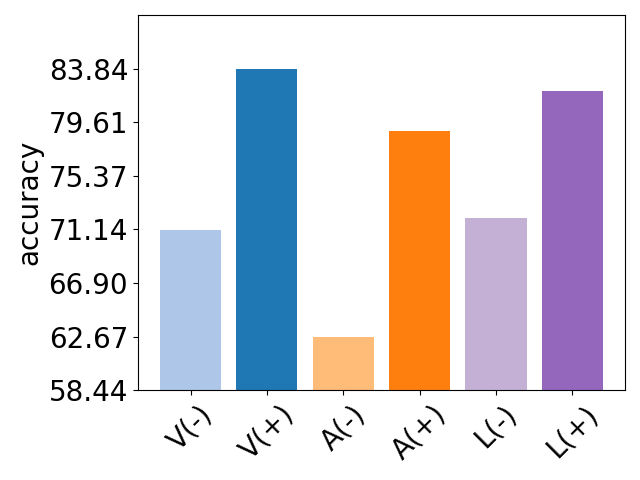}
			}
			\subfigure[\scriptsize IEMOCAP$\rightarrow$MELD]{
				\label{Figure7-13}
				\centering
				\includegraphics[width=0.15\linewidth, trim=20 0 0 0]{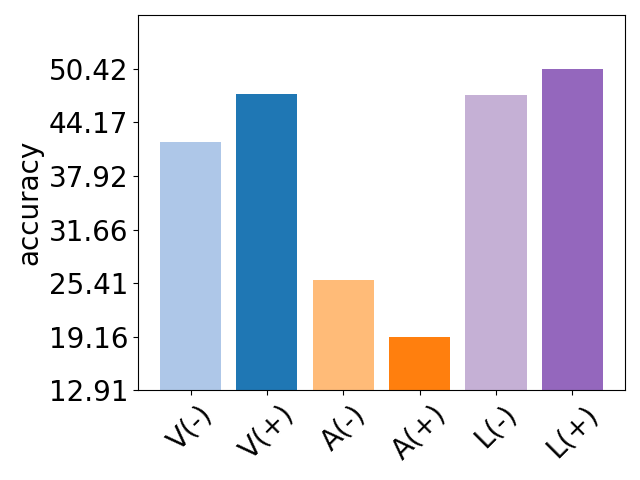}
			} 
			\subfigure[\scriptsize IEMOCAP$\rightarrow$MER]{
				\label{Figure7-14}
				\centering
				\includegraphics[width=0.15\linewidth, trim=20 0 0 0]{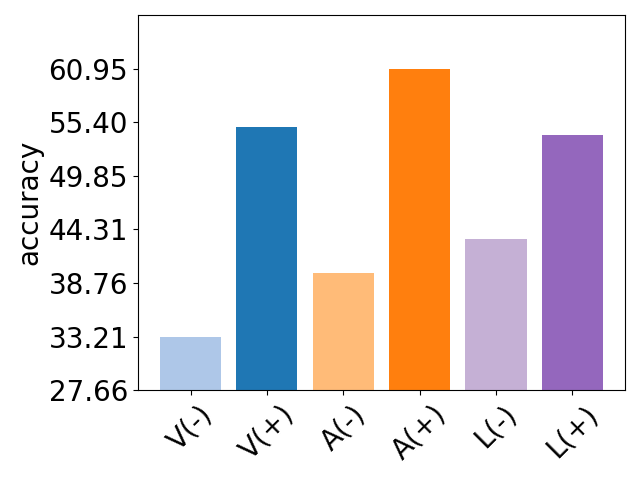}
			}
			\subfigure[\scriptsize MELD$\rightarrow$IEMOCAP]{
				\label{Figure7-15}
				\centering
				\includegraphics[width=0.15\linewidth, trim=20 0 0 0]{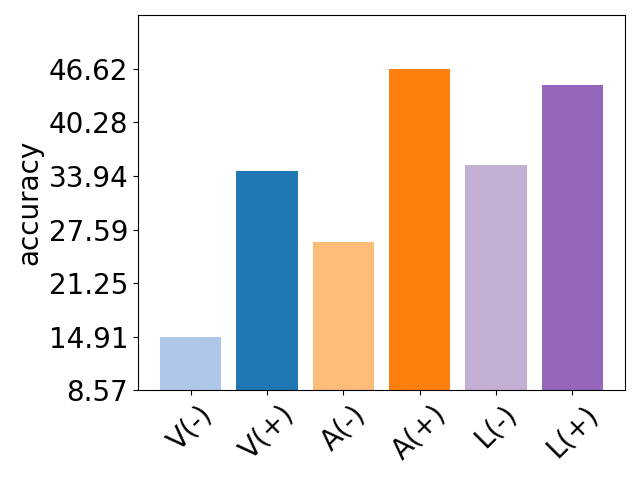}
			}
			\subfigure[\scriptsize MELD$\rightarrow$MER]{
				\label{Figure7-16}
				\centering
				\includegraphics[width=0.15\linewidth, trim=20 0 0 0]{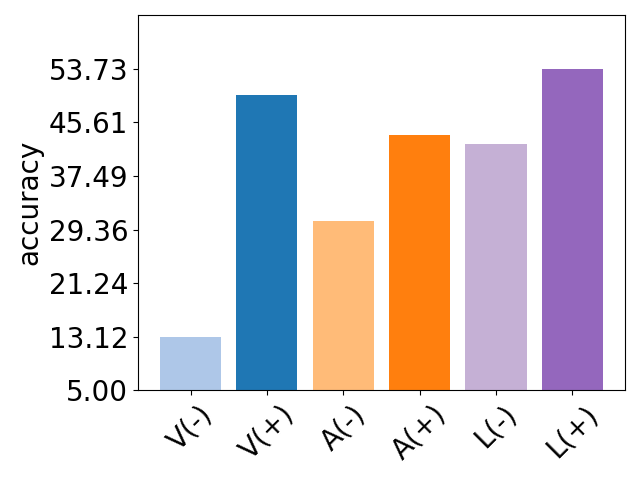}
			}
			\subfigure[\scriptsize MER$\rightarrow$IEMOCAP]{
				\label{Figure7-17}
				\centering
				\includegraphics[width=0.15\linewidth, trim=20 0 0 0]{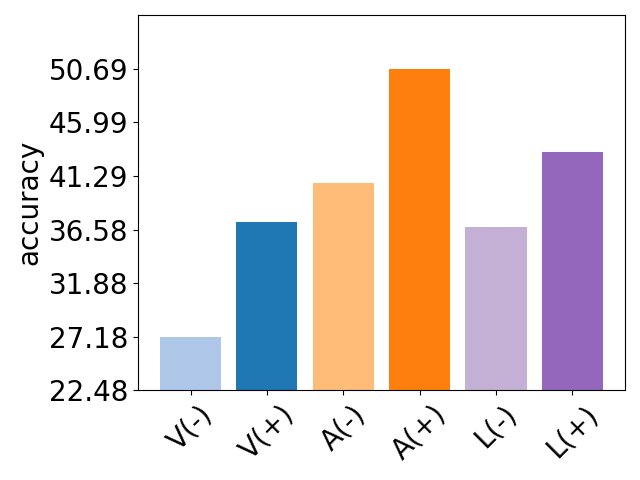}
			} 
			\subfigure[\scriptsize MER$\rightarrow$MELD]{
				\label{Figure7-18}
				\centering
				\includegraphics[width=0.15\linewidth, trim=20 0 0 0]{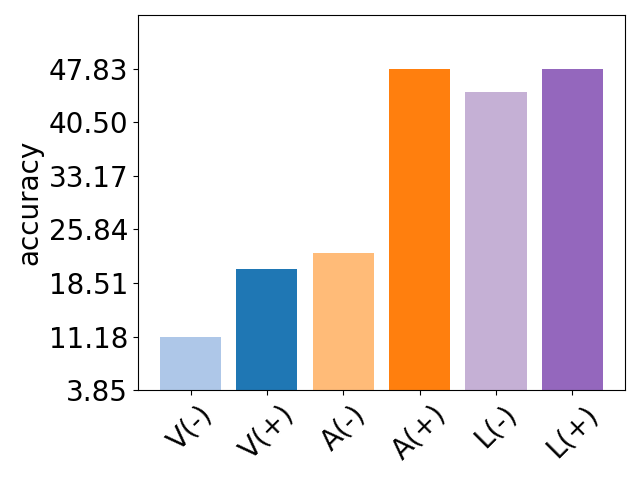}
			}
		\end{center}
		\caption{Cross-corpus results using unimodal features. ``V'', ``A'', and ``L'' represent visual, acoustic, and lexical features. ``(-)'' and ``(+)'' denote poorly- and well-performing features in the within-corpus setup. We observe that good within-corpus features generally lead to good cross-corpus results.}
		\label{Figure7}
	\end{figure*}

	\subsection{Cross-corpus Benchmark}
	This section concentrates on cross-corpus performance. Given that IEMOCAP(four), MELD, and MER-MULTI share four discrete labels (i.e., \emph{happiness}, \emph{sadness}, \emph{anger}, \emph{neutral}), cross-corpus experiments are conducted based on these labels. Due to sample overlap between IEMOCAP(four) and IEMOCAP(six), this section only focuses on IEMOCAP(four). CMU-MOSI, CMU-MOSEI, CH-SIMS, and CH-SIMS v2 provide continuous sentiment scores, but the labels of the first two datasets are in the range [-3, 3], while the labels of the latter two datasets are in the range [-1, 1]. Therefore, we normalize these labels before conducting experiments. Tables \ref{Table19}$\sim$\ref{Table20}  (in Appendix) provide the cross-corpus results. For a clearer presentation, we further summarize these results in the following analysis.
	
	Fig. \ref{Figure7} presents cross-corpus results using unimodal features. We select two features for each modality: one performing well and one performing poorly. Specifically, visual features (ResNet-MSCeleb and CLIP-large), acoustic features (VGGish and HUBERT-large), and lexical features (OPT-13B and Baichuan-13B) are evaluated. In Fig. \ref{Figure7}, good within-corpus features generally lead to good cross-corpus results. Therefore, a straightforward idea is to leverage multimodal fusion to enhance cross-corpus performance.
	
	Table \ref{Table11} evaluates the impact of multimodal fusion on within- and cross-corpus results. We test five fusion algorithms and use the best score as the multimodal result. Since MELD and MER-MULTI have 7 and 6 labels in the within-corpus setup but 4 labels in the cross-corpus setup, their results are excluded. Meanwhile, due to sample overlap, the cross-corpus results between CMU-MOSEI and CMU-MOSI (CH-SIMS and CH-SIMS v2) are also excluded. In general, multimodal fusion can improve both within- and cross-corpus results. However, its utility is relatively lower in the cross-corpus setting. The reason is that different datasets convey emotions in distinct ways (see Fig. \ref{Figure6}). Multimodal fusion may cause the model to overfit the target dataset, resulting in limited improvements in the cross-corpus setup.

	\begin{table}[t]
		\centering
		\renewcommand\tabcolsep{3.6pt}
		\caption{Impact of multimodal fusion on cross- and within-corpus results. In this table, the unimodal results represent the best scores among 6 unimodal features and the multimodal results represent the best scores among 5 fusion algorithms. See Tables \ref{Table19}$\sim$\ref{Table20} (in Appendix) for more details.}
		\label{Table11}
		\begin{tabular}{ll|ccc}
			\hline
			Source & Target & Unimodal & Multimodal & $\Delta$ \\
			
			\hline
			MELD & IEMOCAP(4) & 46.62$\pm$0.98 & 48.86$\pm$0.75 &\textcolor[rgb]{0.0,0.6,0.0}{$\uparrow${2.24}}\\
			MER-MULTI & IEMOCAP(4) & 50.69$\pm$0.16 & 59.58$\pm$1.10 &\textcolor[rgb]{0.0,0.6,0.0}{$\uparrow${8.89}}\\
			\rowcolor{lightgray}
			IEMOCAP(4) & IEMOCAP(4) & 69.67$\pm$0.15 & 78.20$\pm$0.12 &\textcolor[rgb]{0.0,0.6,0.0}{$\uparrow${8.53}}\\
			\hline
			CH-SIMS & CMU-MOSI & 70.15$\pm$0.71 & 70.86$\pm$1.32 &\textcolor[rgb]{0.0,0.6,0.0}{$\uparrow${0.71}}\\
			CH-SIMS v2 & CMU-MOSI & 74.22$\pm$0.21 & 73.85$\pm$1.17 &\textcolor[rgb]{0.5,0.5,0.5}{$\downarrow${0.37}}\\
			\rowcolor{lightgray}
			CMU-MOSI & CMU-MOSI & 79.37$\pm$0.39 & 79.40$\pm$0.20 &\textcolor[rgb]{0.0,0.6,0.0}{$\uparrow${0.03}}\\
			\hline
			CH-SIMS & CMU-MOSEI & 73.02$\pm$0.69 & 72.20$\pm$0.45 &\textcolor[rgb]{0.5,0.5,0.5}{$\downarrow${0.82}}\\
			CH-SIMS v2 & CMU-MOSEI & 75.28$\pm$0.32 & 75.19$\pm$0.90 &\textcolor[rgb]{0.5,0.5,0.5}{$\downarrow${0.09}}\\
			\rowcolor{lightgray}
			CMU-MOSEI & CMU-MOSEI & 84.42$\pm$0.05 & 85.05$\pm$0.23 &\textcolor[rgb]{0.0,0.6,0.0}{$\uparrow${0.63}}\\
			\hline
			CMU-MOSI & CH-SIMS & 75.03$\pm$0.87 & 76.55$\pm$0.83 &\textcolor[rgb]{0.0,0.6,0.0}{$\uparrow${1.52}}\\
			CMU-MOSEI & CH-SIMS & 75.57$\pm$0.73 & 81.64$\pm$1.32 &\textcolor[rgb]{0.0,0.6,0.0}{$\uparrow${6.07}}\\
			\rowcolor{lightgray}
			CH-SIMS & CH-SIMS & 82.62$\pm$0.31 & 90.50$\pm$0.36 &\textcolor[rgb]{0.0,0.6,0.0}{$\uparrow${7.88}}\\
			\hline
			CMU-MOSI & CH-SIMS v2 & 75.80$\pm$0.32 & 75.74$\pm$0.56 &\textcolor[rgb]{0.5,0.5,0.5}{$\downarrow${0.06}}\\
			CMU-MOSEI & CH-SIMS v2 & 77.43$\pm$0.35 & 81.18$\pm$1.03 &\textcolor[rgb]{0.0,0.6,0.0}{$\uparrow${3.75}}\\
			\rowcolor{lightgray}
			CH-SIMS v2 & CH-SIMS v2 & 80.93$\pm$0.19 & 88.48$\pm$0.32 &\textcolor[rgb]{0.0,0.6,0.0}{$\uparrow${7.55}}\\
			
			\hline
			
		\end{tabular}
	\end{table}

	\begin{figure}[t]
		\centering
		\includegraphics[width=\linewidth]{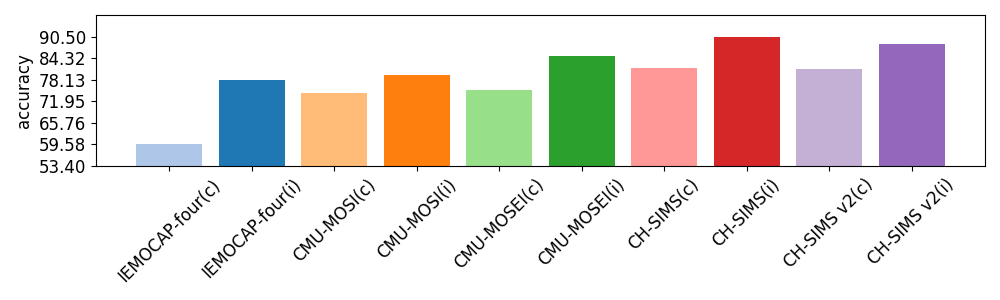}
		\caption{Performance gap between cross-corpus results (``c'') and within-corpus results (``i'') on different datasets.}
		\label{Figure8}
	\end{figure}

	\begin{table}[t]
		\centering
		\renewcommand\tabcolsep{6pt}
		\caption{Performance of GPT-4V. To reduce the cost of API calls, we only evaluate MER-MULTI, CMU-MOSI, and CH-SIMS. We choose the best scores in Table \ref{Table10} as within-corpus results, and the best scores in Tables \ref{Table19}$\sim$\ref{Table20} (in Appendix) as cross-corpus results.}
		\label{Table12}
		\begin{tabular}{c|ccc}
			\hline
			{Methods} & {MER-MULTI} & {CMU-MOSI} & {CH-SIMS} \\
			\hline
			
			Within-corpus 	& 84.75 & 85.96 & 91.75 \\
			Cross-corpus 	& -- 	& 74.22 & 81.64 \\
			\hline
			GPT-4V 			& 63.50 & 80.43 & 81.24 \\
			
			\hline
		\end{tabular}
	\end{table}

	Fig. \ref{Figure8} shows the performance gap between within- and cross-corpus results, highlighting the challenges associated with the cross-corpus setup. Meanwhile, we evaluate the emotion recognition performance of GPT-4V in Table \ref{Table12}. To reduce the cost of API calls, we only evaluate MER-MULTI, CMU-MOSI, and CH-SIMS \cite{lian2023gpt}. Although GPT-4V performs worse than within-corpus results, it performs close to or even better than cross-corpus results. This suggests that multimodal emotion recognition is a complex task that requires integrating diverse knowledge like GPT-4V, including unimodal emotion recognition and some background knowledge (e.g., dark environments may induce negative emotions). Additionally, as suggested in our previous work \cite{lian2023explainable}, it is better to use explainable reasoning processes as labels to enhance consistency and reduce subjectivity in emotion annotations. More uniform and reliable labels contribute to improved cross-corpus performance.

	\begin{table*}[t]
		\centering
		\renewcommand\tabcolsep{3.0pt}
		\caption{Punctuation robustness analysis. This table shows emotion recognition results under complete and incomplete punctuation. Here, ``E'', ``P'', and ``O'' represent emotion-related punctuation (i.e., exclamation mark, question mark, ellipsis), pause-related punctuation (i.e., comma), and other punctuation. Removing punctuation generally results in a performance drop, but this drop is often not noticeable (except for MELD).}
		\label{Table13}
		\begin{tabular}{lccc|cccccccc}
			\hline
			Feature & E & P & O & {MER-MULTI} & {CMU-MOSI} & {CMU-MOSEI} & {CH-SIMS} & {CH-SIMS v2} & {MELD} & {\begin{tabular}[c]{@{}c@{}}IEMOCAP \\ (four-class)\end{tabular}} & {\begin{tabular}[c]{@{}c@{}}IEMOCAP \\ (six-class) \end{tabular}} \\
			\hline
			
			OPT-13B&$\times$ &$\times$&$\times$&46.33 (\textcolor[rgb]{0.0,0.6,0.0}{$\uparrow${0.55}})&73.69 (\textcolor[rgb]{0.5,0.5,0.5}{$\downarrow${0.87}})&79.75 (\textcolor[rgb]{0.0,0.6,0.0}{$\uparrow${0.05}})&73.53 (\textcolor[rgb]{0.0,0.6,0.0}{$\uparrow${0.98}})&69.94 (\textcolor[rgb]{0.5,0.5,0.5}{$\downarrow${0.18}})&48.05 (\textcolor[rgb]{0.5,0.5,0.5}{$\downarrow${9.44}})&61.71 (\textcolor[rgb]{0.0,0.6,0.0}{$\uparrow${0.01}})&48.08 (\textcolor[rgb]{0.0,0.6,0.0}{$\uparrow${0.03}})\\
			OPT-13B&$\times$ &$\times$&$\surd$&45.63 (\textcolor[rgb]{0.5,0.5,0.5}{$\downarrow${0.15}})&73.99 (\textcolor[rgb]{0.5,0.5,0.5}{$\downarrow${0.57}})&79.62 (\textcolor[rgb]{0.5,0.5,0.5}{$\downarrow${0.08}})&74.26 (\textcolor[rgb]{0.0,0.6,0.0}{$\uparrow${1.71}})&70.90 (\textcolor[rgb]{0.0,0.6,0.0}{$\uparrow${0.78}})&49.48 (\textcolor[rgb]{0.5,0.5,0.5}{$\downarrow${8.01}})&62.07 (\textcolor[rgb]{0.0,0.6,0.0}{$\uparrow${0.37}})&48.27 (\textcolor[rgb]{0.0,0.6,0.0}{$\uparrow${0.22}})\\
			OPT-13B&$\times$ &$\surd$&$\surd$&45.12 (\textcolor[rgb]{0.5,0.5,0.5}{$\downarrow${0.66}})&74.09 (\textcolor[rgb]{0.5,0.5,0.5}{$\downarrow${0.47}})&80.11 (\textcolor[rgb]{0.0,0.6,0.0}{$\uparrow${0.41}})&73.02 (\textcolor[rgb]{0.0,0.6,0.0}{$\uparrow${0.47}})&70.19 (\textcolor[rgb]{0.0,0.6,0.0}{$\uparrow${0.07}})&49.55 (\textcolor[rgb]{0.5,0.5,0.5}{$\downarrow${7.94}})&61.80 (\textcolor[rgb]{0.0,0.6,0.0}{$\uparrow${0.10}})&48.01 (\textcolor[rgb]{0.5,0.5,0.5}{$\downarrow${0.04}})\\
			\rowcolor{lightgray}
			OPT-13B&$\surd$&$\surd$&$\surd$&45.78&74.56&79.70&72.55&70.12&57.49&61.70&48.05\\
			\hline
			ALBERT-small&$\times$ &$\times$&$\times$&45.90 (\textcolor[rgb]{0.5,0.5,0.5}{$\downarrow${0.53}})&74.31 (\textcolor[rgb]{0.0,0.6,0.0}{$\uparrow${0.15}})&79.44 (\textcolor[rgb]{0.5,0.5,0.5}{$\downarrow${0.06}})&72.33 (\textcolor[rgb]{0.5,0.5,0.5}{$\downarrow${0.94}})&71.64 (\textcolor[rgb]{0.5,0.5,0.5}{$\downarrow${0.66}})&47.24 (\textcolor[rgb]{0.5,0.5,0.5}{$\downarrow${9.18}})&61.15 (\textcolor[rgb]{0.0,0.6,0.0}{$\uparrow${0.01}})&47.08 (\textcolor[rgb]{0.5,0.5,0.5}{$\downarrow${0.35}})\\
			ALBERT-small&$\times$ &$\times$&$\surd$&46.46 (\textcolor[rgb]{0.0,0.6,0.0}{$\uparrow${0.03}})&73.38 (\textcolor[rgb]{0.5,0.5,0.5}{$\downarrow${0.78}})&79.67 (\textcolor[rgb]{0.0,0.6,0.0}{$\uparrow${0.17}})&74.00 (\textcolor[rgb]{0.0,0.6,0.0}{$\uparrow${0.73}})&71.65 (\textcolor[rgb]{0.5,0.5,0.5}{$\downarrow${0.65}})&48.07 (\textcolor[rgb]{0.5,0.5,0.5}{$\downarrow${8.35}})&61.00 (\textcolor[rgb]{0.5,0.5,0.5}{$\downarrow${0.14}})&47.34 (\textcolor[rgb]{0.5,0.5,0.5}{$\downarrow${0.09}})\\
			ALBERT-small&$\times$ &$\surd$&$\surd$&45.74 (\textcolor[rgb]{0.5,0.5,0.5}{$\downarrow${0.69}})&73.92 (\textcolor[rgb]{0.5,0.5,0.5}{$\downarrow${0.24}})&79.75 (\textcolor[rgb]{0.0,0.6,0.0}{$\uparrow${0.25}})&74.58 (\textcolor[rgb]{0.0,0.6,0.0}{$\uparrow${1.31}})&72.14 (\textcolor[rgb]{0.5,0.5,0.5}{$\downarrow${0.16}})&48.12 (\textcolor[rgb]{0.5,0.5,0.5}{$\downarrow${8.30}})&61.47 (\textcolor[rgb]{0.0,0.6,0.0}{$\uparrow${0.33}})&47.37 (\textcolor[rgb]{0.5,0.5,0.5}{$\downarrow${0.06}})\\
			\rowcolor{lightgray}
			ALBERT-small&$\surd$&$\surd$&$\surd$&46.43&74.16&79.50&73.27&72.30&56.42&61.14&47.43\\
			\hline
			Llama2-13B&$\times$ &$\times$&$\times$&50.39 (\textcolor[rgb]{0.5,0.5,0.5}{$\downarrow${1.24}})&77.35 (\textcolor[rgb]{0.5,0.5,0.5}{$\downarrow${1.27}})&82.86 (\textcolor[rgb]{0.5,0.5,0.5}{$\downarrow${0.76}})&73.74 (\textcolor[rgb]{0.5,0.5,0.5}{$\downarrow${1.72}})&73.73 (\textcolor[rgb]{0.5,0.5,0.5}{$\downarrow${1.64}})&49.42 (\textcolor[rgb]{0.5,0.5,0.5}{$\downarrow${8.67}})&64.58 (\textcolor[rgb]{0.5,0.5,0.5}{$\downarrow${1.16}})&50.05 (\textcolor[rgb]{0.5,0.5,0.5}{$\downarrow${0.70}})\\
			Llama2-13B&$\times$ &$\times$&$\surd$&51.05 (\textcolor[rgb]{0.5,0.5,0.5}{$\downarrow${0.58}})&78.37 (\textcolor[rgb]{0.5,0.5,0.5}{$\downarrow${0.25}})&83.23 (\textcolor[rgb]{0.5,0.5,0.5}{$\downarrow${0.39}})&73.79 (\textcolor[rgb]{0.5,0.5,0.5}{$\downarrow${1.67}})&74.36 (\textcolor[rgb]{0.5,0.5,0.5}{$\downarrow${1.01}})&50.76 (\textcolor[rgb]{0.5,0.5,0.5}{$\downarrow${7.33}})&65.14 (\textcolor[rgb]{0.5,0.5,0.5}{$\downarrow${0.60}})&50.73 (\textcolor[rgb]{0.5,0.5,0.5}{$\downarrow${0.02}})\\
			Llama2-13B&$\times$ &$\surd$&$\surd$&51.70 (\textcolor[rgb]{0.0,0.6,0.0}{$\uparrow${0.07}})&79.16 (\textcolor[rgb]{0.0,0.6,0.0}{$\uparrow${0.54}})&83.52 (\textcolor[rgb]{0.5,0.5,0.5}{$\downarrow${0.10}})&75.36 (\textcolor[rgb]{0.5,0.5,0.5}{$\downarrow${0.10}})&75.33 (\textcolor[rgb]{0.5,0.5,0.5}{$\downarrow${0.04}})&50.96 (\textcolor[rgb]{0.5,0.5,0.5}{$\downarrow${7.13}})&65.48 (\textcolor[rgb]{0.5,0.5,0.5}{$\downarrow${0.26}})&50.80 (\textcolor[rgb]{0.0,0.6,0.0}{$\uparrow${0.05}})\\
			\rowcolor{lightgray}
			Llama2-13B&$\surd$&$\surd$&$\surd$&51.63&78.62&83.62&75.46&75.37&58.09&65.74&50.75\\
			\hline
			PERT-base&$\times$ &$\times$&$\times$&53.61 (\textcolor[rgb]{0.0,0.6,0.0}{$\uparrow${0.86}})&80.61 (\textcolor[rgb]{0.5,0.5,0.5}{$\downarrow${0.58}})&83.05 (\textcolor[rgb]{0.5,0.5,0.5}{$\downarrow${0.31}})&78.34 (\textcolor[rgb]{0.0,0.6,0.0}{$\uparrow${0.32}})&78.01 (\textcolor[rgb]{0.0,0.6,0.0}{$\uparrow${0.25}})&49.72 (\textcolor[rgb]{0.5,0.5,0.5}{$\downarrow${7.48}})&64.42 (\textcolor[rgb]{0.5,0.5,0.5}{$\downarrow${0.31}})&49.53 (\textcolor[rgb]{0.0,0.6,0.0}{$\uparrow${0.09}})\\
			PERT-base&$\times$ &$\times$&$\surd$&53.29 (\textcolor[rgb]{0.0,0.6,0.0}{$\uparrow${0.54}})&81.81 (\textcolor[rgb]{0.0,0.6,0.0}{$\uparrow${0.62}})&83.29 (\textcolor[rgb]{0.5,0.5,0.5}{$\downarrow${0.07}})&77.60 (\textcolor[rgb]{0.5,0.5,0.5}{$\downarrow${0.42}})&77.59 (\textcolor[rgb]{0.5,0.5,0.5}{$\downarrow${0.17}})&50.68 (\textcolor[rgb]{0.5,0.5,0.5}{$\downarrow${6.52}})&64.48 (\textcolor[rgb]{0.5,0.5,0.5}{$\downarrow${0.25}})&49.46 (\textcolor[rgb]{0.0,0.6,0.0}{$\uparrow${0.02}})\\
			PERT-base&$\times$ &$\surd$&$\surd$&52.76 (\textcolor[rgb]{0.0,0.6,0.0}{$\uparrow${0.01}})&81.47 (\textcolor[rgb]{0.0,0.6,0.0}{$\uparrow${0.28}})&83.29 (\textcolor[rgb]{0.5,0.5,0.5}{$\downarrow${0.07}})&77.57 (\textcolor[rgb]{0.5,0.5,0.5}{$\downarrow${0.45}})&77.41 (\textcolor[rgb]{0.5,0.5,0.5}{$\downarrow${0.35}})&50.90 (\textcolor[rgb]{0.5,0.5,0.5}{$\downarrow${6.30}})&64.62 (\textcolor[rgb]{0.5,0.5,0.5}{$\downarrow${0.11}})&49.53 (\textcolor[rgb]{0.0,0.6,0.0}{$\uparrow${0.09}})\\
			\rowcolor{lightgray}
			PERT-base&$\surd$&$\surd$&$\surd$&52.75&81.19&83.36&78.02&77.76&57.20&64.73&49.44\\
			\hline
			RoBERTa-large&$\times$ &$\times$&$\times$&57.40 (\textcolor[rgb]{0.5,0.5,0.5}{$\downarrow${0.82}})&82.87 (\textcolor[rgb]{0.5,0.5,0.5}{$\downarrow${0.30}})&83.82 (\textcolor[rgb]{0.5,0.5,0.5}{$\downarrow${0.29}})&81.83 (\textcolor[rgb]{0.5,0.5,0.5}{$\downarrow${0.73}})&79.72 (\textcolor[rgb]{0.5,0.5,0.5}{$\downarrow${0.32}})&50.80 (\textcolor[rgb]{0.5,0.5,0.5}{$\downarrow${8.40}})&66.12 (\textcolor[rgb]{0.5,0.5,0.5}{$\downarrow${0.09}})&51.14 (\textcolor[rgb]{0.0,0.6,0.0}{$\uparrow${0.16}})\\
			RoBERTa-large&$\times$ &$\times$&$\surd$&57.32 (\textcolor[rgb]{0.5,0.5,0.5}{$\downarrow${0.90}})&83.04 (\textcolor[rgb]{0.5,0.5,0.5}{$\downarrow${0.13}})&84.03 (\textcolor[rgb]{0.5,0.5,0.5}{$\downarrow${0.08}})&81.60 (\textcolor[rgb]{0.5,0.5,0.5}{$\downarrow${0.96}})&79.55 (\textcolor[rgb]{0.5,0.5,0.5}{$\downarrow${0.49}})&51.10 (\textcolor[rgb]{0.5,0.5,0.5}{$\downarrow${8.10}})&66.23 (\textcolor[rgb]{0.0,0.6,0.0}{$\uparrow${0.02}})&51.17 (\textcolor[rgb]{0.0,0.6,0.0}{$\uparrow${0.19}})\\
			RoBERTa-large&$\times$ &$\surd$&$\surd$&58.20 (\textcolor[rgb]{0.5,0.5,0.5}{$\downarrow${0.02}})&83.30 (\textcolor[rgb]{0.0,0.6,0.0}{$\uparrow${0.13}})&84.09 (\textcolor[rgb]{0.5,0.5,0.5}{$\downarrow${0.02}})&82.26 (\textcolor[rgb]{0.5,0.5,0.5}{$\downarrow${0.30}})&80.16 (\textcolor[rgb]{0.0,0.6,0.0}{$\uparrow${0.12}})&51.20 (\textcolor[rgb]{0.5,0.5,0.5}{$\downarrow${8.00}})&66.22 (\textcolor[rgb]{0.0,0.6,0.0}{$\uparrow${0.01}})&51.00 (\textcolor[rgb]{0.0,0.6,0.0}{$\uparrow${0.02}})\\
			\rowcolor{lightgray}
			RoBERTa-large&$\surd$&$\surd$&$\surd$&58.22&83.17&84.11&82.56&80.04&59.20&66.21&50.98\\
			\hline
			Baichuan-13B&$\times$ &$\times$&$\times$&59.65 (\textcolor[rgb]{0.5,0.5,0.5}{$\downarrow${2.89}})&78.20 (\textcolor[rgb]{0.5,0.5,0.5}{$\downarrow${1.17}})&83.77 (\textcolor[rgb]{0.5,0.5,0.5}{$\downarrow${0.65}})&79.95 (\textcolor[rgb]{0.5,0.5,0.5}{$\downarrow${2.04}})&79.44 (\textcolor[rgb]{0.5,0.5,0.5}{$\downarrow${1.49}})&50.32 (\textcolor[rgb]{0.5,0.5,0.5}{$\downarrow${7.95}})&66.28 (\textcolor[rgb]{0.5,0.5,0.5}{$\downarrow${0.73}})&51.39 (\textcolor[rgb]{0.5,0.5,0.5}{$\downarrow${0.31}})\\
			Baichuan-13B&$\times$ &$\times$&$\surd$&60.64 (\textcolor[rgb]{0.5,0.5,0.5}{$\downarrow${1.90}})&79.30 (\textcolor[rgb]{0.5,0.5,0.5}{$\downarrow${0.07}})&84.45 (\textcolor[rgb]{0.0,0.6,0.0}{$\uparrow${0.03}})&81.53 (\textcolor[rgb]{0.5,0.5,0.5}{$\downarrow${0.46}})&79.70 (\textcolor[rgb]{0.5,0.5,0.5}{$\downarrow${1.23}})&51.47 (\textcolor[rgb]{0.5,0.5,0.5}{$\downarrow${6.80}})&66.98 (\textcolor[rgb]{0.5,0.5,0.5}{$\downarrow${0.03}})&52.03 (\textcolor[rgb]{0.0,0.6,0.0}{$\uparrow${0.33}})\\
			Baichuan-13B&$\times$ &$\surd$&$\surd$&62.00 (\textcolor[rgb]{0.5,0.5,0.5}{$\downarrow${0.54}})&79.36 (\textcolor[rgb]{0.5,0.5,0.5}{$\downarrow${0.01}})&84.36 (\textcolor[rgb]{0.5,0.5,0.5}{$\downarrow${0.06}})&81.65 (\textcolor[rgb]{0.5,0.5,0.5}{$\downarrow${0.34}})&81.06 (\textcolor[rgb]{0.0,0.6,0.0}{$\uparrow${0.13}})&51.45 (\textcolor[rgb]{0.5,0.5,0.5}{$\downarrow${6.82}})&66.82 (\textcolor[rgb]{0.5,0.5,0.5}{$\downarrow${0.19}})&51.73 (\textcolor[rgb]{0.0,0.6,0.0}{$\uparrow${0.03}})\\
			\rowcolor{lightgray}
			Baichuan-13B&$\surd$&$\surd$&$\surd$&62.54&79.37&84.42&81.99&80.93&58.27&67.01&51.70\\
			\hline

		\end{tabular}
	\end{table*}

	\subsection{Robustness to Punctuation}
	Punctuation can also convey emotions. For example, the exclamation mark can express \emph{surprise} or \emph{excitement}, and the question mark can express \emph{confusion}. In this section, we evaluate the punctuation robustness of different lexical encoders. We classify all punctuation into three categories: emotion-related punctuation (i.e., exclamation mark, question mark, ellipsis), pause-related punctuation (i.e., comma), and other punctuation. Table \ref{Table13} reveals the impact of different punctuation on emotion recognition. Experimental results demonstrate that removing punctuation generally results in a performance drop, but this drop is often not noticeable (except for MELD). These results indicate that the lexical encoder is somewhat robust to missing punctuation.

\begin{figure}[t]
		\centering
		\includegraphics[width=\linewidth]{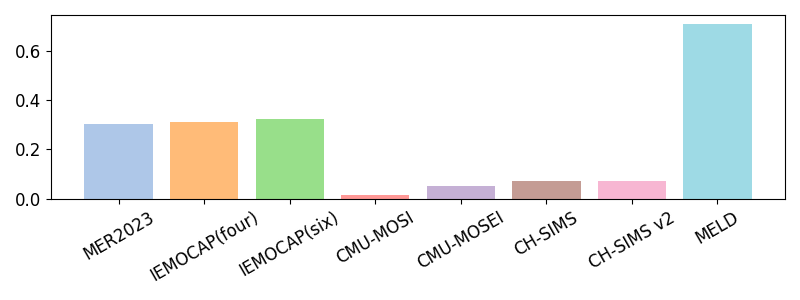}
		\caption{Average number of emotion-related punctuation per sample. MELD has more emotion-related punctuation than other datasets.}
		\label{Figure9}
	\end{figure}

	To explain why MELD is sensitive to missing punctuation, we calculate the average number of emotion-related punctuation per sample for different datasets (see Fig. \ref{Figure9}). We observe that MELD has more emotion-related punctuation than other datasets, possibly because the subtitles of this dataset are collected from original Friends TV scripts rather than ASR outputs. By further conducting case studies, we observe that many samples express emotions primarily through punctuation. For example, ``Hi Joey! What are you doing here?'' expresses \emph{joy}, while ``Rachel!'' conveys \emph{angry} \cite{chen2018emotionlines, poria2019meld}. Therefore, although the lexical encoder is somewhat robust to missing punctuation, we still need to consider punctuation for some punctuation-rich datasets. From another perspective, if we can predict detailed punctuation using ASR, we can better understand emotional states.

	\subsection{Robustness to Additive Noise}
	This section assesses the noise robustness of acoustic encoders. In the experiments, we train the model on a clean corpus and evaluate its performance on a noisy dataset. We focus on additive noise and choose the SNR from \{5dB, 10dB, 15dB\}. Experimental results are shown in Table \ref{Table14}. From these results, we observe that emotion recognition performance degrades as additive noise becomes more intense, highlighting the challenges in noisy conditions. Although eGeMAPS exhibits the smallest decrease among all features, its overall performance is relatively poor.

	Meanwhile, features performing well on clean corpora generally exhibit good results under noisy conditions (see Table \ref{Table14}). This suggests that noise robustness can be enhanced by using more powerful acoustic encoders. However, there are some exceptions to this trend. Although Whisper-large performs worse than HUBERT-large on clean corpora, it achieves better performance under noisy conditions. This is attributed to the fact that Whisper-large is trained on a highly diverse dataset, encompassing a broad distribution of audio from various environments, recording setups, speakers, and languages \cite{radford2023robust}. These findings indicate a pathway to build more robust acoustic encoders by increasing the diversity of audio quality during training.

	\begin{table}[t]
		\centering
		\renewcommand\tabcolsep{3pt}
		\caption{Noise robustness analysis. This table shows the results of various acoustic features under noisy conditions. We train the model on a clean corpus and evaluate its performance on a noisy test set with SNR in \{5dB, 10dB, 15dB\}. We present average results across all datasets.}
		\label{Table14}
		\begin{tabular}{l|c|c|c|>{\columncolor{lightgray}}c}
			\hline
			\multirow{2}{*}{Feature} & \multicolumn{4}{c}{Test SNR} \\
			& 5dB & 10dB & 15dB & Clean\\
			\hline
			eGeMAPS & 44.20(\textcolor[rgb]{0.5,0.5,0.5}{$\downarrow${1.65}}) & 44.39(\textcolor[rgb]{0.5,0.5,0.5}{$\downarrow${1.46}}) & 45.16(\textcolor[rgb]{0.5,0.5,0.5}{$\downarrow${0.69}}) & 45.85 \\
			
			data2vec-base & 48.01(\textcolor[rgb]{0.5,0.5,0.5}{$\downarrow${10.2}}) & 50.65(\textcolor[rgb]{0.5,0.5,0.5}{$\downarrow${7.59}}) & 53.33(\textcolor[rgb]{0.5,0.5,0.5}{$\downarrow${4.91}}) & 58.24 \\
			
			wav2vec 2.0-large & 51.96(\textcolor[rgb]{0.5,0.5,0.5}{$\downarrow${7.02}}) & 54.53(\textcolor[rgb]{0.5,0.5,0.5}{$\downarrow${4.45}}) & 56.23(\textcolor[rgb]{0.5,0.5,0.5}{$\downarrow${2.75}}) & 58.98\\
			
			WavLM-large & 53.19(\textcolor[rgb]{0.5,0.5,0.5}{$\downarrow${11.9}}) & 57.27(\textcolor[rgb]{0.5,0.5,0.5}{$\downarrow${7.83}}) & 59.23(\textcolor[rgb]{0.5,0.5,0.5}{$\downarrow${5.87}}) & 65.10\\
			
			Whisper-large & 57.92(\textcolor[rgb]{0.5,0.5,0.5}{$\downarrow${7.27}}) & \textbf{61.03}(\textcolor[rgb]{0.5,0.5,0.5}{$\downarrow${4.16}}) & \textbf{62.90}(\textcolor[rgb]{0.5,0.5,0.5}{$\downarrow${2.29}}) & 65.19 \\
			
			HUBERT-large & \textbf{58.02}(\textcolor[rgb]{0.5,0.5,0.5}{$\downarrow${8.60}}) & 60.79(\textcolor[rgb]{0.5,0.5,0.5}{$\downarrow${5.83}}) & 61.76(\textcolor[rgb]{0.5,0.5,0.5}{$\downarrow${4.86}}) & \textbf{66.62} \\
			\hline
			
		\end{tabular}
	\end{table}

	Table \ref{Table15} reveals the role of data augmentation under noisy conditions. We choose Whisper-large as the acoustic encoder due to its noise robustness, and report results under various training and test SNR combinations. In noisy conditions, data augmentation consistently enhances performance compared to models trained on clean data. Additionally, using the same training and test SNR generally results in improved performance, except when the training SNR is 5dB. This suggests that low training SNR may impair the model's ability to recognize emotions. Therefore, we should choose an appropriate SNR for data augmentation.

	\begin{table}[t]
		\centering
		\renewcommand\tabcolsep{10pt}
		\caption{Impact of data augmentation. We select Whisper-large as the acoustic encoder and report results under various training and test SNR combinations. We present average results across all datasets.}
		\label{Table15}
		\begin{tabular}{c|ccc}
			\hline
			\multirow{2}{*}{Training SNR} & \multicolumn{3}{c}{Test SNR} \\
			& 5dB & 10dB & 15dB \\
			\hline
			5dB	& 59.86 		& 60.61 		& 61.49 \\
			10dB 	&\textbf{60.49}	&\textbf{62.60}	& 63.29 \\
			15dB 	&60.20			&62.10			& \textbf{63.68}\\
			\rowcolor{lightgray}
			Clean & 57.92 & 61.03 & 62.90 \\
			\hline
		\end{tabular}
	\end{table}

	\begin{table*}[t]
		\centering
		\renewcommand\tabcolsep{3.6pt}
		\caption{Fine-tuning necessity analysis. In this table, we select two feature encoders for each modality and study the effect of fine-tuning. The row with $\Delta$ demonstrates the gap between results with and without fine-tuning. From this table, we observe that fine-tuning performs differently on distinct feature and dataset combinations.}
		\label{Table16}
		\begin{tabular}{lc|cccccccc}
			\hline
			
			{Feature} & Finetune & {MER-MULTI} & {CMU-MOSI} & {CMU-MOSEI} & {CH-SIMS} & {CH-SIMS v2} & {MELD} & {\begin{tabular}[c]{@{}c@{}}IEMOCAP \\ (four-class)\end{tabular}} & {\begin{tabular}[c]{@{}c@{}}IEMOCAP \\ (six-class) \end{tabular}}  \\
			
			\hline
			
			VideoMAE-base&$\times$&46.72$\pm$0.75& 56.97$\pm$0.64 & 70.40$\pm$0.05 &68.95$\pm$0.69&70.11$\pm$0.26&36.74$\pm$0.20&39.15$\pm$0.18&26.76$\pm$0.17 \\
			VideoMAE-base&$\surd$&51.20$\pm$0.01&56.17$\pm$0.71&69.79$\pm$0.07&75.38$\pm$1.27&72.31$\pm$0.59&34.50$\pm$0.06&41.45$\pm$0.23&28.74$\pm$0.24  \\
			VideoMAE-base&$\Delta$&\textcolor[rgb]{0.0,0.6,0.0}{$\uparrow${4.48}}&\textcolor[rgb]{0.5,0.5,0.5}{$\downarrow${0.80}}&\textcolor[rgb]{0.5,0.5,0.5}{$\downarrow${0.61}}&\textcolor[rgb]{0.0,0.6,0.0}{$\uparrow${6.43}}&\textcolor[rgb]{0.0,0.6,0.0}{$\uparrow${2.20}}&\textcolor[rgb]{0.5,0.5,0.5}{$\downarrow${2.24}}&\textcolor[rgb]{0.0,0.6,0.0}{$\uparrow${2.30}}&\textcolor[rgb]{0.0,0.6,0.0}{$\uparrow${1.98}}\\
			
			\hline
			
			CLIP-base&$\times$ & 60.12$\pm$0.85 &55.04$\pm$0.97& 70.24$\pm$0.07 &80.13$\pm$0.11& 77.97$\pm$0.19 & 38.59$\pm$0.33 &44.52$\pm$0.21&28.70$\pm$0.08 \\
			CLIP-base&$\surd$&57.07$\pm$0.09&56.69$\pm$1.49&69.80$\pm$1.27&79.83$\pm$0.58&78.52$\pm$0.16&36.54$\pm$0.27&43.41$\pm$0.56&26.98$\pm$1.35 \\
			CLIP-base&$\Delta$&\textcolor[rgb]{0.5,0.5,0.5}{$\downarrow${3.05}}&\textcolor[rgb]{0.0,0.6,0.0}{$\uparrow${1.65}}&\textcolor[rgb]{0.5,0.5,0.5}{$\downarrow${0.44}}&\textcolor[rgb]{0.5,0.5,0.5}{$\downarrow${0.30}}&\textcolor[rgb]{0.0,0.6,0.0}{$\uparrow${0.55}}&\textcolor[rgb]{0.5,0.5,0.5}{$\downarrow${2.05}}&\textcolor[rgb]{0.5,0.5,0.5}{$\downarrow${1.11}}&\textcolor[rgb]{0.5,0.5,0.5}{$\downarrow${1.72}}\\
			
			\hline
			
			data2vec-base&$\times$&45.13$\pm$0.39& 66.24$\pm$0.45 &73.11$\pm$0.12&65.00$\pm$0.30&59.87$\pm$0.57&44.30$\pm$0.34&64.08$\pm$0.12&48.22$\pm$0.28 \\
			data2vec-base&$\surd$&46.72$\pm$0.48&62.30$\pm$0.33&76.47$\pm$0.04&68.37$\pm$0.08&66.93$\pm$0.29&43.86$\pm$0.02&66.93$\pm$0.10&51.27$\pm$0.35  \\
			data2vec-base&$\Delta$&\textcolor[rgb]{0.0,0.6,0.0}{$\uparrow${1.59}}&\textcolor[rgb]{0.5,0.5,0.5}{$\downarrow${3.94}}&\textcolor[rgb]{0.0,0.6,0.0}{$\uparrow${3.36}}&\textcolor[rgb]{0.0,0.6,0.0}{$\uparrow${3.37}}&\textcolor[rgb]{0.0,0.6,0.0}{$\uparrow${7.06}}&\textcolor[rgb]{0.5,0.5,0.5}{$\downarrow${0.44}}&\textcolor[rgb]{0.0,0.6,0.0}{$\uparrow${2.85}}&\textcolor[rgb]{0.0,0.6,0.0}{$\uparrow${3.05}} \\
			
			\hline
			
			HUBERT-base&$\times$& 67.48$\pm$0.18 &58.05$\pm$0.51& 73.50$\pm$0.11 & 78.72$\pm$0.21 & 75.36$\pm$0.31 & 48.16$\pm$0.21 &67.23$\pm$0.11&51.83$\pm$0.19 \\
			HUBERT-base&$\surd$&65.58$\pm$0.67&62.88$\pm$0.15&73.33$\pm$0.12&78.27$\pm$0.55&77.75$\pm$0.27&46.94$\pm$0.20&67.36$\pm$0.50&51.21$\pm$0.10 \\
			HUBERT-base&$\Delta$&\textcolor[rgb]{0.5,0.5,0.5}{$\downarrow${1.90}}&\textcolor[rgb]{0.0,0.6,0.0}{$\uparrow${4.83}}&\textcolor[rgb]{0.5,0.5,0.5}{$\downarrow${0.17}}&\textcolor[rgb]{0.5,0.5,0.5}{$\downarrow${0.45}}&\textcolor[rgb]{0.0,0.6,0.0}{$\uparrow${2.39}}&\textcolor[rgb]{0.5,0.5,0.5}{$\downarrow${1.22}}&\textcolor[rgb]{0.0,0.6,0.0}{$\uparrow${0.13}}&\textcolor[rgb]{0.5,0.5,0.5}{$\downarrow${0.62}} \\

			\hline
			
			ALBERT-small&$\times$&46.43$\pm$0.29&74.16$\pm$0.63&79.50$\pm$0.24&73.27$\pm$0.23&72.30$\pm$0.22&56.42$\pm$0.12&61.14$\pm$0.14&47.43$\pm$0.23  \\
			ALBERT-small&$\surd$&44.52$\pm$1.28&79.75$\pm$0.39&81.92$\pm$0.17&72.28$\pm$0.31&72.02$\pm$0.11&56.64$\pm$0.10&62.73$\pm$0.19&47.42$\pm$0.45   \\
			ALBERT-small&$\Delta$&\textcolor[rgb]{0.5,0.5,0.5}{$\downarrow${1.91}}&\textcolor[rgb]{0.0,0.6,0.0}{$\uparrow${5.59}}&\textcolor[rgb]{0.0,0.6,0.0}{$\uparrow${2.42}}&\textcolor[rgb]{0.5,0.5,0.5}{$\downarrow${0.99}}&\textcolor[rgb]{0.5,0.5,0.5}{$\downarrow${0.28}}&\textcolor[rgb]{0.0,0.6,0.0}{$\uparrow${0.22}}&\textcolor[rgb]{0.0,0.6,0.0}{$\uparrow${1.59}}&\textcolor[rgb]{0.5,0.5,0.5}{$\downarrow${0.01}} \\
			
			\hline
			
			MacBERT-base&$\times$&56.38$\pm$0.32& 83.89$\pm$0.19 &84.03$\pm$0.25&80.24$\pm$0.26&79.73$\pm$0.10&57.04$\pm$0.14&65.41$\pm$0.04&50.18$\pm$0.08  \\
			MacBERT-base&$\surd$&55.91$\pm$0.23&84.47$\pm$0.16&85.28$\pm$0.16&80.22$\pm$0.13&79.61$\pm$0.14&58.34$\pm$0.31&66.30$\pm$0.07&51.09$\pm$0.08   \\
			MacBERT-base&$\Delta$&\textcolor[rgb]{0.5,0.5,0.5}{$\downarrow${0.47}}&\textcolor[rgb]{0.0,0.6,0.0}{$\uparrow${0.58}}&\textcolor[rgb]{0.0,0.6,0.0}{$\uparrow${1.25}}&\textcolor[rgb]{0.5,0.5,0.5}{$\downarrow${0.02}}&\textcolor[rgb]{0.5,0.5,0.5}{$\downarrow${0.12}}&\textcolor[rgb]{0.0,0.6,0.0}{$\uparrow${1.30}}&\textcolor[rgb]{0.0,0.6,0.0}{$\uparrow${0.89}}&\textcolor[rgb]{0.0,0.6,0.0}{$\uparrow${0.91}} \\
			
			\hline
			
		\end{tabular}
	\end{table*}

	\begin{table}[t]
		\centering
		\renewcommand\tabcolsep{3.6pt}
		\caption{Complexity comparison with and without fine-tuning. This table compares MACs, number of parameters, and training time per sample. For models without fine-tuning, we freeze the pre-trained feature encoders and only optimize shallow classifiers. Thus, we calculate the computational complexity on the classifiers. For models with fine-tuning, both pre-trained models and shallow classifiers are optimized. Thus, we calculate the computational complexity for all trainable parameters.}
		\label{Table17}
		\begin{tabular}{lc|c|c|c}
			\hline
			
			{Feature} & Finetune & {\begin{tabular}[c]{@{}c@{}}MACs \\ (G)\end{tabular}} & {\begin{tabular}[c]{@{}c@{}}\# Params \\ (M)\end{tabular}} & {\begin{tabular}[c]{@{}c@{}}Training Time \\ (ms)\end{tabular}} \\
			
			\hline
			
			VideoMAE-base&$\times$ & 0.02 & 0.48 & 0.03 \\
			VideoMAE-base&$\surd$ & 407.41 & 65.11 & 49.83 \\ 
			
			\hline
			
			CLIP-base&$\times$ & 0.01 & 0.38 & 0.02 \\
			CLIP-base&$\surd$ & 1047.98 & 87.91 & 140.78 \\
			
			\hline
			
			data2vec-base&$\times$ & 0.02 & 0.48 & 0.03 \\ 
			data2vec-base&$\surd$ &878.92 & 93.30 & 26.52 \\

			\hline
			
			HUBERT-base&$\times$ &0.02 & 0.48 & 0.03 \\
			HUBERT-base&$\surd$ &885.22 & 94.50 & 21.13 \\

			\hline
			
			ALBERT-small&$\times$ &0.01 &0.33 & 0.03 \\ 
			ALBERT-small&$\surd$ &21.55 & 2.06 & 1.12 \\ 
			
			\hline
			
			MacBERT-base&$\times$ &0.02 & 0.48 & 0.02 \\
			MacBERT-base&$\surd$ &58.50 &85.78 & 1.28 \\ 
			
			\hline
		\end{tabular}
	\end{table}

	\subsection{Necessity of Fine-tuning}
	The above experiments primarily utilize pre-trained models as feature extractors and then use shallow classifiers for emotion recognition. In other words, we freeze the weights of pre-trained models and only optimize the shallow classifiers. However, some recent studies have highlighted the effectiveness of jointly optimizing feature extractors and classifiers \cite{siriwardhana2020jointly}. Therefore, we further investigate the role of fine-tuning the pre-trained feature extractor.
	
	In Table \ref{Table16}, we select two representative feature encoders for each modality and study the effect of fine-tuning. Experimental results show that fine-tuning performs differently on distinct feature and dataset combinations. The reason may be twofold. On the one hand, fine-tuning improves performance when the encoder is incompatible with the emotion dataset. On the other hand, due to the small size of emotion datasets, fine-tuning increases the trainable parameters, causing the model to overfit training data and perform poorly on unseen data. Therefore, we should consider the trade-off between compatibility and overfitting.
	
	Table \ref{Table17} provides a comprehensive comparison of computational complexity with and without fine-tuning. We observe that fine-tuning increases the MACs and trainable parameters, resulting in longer training times. In practice, fine-tuning is required for each downstream dataset, which further increases the computational cost. In contrast, pre-training only needs to be conducted once, and we can freeze the pre-trained weights and optimize the shallow classifier, which is more computationally efficient. Moreover, according to the results in Table \ref{Table6}, the improvement of fine-tuning is smaller than that of pre-training. Therefore, for tasks with limited samples (such as emotion recognition), we recommend using pre-training instead of fine-tuning.

	\section{Conclusions}
	\label{sec6-conclusion}
	This paper builds MERBench, a unified evaluation benchmark for multimodal emotion recognition. We aim to reveal the role of some key techniques in this field and point out future research directions: (1) For visual encoders, we note the advantages of weakly-supervised and self-supervised models. To obtain better results, it is necessary to further train them on human-centric videos to reduce the domain gap; (2) Lexical encoders are language-sensitive. Although we can translate the source language into the target language, this translation process often causes the loss of emotion-related information; (3) Acoustic encoders can implicitly capture language-sensitive linguistic information from audio, resulting in their language sensitivity. To obtain more powerful and universal acoustic encoders, we suggest training acoustic encoders on expressive multilingual audio; (4) For multimodal fusion, the attention mechanism can achieve relatively good performance among all fusion algorithms. Meanwhile, not all datasets are suitable for multimodal fusion research, as some datasets primarily convey emotions through a single modality; (5) For cross-corpus settings, more powerful within-corpus models generally achieve better cross-corpus results. But to truly solve the cross-corpus problem, we should combine knowledge from different tasks and use explainable reasoning processes to increase label consistency across different datasets; (6) Lexical encoders are somewhat robust to missing punctuation, but they cannot handle missing cases in punctuation-rich samples. From another perspective, it will help us to recognize emotions if we can predict detailed punctuation from audio; (7) To improve noise robustness of acoustic encoders, we should increase the diversity of audio quality during training. At the same time, data augmentation can effectively handle noisy data, but an appropriate SNR should be selected; (8) Fine-tuning requires more computational costs but achieves less improvement than pre-training. Therefore, we recommend that follow-up researchers pay more attention to pre-training, especially pre-training on datasets that are compatible with downstream tasks. Furthermore, this paper introduces a new dataset MER2023, aiming to provide a benchmark dataset for research on multi-label learning, noise robustness, and semi-supervised learning.
	
	Beyond multimodal emotion recognition, we plan to incorporate more emotion-related tasks into our benchmark, such as stress, humor, sarcasm, and depression detection. Additionally, we aim to broaden the evaluation scope by including more features and multimodal fusion strategies. We hope that MERBench can provide guidance for developing robust and powerful emotion recognition systems.

	\section*{Acknowledgements}
	This work is supported by the National Natural Science Foundation of China (NSFC) (No.62201572, No.61831022, No.62276259, No.U21B2010), Beijing Municipal Science \& Technology Commission, Administrative Commission of Zhongguancun Science Park No.Z211100004821013, Open Research Projects of Zhejiang Lab (No. 2021KH0AB06), and CCF-Baidu Open Fund (No.OF2022025).

	%% add reference
	\bibliographystyle{IEEEtran}
	\bibliography{mybib}
	
	%% add author information
	\begin{IEEEbiography}[{\includegraphics[width=1.1in,height=1.25in,clip,keepaspectratio]{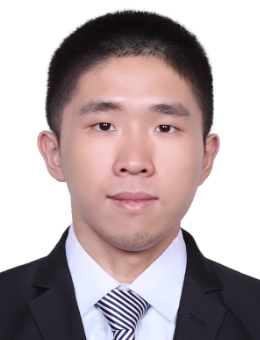}}]{Zheng Lian}
		received the B.S. degree from the Beijing University of Posts and Telecommunications, Beijing, China, in 2016, and the Ph.D degree from the Institute of Automation, Chinese Academy of Sciences, Beijing, China, in 2021. He is currently an Assistant Professor at State Key Laboratory of Multimodal Artificial Intelligence Systems, Institute of Automation, Chinese Academy of Sciences, Beijing, China. His current research interests include affective computing and multimodal learning.
	\end{IEEEbiography}
	\begin{IEEEbiography}[{\includegraphics[width=1.1in,height=1.25in,clip,keepaspectratio]{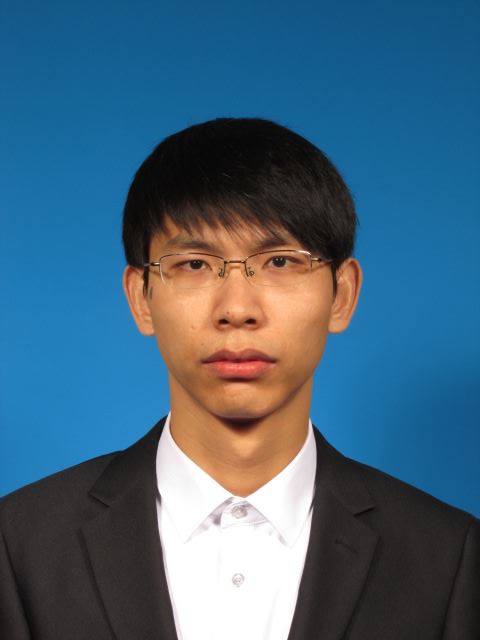}}]{Licai Sun}
		received the B.S. degree from Beijing Forestry University, Beijing, China, in 2016, and the M.S. degree from University of Chinese Academy of Sciences, Beijing, China, in 2019. He is currently working toward the Ph.D degree with the School of Artificial Intelligence, University of Chinese Academy of Sciences, Beijing, China. His current research interests include affective computing, deep learning, and multimodal representation learning.
	\end{IEEEbiography}
	\begin{IEEEbiography}[{\includegraphics[width=1.1in,height=1.25in,clip,keepaspectratio]{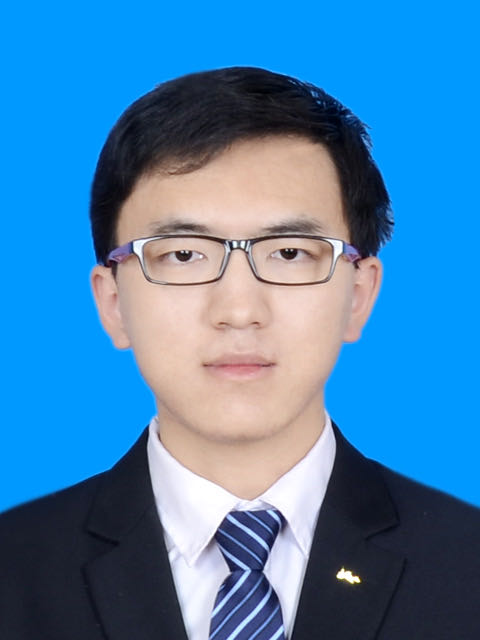}}]{Yong Ren}
		received the B.S. degree from the Department of Automation, Tsinghua University, Beijing, China, in 2020. He is currently working toward the Ph.D degree with the Institute of Automation, Chinese Academy of Sciences, Beijing, China. His current research interests include self-supervised learning and speech synthesis.
	\end{IEEEbiography}
	\begin{IEEEbiography}[{\includegraphics[width=1.1in,height=1.25in,clip,keepaspectratio]{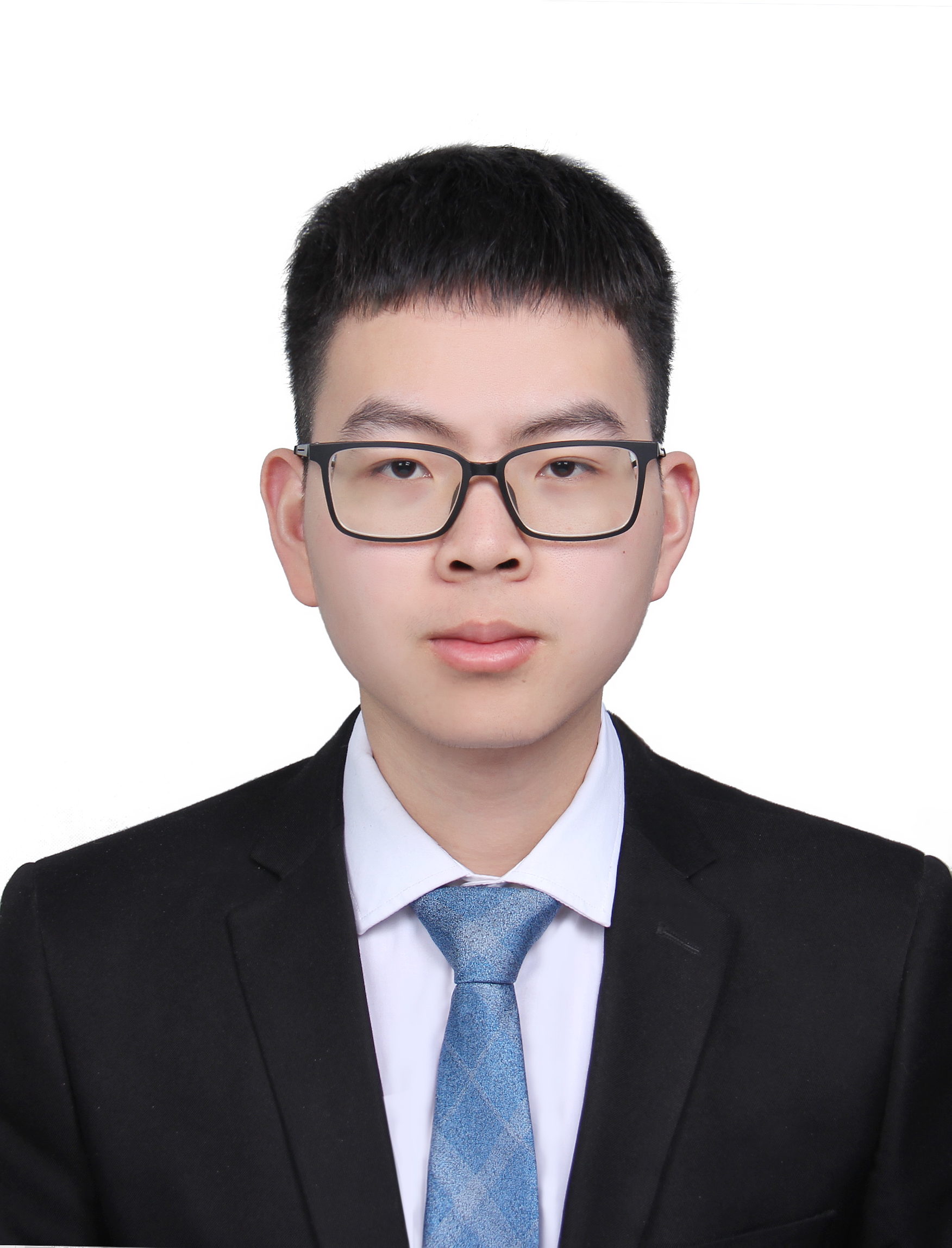}}]{Hao Gu}
		received the B.S. degree from Harbin Institute of Technology, China, in 2022. He is currently pursuing his M.S. degree at Institute of Automation, Chinese Academy of Sciences in Beijing, China. His current research interest include self-supervised learning and audio signal processing.
	\end{IEEEbiography}
	\begin{IEEEbiography}[{\includegraphics[width=1.1in,height=1.25in,clip,keepaspectratio]{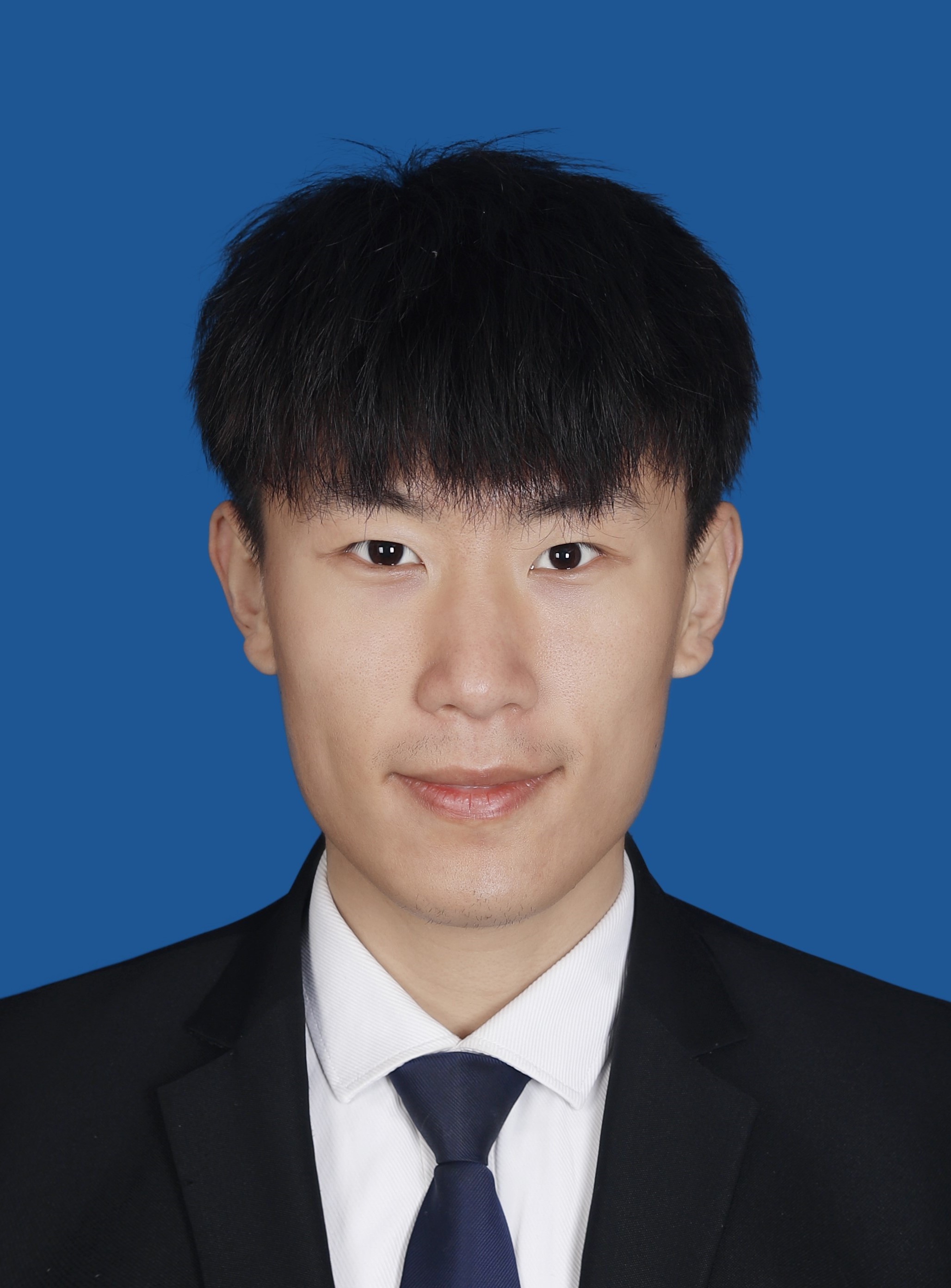}}]{Haiyang Sun}
		received the B.E. degree from Shandong University of Science and Technology, China, in 2021. He is currently working toward the M.S. degree with the Institute of Automation, China Academy of Sciences, Beijing, China. His current research interests include multimodal emotion recognition and neural architecture search.
	\end{IEEEbiography}
	\begin{IEEEbiography}[{\includegraphics[width=1.1in,height=1.25in,clip,keepaspectratio]{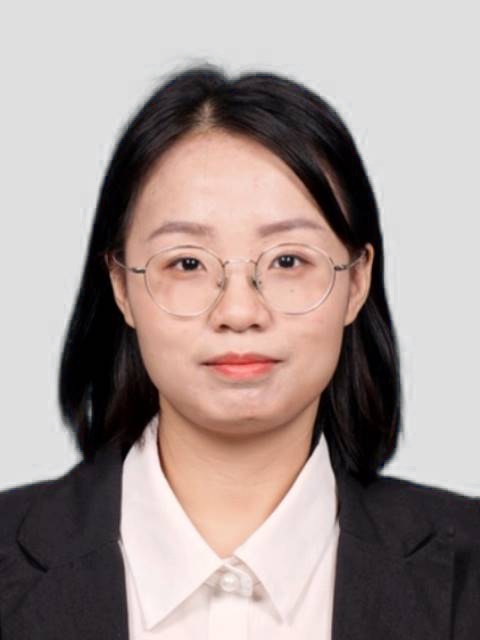}}]{Lan Chen}
		received the B.S. degree from the China University of Petroleum, Beijing, China, in 2016, and the Ph.D degree from the Institute of Automation, Chinese Academy of Sciences, Beijing, China, in 2022. Her current research interests include noisy label learning and image processing.
	\end{IEEEbiography}
	\begin{IEEEbiography}[{\includegraphics[width=1in,height=1.25in,clip]{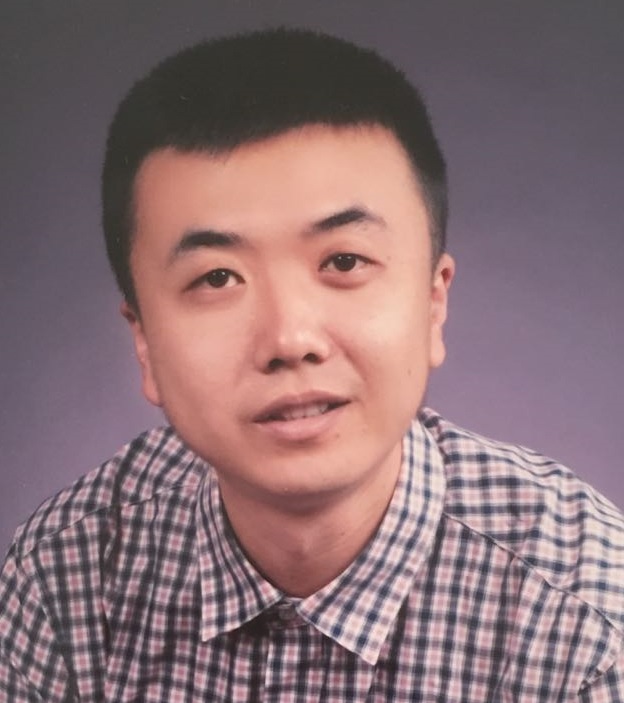}}]{Bin Liu}
		received his the B.S. degree and the M.S. degree from Beijing institute of technology, Beijing, China, in 2007 and 2009 respectively. He received Ph.D. degree from the Institute of Automation, Chinese Academy of Sciences, Beijing, China, in 2015. He is currently an Associate Professor in the State Key Laboratory of Multimodal Artificial Intelligence Systems, Institute of Automation, Chinese Academy of Sciences, Beijing, China. His current research interests include affective computing and signal processing.
	\end{IEEEbiography}
	\begin{IEEEbiography}[{\includegraphics[width=1.1in,height=1.25in,clip,keepaspectratio]{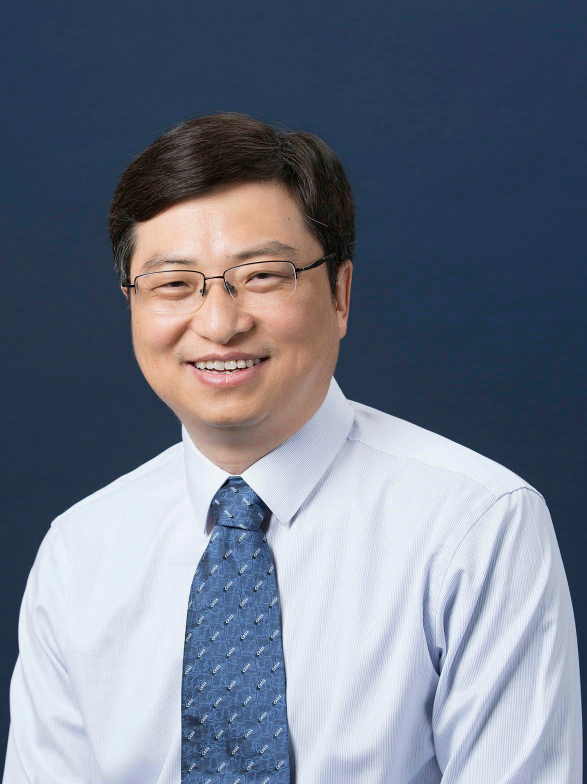}}]{Jianhua Tao}
		received the Ph.D. degree from Tsinghua University, Beijing, China, in 2001, and the M.S. degree from Nanjing University, Nanjing, China, in 1996. He is currently a Professor with Department of Automation, Tsinghua University, Beijing, China. He has authored or coauthored more than eighty papers on major journals and proceedings. His current research interests include speech recognition, speech synthesis and coding methods, human–computer interaction, multimedia information processing, and pattern recognition. He is the Chair or Program Committee Member for several major conferences, including ICPR, ACII, ICMI, ISCSLP, etc. He is also the Steering Committee Member for the IEEE Transactions on Affective Computing, an Associate Editor for Journal on Multimodal User Interface and International Journal on Synthetic Emotions, and the Deputy Editor-in-Chief for Chinese Journal of Phonetics.
	\end{IEEEbiography}

	\newpage

	\begin{table*}[t]
		\centering
		\renewcommand\tabcolsep{3.6pt}
		\caption{Model cards for different feature extractors. ``CH'', ``EN'', ``FR'', and ``MULTI'' are the abbreviations of Chinese, English, French, and multilingualism.}
		\label{Table18}
		\begin{tabular}{l|l|l}
			\hline
			Feature & Training Data (Language) & Link \\
			\hline
			\multicolumn{3}{c}{Visual Modality} \\
			\hline
			
			VideoMAE-base \cite{tong2022videomae} 	& Kinetics-400 & \textcolor[rgb]{0.93,0.0,0.47}{huggingface.co/MCG-NJU/videomae-base} \\
			VideoMAE-large \cite{tong2022videomae} 	& Kinetics-400 & \textcolor[rgb]{0.93,0.0,0.47}{huggingface.co/MCG-NJU/videomae-large} \\
			CLIP-base \cite{radford2021learning}		& WIT & \textcolor[rgb]{0.93,0.0,0.47}{huggingface.co/openai/clip-vit-base-patch32} \\
			CLIP-large \cite{radford2021learning}		& WIT & \textcolor[rgb]{0.93,0.0,0.47}{huggingface.co/openai/clip-vit-large-patch14} \\
			EVA-02-base \cite{fang2023eva}	& ImageNet-22k & \textcolor[rgb]{0.93,0.0,0.47}{huggingface.co/timm/eva02\_base\_patch14\_224.mim\_in22k} \\
			DINOv2-large \cite{oquab2023dinov2} 	& LVD-142M & \textcolor[rgb]{0.93,0.0,0.47}{huggingface.co/facebook/dinov2-large} \\
			
			\hline
			\multicolumn{3}{c}{Acoustic Modality} \\
			\hline
			
			data2vec-base \cite{baevski2022data2vec} 		& Librispeech (EN) & \textcolor[rgb]{0.93,0.0,0.47}{huggingface.co/facebook/data2vec-audio-base-960h} \\
			wav2vec-large \cite{schneider2019wav2vec}		& Librispeech (EN) & \textcolor[rgb]{0.93,0.0,0.47}{github.com/pytorch/fairseq/tree/master/examples/wav2vec} \\
			WavLM-base \cite{chen2022wavlm} 		& Librispeech (EN) & \textcolor[rgb]{0.93,0.0,0.47}{huggingface.co/microsoft/wavlm-base} \\
			WavLM-large \cite{chen2022wavlm}		& Mix (MULTI, mainly EN) & \textcolor[rgb]{0.93,0.0,0.47}{huggingface.co/microsoft/wavlm-large} \\
			Whisper-base \cite{radford2023robust}		& Internet (MULTI, mainly EN)& \textcolor[rgb]{0.93,0.0,0.47}{huggingface.co/openai/whisper-base} \\
			Whisper-large \cite{radford2023robust} 		& Internet (MULTI, mainly EN) & \textcolor[rgb]{0.93,0.0,0.47}{huggingface.co/openai/whisper-large-v2} \\
			wav2vec 2.0-base \cite{baevski2020wav2vec}	& WenetSpeech (CH) & \textcolor[rgb]{0.93,0.0,0.47}{huggingface.co/TencentGameMate/chinese-wav2vec2-base} \\
			wav2vec 2.0-large \cite{baevski2020wav2vec}	& WenetSpeech (CH) & \textcolor[rgb]{0.93,0.0,0.47}{huggingface.co/TencentGameMate/chinese-wav2vec2-large} \\
			HUBERT-base \cite{hsu2021hubert} 	& WenetSpeech (CH) & \textcolor[rgb]{0.93,0.0,0.47}{huggingface.co/TencentGameMate/chinese-hubert-base} \\
			HUBERT-large \cite{hsu2021hubert} 	& WenetSpeech (CH) & \textcolor[rgb]{0.93,0.0,0.47}{huggingface.co/TencentGameMate/chinese-hubert-large} \\
			
			\hline
			\multicolumn{3}{c}{Lexical Modality} \\
			\hline
			ALBERT-small \cite{lan2020albert}	& CLUECorpusSmall (CH) & \textcolor[rgb]{0.93,0.0,0.47}{huggingface.co/clue/albert\_chinese\_small} \\
			PERT-base \cite{cui2022pert}		& EXT Data (CH) & \textcolor[rgb]{0.93,0.0,0.47}{huggingface.co/hfl/chinese-pert-base} \\
			LERT-base \cite{cui2022lert} 		& EXT Data (CH) & \textcolor[rgb]{0.93,0.0,0.47}{huggingface.co/hfl/chinese-lert-base} \\
			XLNet-base \cite{yang2019xlnet} 		& EXT Data (CH) & \textcolor[rgb]{0.93,0.0,0.47}{huggingface.co/hfl/chinese-xlnet-base} \\
			MacBERT-base \cite{cui2020revisiting} 	& EXT Data (CH) & \textcolor[rgb]{0.93,0.0,0.47}{huggingface.co/hfl/chinese-macbert-base} \\
			RoBERTa-large \cite{liu2019roberta} 	& EXT Data (CH) & \textcolor[rgb]{0.93,0.0,0.47}{huggingface.co/hfl/chinese-roberta-wwm-ext-large} \\
			ELECTRA-base \cite{clark2020electra}	& EXT Data (CH) & \textcolor[rgb]{0.93,0.0,0.47}{huggingface.co/hfl/chinese-electra-180g-base-discriminator} \\
			Sentence-BERT \cite{reimers2019sentence}	& Mix (MULTI) & \textcolor[rgb]{0.93,0.0,0.47}{huggingface.co/sentence-transformers/paraphrase-multilingual-mpnet-base-v2} \\
			
			Llama-13B \cite{touvron2023llama} 		& Mix (MULTI, mainly EN) & \textcolor[rgb]{0.93,0.0,0.47}{huggingface.co/decapoda-research/llama-13b-hf} \\
			Llama2-13B \cite{touvron2023llama2} 	& Mix (MULTI, mainly EN) & \textcolor[rgb]{0.93,0.0,0.47}{huggingface.co/meta-llama/Llama-2-13b-hf} \\
			Vicuna-13B \cite{vicuna2023} 	& Mix (MULTI, mainly EN) & \textcolor[rgb]{0.93,0.0,0.47}{huggingface.co/CarperAI/stable-vicuna-13b-delta} \\
			StableLM-7B \cite{StableLMAlphaV2Models}	& Mix (MULTI, mainly EN) & \textcolor[rgb]{0.93,0.0,0.47}{huggingface.co/stabilityai/stablelm-base-alpha-7b-v2} \\
			OPT-13B \cite{zhang2022opt} 		& Mix (MULTI, mainly EN) & \textcolor[rgb]{0.93,0.0,0.47}{huggingface.co/facebook/opt-13b} \\
			BLOOM-7B \cite{workshop2022bloom} 		& ROOTS (MULTI) & \textcolor[rgb]{0.93,0.0,0.47}{huggingface.co/bigscience/bloom-7b1} \\
			Falcon-7B \cite{penedo2023refinedweb} 		& Mix (MULTI, mainly EN and FR) & \textcolor[rgb]{0.93,0.0,0.47}{huggingface.co/tiiuae/falcon-7b} \\
			MOSS-7B 		& Mix (MULTI, mainly EN and CH) & \textcolor[rgb]{0.93,0.0,0.47}{huggingface.co/fnlp/moss-base-7b} \\
			ChatGLM2-6B \cite{du2022glm}	& Mix (MULTI, mainly EN and CH) & \textcolor[rgb]{0.93,0.0,0.47}{huggingface.co/THUDM/chatglm2-6b} \\
			Baichuan-13B \cite{yang2023baichuan}	& Mix (MULTI, mainly EN and CH) & \textcolor[rgb]{0.93,0.0,0.47}{huggingface.co/baichuan-inc/Baichuan-13B-Base} \\
			
			\hline
			
		\end{tabular}
	\end{table*}

	\begin{table*}[t]
		\centering
		\caption{Cross-corpus results for IEMOCAP(four), MELD, and MER-MULTI. For multi-modalities, we use the high-performance set (CLIP-large, HUBERT-large, Baichuan-13B) and study different fusion algorithms.}
		\label{Table19}
		\begin{tabular}{c|l|l|ccc}
			\hline
			Modality & Feature/Fusion & Source & IEMOCAP(four) & MELD & MER-MULTI \\
			\hline \hline
			
			\multirow{6}{*}{Video}&\multirow{3}{*}{ResNet-MSCeleb} & IEMOCAP(four)&--&41.87$\pm$0.36&33.21$\pm$0.42 \\
			&& MELD&14.91$\pm$0.42&--&13.12$\pm$0.64 \\
			&& MER-MULTI&27.18$\pm$0.24&11.18$\pm$0.39&-- \\
			\cline{2-6} 
			&\multirow{3}{*}{CLIP-large} & IEMOCAP(four)&--&47.46$\pm$0.70&54.96$\pm$0.35 \\
			&& MELD&34.49$\pm$0.56&--&49.68$\pm$1.64 \\
			&& MER-MULTI&37.22$\pm$0.79&20.39$\pm$0.52&-- \\
			\hline \hline
			\multirow{6}{*}{Audio} & \multirow{3}{*}{VGGish} & IEMOCAP(four)&--&25.76$\pm$0.64&39.81$\pm$0.74 \\
			&& MELD&26.14$\pm$0.64&--&30.68$\pm$1.10 \\
			&& MER-MULTI&40.66$\pm$0.09&22.62$\pm$0.88&-- \\
			\cline{2-6} 
			&\multirow{3}{*}{HUBERT-large} & IEMOCAP(four)&--&19.16$\pm$4.03&60.95$\pm$0.76 \\
			&& MELD&46.62$\pm$0.98&--&43.62$\pm$1.12 \\
			&& MER-MULTI&50.69$\pm$0.16&47.83$\pm$0.19&-- \\
			\hline \hline
			\multirow{6}{*}{Text} & \multirow{3}{*}{OPT-13B} & IEMOCAP(four)&--&47.40$\pm$0.16&43.32$\pm$0.69 \\
			&& MELD&35.27$\pm$0.40&--&42.30$\pm$0.50 \\
			&& MER-MULTI&36.81$\pm$0.23&44.65$\pm$0.75&-- \\
			\cline{2-6} 
			&\multirow{3}{*}{Baichuan-13B} & IEMOCAP(four)&--&50.42$\pm$0.32&54.05$\pm$0.54 \\
			&& MELD&44.73$\pm$0.21&--&53.73$\pm$0.50 \\
			&& MER-MULTI&43.37$\pm$0.55&47.79$\pm$0.85&-- \\
			\hline \hline
			\multirow{15}{*}{Multi-modalities} & \multirow{3}{*}{MISA} & IEMOCAP(four)&--&47.90$\pm$2.41&69.60$\pm$0.93 \\
			&& MELD&47.09$\pm$0.91&--&57.83$\pm$0.94 \\
			&& MER-MULTI&58.00$\pm$0.36&48.04$\pm$2.50&-- \\
			\cline{2-6}
			&\multirow{3}{*}{MMIM} & IEMOCAP(four)&--&42.40$\pm$2.44&68.30$\pm$0.43 \\
			&& MELD&47.77$\pm$0.78&--&60.16$\pm$2.09 \\
			&& MER-MULTI&59.58$\pm$1.10&48.15$\pm$1.13&-- \\
			\cline{2-6}
			&\multirow{3}{*}{LMF} & IEMOCAP(four)&--&38.92$\pm$3.52&70.15$\pm$0.44 \\
			&& MELD&47.22$\pm$0.29&--&59.97$\pm$0.82 \\
			&& MER-MULTI&57.57$\pm$0.54&47.52$\pm$1.56&-- \\
			\cline{2-6}
			&\multirow{3}{*}{TFN} & IEMOCAP(four)&--&37.34$\pm$1.52&70.40$\pm$0.33 \\
			&& MELD&48.57$\pm$0.43&--&61.86$\pm$0.29 \\
			&& MER-MULTI&58.35$\pm$0.87&45.90$\pm$2.15&-- \\
			\cline{2-6}
			&\multirow{3}{*}{Attention} & IEMOCAP(four)&--&39.48$\pm$3.59&69.09$\pm$0.26 \\
			&& MELD&48.86$\pm$0.75&--&62.57$\pm$1.06 \\
			&& MER-MULTI&58.59$\pm$0.43&42.95$\pm$1.58&-- \\
			\hline
		\end{tabular}
	\end{table*}

	\begin{table*}[t]
		\centering
		\caption{Cross-corpus results for CMU-MOSI, CMU-MOSEI, CH-SIMS, and CH-SIMS v2. For multi-modalities, we use the high-performance set (CLIP-large, HUBERT-large, Baichuan-13B) and study different fusion algorithms.}
		\label{Table20}
		\begin{tabular}{c|l|l|cccc}
			\hline
			Modality& Feature/Fusion & Source & CMU-MOSI & CMU-MOSEI &CH-SIMS &CH-SIMS v2 \\
			\hline \hline
			
			\multirow{8}{*}{Video} & \multirow{4}{*}{ResNet-MSCeleb} & CMU-MOSI&--&56.64$\pm$0.28&52.68$\pm$0.83&50.72$\pm$0.21 \\
			&& CMU-MOSEI&51.11$\pm$0.34&--&49.73$\pm$0.44&53.62$\pm$1.69 \\
			&& CH-SIMS&51.72$\pm$0.39&58.02$\pm$0.14&--&64.31$\pm$0.42 \\
			&& CH-SIMS v2&47.43$\pm$0.25&57.97$\pm$0.14&71.09$\pm$0.48&-- \\
			\cline{2-7}
			&\multirow{4}{*}{CLIP-large} & CMU-MOSI&--&58.80$\pm$0.30&63.48$\pm$2.62&60.44$\pm$2.05 \\
			&& CMU-MOSEI&57.66$\pm$0.20&--&65.79$\pm$1.86&67.38$\pm$1.32 \\
			&& CH-SIMS&59.31$\pm$0.50&53.86$\pm$1.14&--&78.74$\pm$0.09 \\
			&& CH-SIMS v2&59.03$\pm$0.26&51.73$\pm$2.73&83.84$\pm$0.25&-- \\
			\hline \hline
			\multirow{8}{*}{Audio} & \multirow{4}{*}{VGGish} & CMU-MOSI&--&59.49$\pm$0.48&58.13$\pm$0.26&54.16$\pm$1.38 \\
			& & CMU-MOSEI&52.65$\pm$0.22&--&22.49$\pm$0.35&34.33$\pm$3.60 \\
			& & CH-SIMS&51.20$\pm$0.17&59.42$\pm$0.76&--&58.89$\pm$0.14 \\
			& & CH-SIMS v2&55.79$\pm$0.49&62.57$\pm$0.29&62.67$\pm$0.32&-- \\
			\cline{2-7}
			& \multirow{4}{*}{HUBERT-large} & CMU-MOSI&--&61.64$\pm$0.50&56.49$\pm$1.97&42.64$\pm$9.59\\
			& & CMU-MOSEI&61.17$\pm$0.50&--&59.25$\pm$1.83&35.59$\pm$1.15\\
			& & CH-SIMS&54.17$\pm$1.24&46.31$\pm$3.77&--&65.39$\pm$3.17 \\
			& & CH-SIMS v2&54.53$\pm$1.91&37.84$\pm$10.35&78.90$\pm$0.27&-- \\
			\hline \hline
			\multirow{8}{*}{Text} & \multirow{4}{*}{OPT-13B} & CMU-MOSI&--&68.11$\pm$0.55&66.63$\pm$0.81&65.62$\pm$0.45 \\
			& & CMU-MOSEI&75.85$\pm$0.13&--&68.02$\pm$0.15&68.57$\pm$0.36 \\
			& & CH-SIMS&68.34$\pm$1.29&67.11$\pm$0.53&--&68.75$\pm$0.34 \\
			& & CH-SIMS v2&67.13$\pm$0.86&68.61$\pm$0.17&72.02$\pm$0.31&-- \\
			\cline{2-7}
			& \multirow{4}{*}{Baichuan-13B} & CMU-MOSI&--&74.87$\pm$0.44&75.03$\pm$0.87&75.80$\pm$0.32 \\
			& & CMU-MOSEI&79.15$\pm$0.45&--&75.57$\pm$0.73&77.43$\pm$0.35 \\
			& & CH-SIMS&70.15$\pm$0.71&73.02$\pm$0.69&--&79.80$\pm$0.29 \\
			& & CH-SIMS v2&74.22$\pm$0.21&75.28$\pm$0.32&82.12$\pm$0.16&-- \\
			\hline \hline
			\multirow{20}{*}{Multi-modalities} & \multirow{4}{*}{MISA} & CMU-MOSI&--&75.39$\pm$1.35&75.43$\pm$1.14&75.72$\pm$0.31 \\
			& & CMU-MOSEI&79.14$\pm$0.17&--&79.17$\pm$1.31&79.31$\pm$1.05\\
			& & CH-SIMS&70.05$\pm$1.18&71.16$\pm$0.61&--&85.97$\pm$0.60\\
			& & CH-SIMS v2&72.40$\pm$0.80&74.44$\pm$0.62&89.72$\pm$0.45&-- \\
			\cline{2-7}
			& \multirow{4}{*}{MMIM} & CMU-MOSI&--&76.17$\pm$1.19&75.56$\pm$0.53&75.70$\pm$0.73\\
			& & CMU-MOSEI&79.03$\pm$0.41&--&81.64$\pm$1.32&78.54$\pm$1.93 \\
			& & CH-SIMS&70.86$\pm$1.32&72.04$\pm$0.79&--&85.07$\pm$0.55 \\
			& & CH-SIMS v2&72.17$\pm$0.86&74.32$\pm$0.77&89.73$\pm$1.08&-- \\
			\cline{2-7}
			& \multirow{4}{*}{LMF} & CMU-MOSI&--&76.07$\pm$0.71&75.86$\pm$0.78&75.52$\pm$0.49 \\
			& & CMU-MOSEI&80.07$\pm$0.16&--&80.16$\pm$0.62&81.00$\pm$1.03 \\
			& & CH-SIMS&68.55$\pm$1.30&72.20$\pm$0.45&--&86.91$\pm$0.51 \\
			& & CH-SIMS v2&72.53$\pm$1.65&74.64$\pm$0.41&90.07$\pm$0.43&-- \\
			\cline{2-7}
			& \multirow{4}{*}{TFN} & CMU-MOSI&--&74.09$\pm$1.25&76.55$\pm$0.83&75.74$\pm$0.56 \\
			& & CMU-MOSEI&80.19$\pm$0.55&--&81.44$\pm$0.76&80.33$\pm$0.31 \\
			& & CH-SIMS&70.60$\pm$0.76&71.78$\pm$0.22&--&86.47$\pm$0.30 \\
			& & CH-SIMS v2&73.85$\pm$1.17&75.19$\pm$0.90&90.84$\pm$0.56&-- \\
			\cline{2-7}
			& \multirow{4}{*}{Attention} & CMU-MOSI&--&74.13$\pm$1.72&75.11$\pm$0.34&75.41$\pm$0.20 \\
			& & CMU-MOSEI&79.84$\pm$0.28&--&80.56$\pm$0.39&81.18$\pm$1.03 \\
			& & CH-SIMS&70.85$\pm$0.45&70.94$\pm$1.12&--&86.66$\pm$0.94 \\
			& & CH-SIMS v2&73.77$\pm$0.31&74.91$\pm$0.33&91.13$\pm$0.45&-- \\
			\hline
		\end{tabular}
	\end{table*}

\end{document}